\documentclass{aa}

\usepackage[varg]{txfonts} 
\usepackage{xcolor} 
\usepackage{natbib}
\bibpunct{(}{)}{;}{a}{}{,} 
\usepackage{amssymb, amsmath}
\usepackage{wasysym}
\usepackage{hyperref}
\usepackage{lipsum}
\usepackage{float} 

\usepackage{siunitx}
\DeclareSIUnit\erg{erg}
\DeclareSIUnit\year{yr}
\DeclareSIUnit\au{au}

\newcommand*{\fullref}[1]{\hyperref[{#1}]{\autoref*{#1} \nameref*{#1}}}

\defcitealias{2013A&AAlibert}{A13}

\defcitealias{Emsenhuber2020A}{Paper I}
\def\paperone/{\citetalias{Emsenhuber2020A}}

\defcitealias{Emsenhuber2020B}{Paper II}
\def\papertwo/{\citetalias{Emsenhuber2020B}}

\defcitealias{Weiss2018}{W18}
\def\W18{\citetalias{Weiss2018}}

\def\kobe/{\texttt{KOBE}}
\def\kobeshadows/{\texttt{\kobe/-Shadows}}
\def\kobetransits/{\texttt{\kobe/-Transits}}
\def\kobevetter/{\texttt{\kobe/-Vetter}}
\def\kmps/{\editbold{KMPS}}

\newcommand{\editmajor}[1]{#1}
\newcommand{\editminor}[1]{#1}
\newcommand{\editbold}[1]{#1}
\newcommand{\editle}[1]{#1}

\def\mearth{M_\oplus}
\def\rearth{R_\oplus}
\def\rjupiter{R_\text{J}}
\def\msun{M_\odot}
\def\rsun{R_\odot}

\def\mstar{M_\star}
\def\lstar{L_\star}
\def\rstar{R_\star}
\def\tstar{T_\star}

\def\mcore{M_{\rm core}}
\def\mdotcore{\dot{M}_{\rm core}}

\def\menv{M_\mathrm{env}}

\def\mplanet{M_\mathrm{planet}}

\def\rplanet{R_\mathrm{planet}}

\def\rhill{R_\mathrm{H}}

\def\sigmag{\Sigma_\mathrm{g}}
\def\sigmadotg{\dot{\Sigma}_\mathrm{g}}
\def\sigmadotgplan{\dot{\Sigma}_\mathrm{g,planet}}
\def\sigmadotgphoto{\dot{\Sigma}_\mathrm{g,photo}}

\def\betag{\beta_\mathrm{g}}

\def\sigmanorm{\Sigma_{g,0}}
\def\rin{r_\mathrm{in}}
\def\rcutg{r_\mathrm{cut,g}}

\def\mwind{\dot{M}_\mathrm{wind}}
\def\mgasdisk{M_\mathrm{g}}

\def\sigmasol{\Sigma_\mathrm{s}}
\def\sigmas0{\Sigma_\mathrm{s,0}}
\def\rcuts{r_\mathrm{cut,s}}
\def\betas{\beta_\mathrm{s}}
\def\msoliddisk{M_\mathrm{s}}

\def\fpg{f_{\rm D/G}}
\def\fpgsun{f_\mathrm{D/G,\odot}}

\def\vrel{v_\mathrm{rel}}
\def\rplan{R_\mathrm{plan}}
\def\rhoplan{\rho_\mathrm{plan}}

\def\sigmamean{\bar{\Sigma}_\mathrm{s}}

\def\pcoll{p_\mathrm{coll}}

\def\rplanetstardistance{r_\mathrm{planet}}

\def\cdppeff{\mathrm{CDPP}_\mathrm{eff}}
\def\ttrial{t_\mathrm{trial}}
\def\tdur{t_\mathrm{dur}}
\def\tkepler{t_\mathrm{kepler}}
\def\ntransit{n_\mathrm{tra}}
\def\periodictce/{\text{\textit{p}TCE}}

\def\rin{R_\mathrm{inner}}
\def\rout{R_\mathrm{outer}}
\def\pin{P_\mathrm{inner}}
\def\pout{P_\mathrm{outer}}

\hypersetup{pdfauthor={L. Mishra}, 
        pdftitle=NGPPS V: Peas in a pod, 
        breaklinks=true, 
        colorlinks=true, 
        urlcolor=blue, 
        linkcolor=blue,  
        citecolor=blue, 
        bookmarksopen=true,}
\usepackage{orcidlink}

\title{The New Generation Planetary Population Synthesis (NGPPS) \\VI. Introducing \kobe/: Kepler Observes Bern Exoplanets \thanks{\kobe/ is available at: \url{https://github.com/exomishra/kobe}.}}
\subtitle{Theoretical perspectives on the architecture of planetary systems: Peas in a pod}

\authorrunning{Lokesh Mishra et al.}
\titlerunning{Introducing KOBE -- Theoretical perspectives on the architecture of planetary systems: Peas in a pod}

\author{Lokesh Mishra\inst{\ref{unibe},\ref{unige}}\thanks{Correspondence: Lokesh Mishra  (\hyperref{mailto:exomishra@gmail.com}{}{}{exomishra@gmail.com})}\orcidlink{0000-0002-1256-7261}
        \and Yann Alibert\inst{\ref{unibe}}\orcidlink{0000-0002-4644-8818}
        \and Adrien Leleu\inst{\ref{unige}}\orcidlink{0000-0003-2051-7974}
        \and Alexandre Emsenhuber\inst{\ref{unibe},\ref{uofa}} \orcidlink{0000-0002-8811-1914}
        \and Christoph Mordasini\inst{ \ref{unibe}}\orcidlink{0000-0002-1013-2811}
        \and \\ Remo Burn\inst{\ref{unibe}}\orcidlink{0000-0002-9020-7309}
        \and St\'ephane Udry\inst{\ref{unige}} \orcidlink{0000-0001-7576-6236}
        \and Willy Benz {\inst{\ref{unibe}}}\orcidlink{0000-0001-7896-6479}
}

\institute{
        Institute of Physics, University of Bern, Gesellschaftsstrasse 6, 3012 Bern, Switzerland\label{unibe}
        \and
        Geneva Observatory, University of Geneva, Chemin Pegasi 51b, 1290 Versoix, Switzerland\label{unige}       
        \and
        Lunar and Planetary Laboratory, University of Arizona, 1629 E. University Blvd., Tucson, AZ 85721, USA\label{uofa}
}

\date{Received 9 March 2021;}

\abstract
{Observations of exoplanets indicate the existence of several correlations in the architecture of planetary systems. Exoplanets within a system tend to be of similar size and mass, evenly spaced, and are often ordered in size and mass. Small planets are frequently packed in tight configurations, while large planets often have wider orbital spacing. Together, these correlations are called the peas in a pod trends in the architecture of planetary systems.} 
{In this paper these trends are investigated in theoretically simulated planetary systems and compared with observations. Whether these correlations emerge from astrophysical processes or the detection biases of the transit method is examined.}
{Synthetic planetary system were simulated using the Generation III Bern Model. \kobe/, a new computer code, simulates the geometrical limitations of the transit method and applies the detection biases and completeness of the Kepler survey. This allows simulated planetary systems to be compared  with observations.}
{The architecture of synthetic planetary systems, observed via \kobe/, show the peas in a pod trends in good agreement with observations. These correlations are also present in the theoretical underlying population, from the Bern Model, indicating that these trends are probably of astrophysical origin.}
{The physical processes involved in planet formation are responsible for the emergence of evenly spaced planets with similar sizes and masses. The size--mass similarity trends are primordial and originate from the oligarchic growth of protoplanetary embryos and the uniform growth of planets at early times. Later stages in planet formation allows planets within a system to grow at different rates, thereby decreasing these correlations. The spacing and packing correlations are absent at early times and arise from dynamical interactions.} 


\begin{document}

        \maketitle
        
        
        \section{Introduction}
        \label{sec:introduction}
                
        Since \cite{Mayor1995}  discovered  51 Pegasi b, the first planet found to orbit another main-sequence star, technological advancements have \editle{engendered} the possibility to address the  question of how common Earth-like worlds are in the habitable zone of Sun-like stars. Addressing this question, the NASA space telescope, the \textit{Kepler} \editle{mission},  measured the brightness of 198,709 stars for $\sim 3.5$ years with a fixed field-of-view pointing towards the Milky Way Galactic plane (near the Cygnus-Lyra constellation) \citep{Borucki2016, Twicken2016}. As a planet passes in front of a star, it can result in a measurable and periodic reduction in the flux coming from this star. Utilizing this method, called the transit method, Kepler discovered and characterized thousands of exoplanets \citep{Borucki1984,Borucki2010, Borucki2011,Thompson2018}. With over 4,000 \editminor{planetary candidates} around over 3,000 stars\footnote{Based on a September 2020 query of the Extrasolar Planets Encyclopaedia \citep{Schneider2011}.}, observations have revealed a staggering diversity in the nature of exoplanets          \citep{Armstrong2020, Hoeijmakers2018,Winn2018, Kreidberg2014, Sing2016, Santerne2019, Espinoza2020, Demory2016, Evans2016, Udry2007}. 
        The rich diversity observed in exoplanets, fortuitously, also extends to the architecture of multi-planetary systems \citep{Lissauer2011,Fabrycky2014,Winn2015}.
        
        The arrangement of multiple planets and the collective distribution of their physical properties around host star(s) characterizes the architecture of a planetary system. This implies that the architecture of any planetary system is an outcome of all the physical processes that lead the system to its present state.  The architecture of a planetary system may reflect several simultaneous processes: (a) the specific formation pathways of individual planets, (b) secular and/or chaotic dynamical interactions, (c) configurations that are stable over \editminor{billions of years}, (d) initial conditions coming from the star and the protoplanetary disks,  (e) the astrophysical environment surrounding the star-forming region.  Specifically, the extent to which planet formation is stochastic remains unknown. Explaining the wide diversity observed in the system architecture remains an open problem \citep{Winn2015}. It is possible that planet formation is dominated by the same physical processes, but the large \editminor{diversity in} initial conditions leads to a wide variety of exoplanets and system architectures (\cite{Benz2014, Mordasini2018}). 
        
        \editminor{Amidst the observed diversity in the architecture of exoplanetary systems several trends have emerged \citep{Ciardi2013}.} \editminor{One such trend} is  called `peas in a pod', which is the subject of this paper. 
        Empirical trends in the system architecture serve two key purposes. Firstly, these trends provide hints about underlying physical processes. Thus, these trends posit additional constraints on theory. Secondly, as the understanding of exoplanetary astrophysics matures, reproduction of these trends in numerical calculations becomes a crucial testing ground amongst competing models. Perhaps several of the observed correlations in planetary system architecture are unifiable, facilitating simpler physical explanations to emerge.
             
        The California-Kepler Survey (CKS) improved the characteristics of 1,305 planet-hosting stars found by Kepler \citep{Petigura2017} leading to an improvement in the characteristics of 2,025 planets transiting these stars \citep{Johnson2017}. Analysing 355 multi-planetary systems from the CKS dataset (CKS-Multis or CKSM), hosting 909 planets, \cite{Weiss2018} (hereafter \W18) reported several correlations in the properties of adjacent planets \editle{akin to} peas in a pod. They find that adjacent planets in a system tend to be similar in size, $\rout/\rin \approx 1$\footnote{Lowercase $r$ is used for radial distance of an object (e.g. distance from the star), while capital $R$ is used to denote the radius of the object itself.}.
        This trend was already suggested by \cite{Lissauer2011} based on 
        the first four months of Kepler's observations. In addition, \W18 report that $\sim 65 \%$ of adjacent pairs in their dataset are size-ordered,  the outer planet \editminor{being} larger than the inner planet. This trend was also hinted at by \cite{Ciardi2013}.     For planetary systems with three or more planets, \W18 find that the orbital spacing ($P_\mathrm{outer}/P_\mathrm{inner}$) of the first pair of planets is similar to the orbital spacing of the next pair of planets. They also report a correlation in the packing of planets within a system: smaller planets often have \editminor{smaller} orbital spacing, while larger planets tend to have larger orbital spacing.
        
        Using transit timing variations, \cite{Hadden2017} inferred the masses and eccentricity of 145 planets hosted in 55 Kepler planetary systems. Studying this dataset, \cite{Millholland2017} show that planets within a system tend to have similar masses and are often ordered in mass, the outer planet being more massive than the inner planet. Additionally, \cite{Wang2017} also reports similarity in mass in 29 systems detected by the radial-velocity method. 
        
        Pertaining to these trends, two kinds of studies have emerged. While some studies have explored theoretical aspects to better explain the observations \citep[e.g.][]{Mulders2020}, \editbold{other authors question the evidence} for peas in a pod in their analysis \citep{Zhu2019, Murchikova2020}. 
        \editmajor{For size-ordering, \cite{Kipping2018} investigated whether traces of initial conditions of planet formation are removed by dynamical evolution. A tally score $T = \Sigma_\mathrm{pairs} t_i$ is defined that tracks whether the radius of an outer adjacent planet is more ($t_i = +1$) or less ($t_i = -1$) than its inner neighbour. The number of different ways
        for a planetary system to obtain the same tally score $T$, is interpreted as the entropy of the system.}\footnote{\editmajor{Alternative definitions of entropy are also explored by incorporating a memory-like term to include size ordering from one adjacent pair of planets to the next adjacent pair of planets.}}   
        \editbold{He finds} that  Kepler systems have lower entropy than expected from randomly constructed systems, implying that the initial conditions for Kepler systems and their formation pathways could be potentially inferred. \cite{Adams2019} finds that energy optimization in planetary pairs, assuming fixed total angular momentum and total mass for a given orbital spacing, leads to planets in circular orbits with no mutual inclination and nearly equal masses. However, when the total mass in the planetary pair exceeds a threshold ($\sim 40 \ M_\oplus$ for $a \sim 0.1$ AU), energy optimization can cause one planet to gain most of the mass \citep{Adams2020}. \cite{Xu2018} suggest that ejection of small planets, caused by dynamical interactions, provides a possible explanation for the observed correlations. \cite{MacDonald2020} find that in situ formation of $1-4 \ R_\oplus$  planets, while varying the amount of solids present in the inner region of the protoplanetary disk, can lead to systems with similarly sized planets with correlated orbital spacings. \editbold{\cite{Chevance2021} examine the effect of stellar clustering on these architecture trends and find that the peas in a pod correlations are persistent in systems irrespective of the influence from stellar neighbours.} 
        
        Although highly successful in discovering exoplanets, the transit method suffers from inherent geometric limitations (only planets whose orbits are, serendipitously, edge-on can transit) and detection biases (large planets close to a small quiet star are \editle{strongly} favoured). This strongly limits our knowledge of the underlying `{ground-truth}' distribution of exoplanets \citep{Borucki1984, Barnes2007, Kipping2016}\footnote{For example, \cite{Sandford2019} estimate that around $\simeq$ 2,400 more exoplanets reside in 1537 planet hosting FGK stars observed by Kepler.}. \editbold{It is therefore unclear whether the peas in a pod trend is arising from an incomplete knowledge convolved with the limitations of the transit method or if this trend reflects an actual property of nature.}
        
        In \W18 (and other similar studies) the origins of the peas in a pod trend was investigated using null hypothesis bootstrap tests. The basic idea behind these tests is that if detection biases of the transit method are responsible for the observed trends, then these correlations should also be present in a mock exoplanetary population that does not possess these trends, inherently through a null hypothesis, but suffers from the same detection biases. For example, the null hypothesis used for testing the  size-similarity trend was that the size of a planet is random and independent of the size of its neighbour \citep[\W18, ][]{Weiss2019}. \W18 performed $1~000$ bootstrap trials in which the detection biases of Kepler was applied to mock populations satisfying the null hypothesis stated above. They found that none of their bootstrap trials  lead to a population that showed the size-similarity correlation. Since the detection biases convolved with the null hypothesis did not result in size similarity, they concluded that this trend is not due to detection biases and must be of astrophysical origin.
        
        \editbold{\cite{Zhu2019}  extensively challenged the method used by \W18 for constructing the mock exoplanetary population. For the observed CKSM planets, he argues that since the radius distribution depends on the stellar noise (see Fig. 2 (left) in \cite{Zhu2019}), resampling the observed radius distribution to construct a mock population (as done in \W18) is not sufficient. Instead, he proposes that mock populations should be created by resampling the transit signal-to-noise ratio (S/N, defined in eq.  \ref{eq:transitsnr})\footnote{\editbold{\cite{Zhu2019} point out that resampling transit S/N is a `shortcut' for a `more appropriate way'. This appropriate way includes an intrinsic planetary distribution with a full-forward model of the transit detection bias. In this paper the Bern Model provides an intrinsic, albeit synthetic, population of exoplanets, and \kobe/ models the Kepler transit survey.}}. In doing so, he finds that his mock populations convolved with the detection biases of the transit method show size-similarity correlation. He interprets that the size-similarity correlation is explained by detection biases. However, \cite{Weiss2019} show that the mock population created by Zhu do not satisfy the null-hypothesis (see Figs. 2 and 3 of \cite{Weiss2019}) and are therefore unsuitable for bootstrap testing. \cite{Murchikova2020} argue that the \W18 bootstrap procedure as well as an improved `balanced bootstrap' procedure does not reveal the correct statistics for the CKSM sample. They find that a scenario in which the planet sizes depend on system properties and planet locations can give rise to the size-similarity correlation. They suggest that size similarity in exoplanetary neighbours can arise even when a planet's size is not influenced by its neighbour. However, \editminor{using a parametrized model of planetary systems, \cite{He2019} find that clusters} of similarly sized and spaced planets provide a better fit to Kepler observations. }
          
        \editle{Now we describe how in this paper theory meets observations.} 
        Planet formation begins in protoplanetary disks around young pre-main-sequence stars. The physical environment in and around these disks sets the initial condition for planet formation. The theory of planet formation and evolution describes the physical processes that  link these initial conditions to the resultant planets. 
        In Sect. \ref{sec:bernmodel} the planet formation model used in this work, the Generation III \textit{Bern Model}, is described. Next, in order to compare theory with observations at the population level, theoretical planetary populations are required\editminor{\footnote{A set of hundreds or thousands of individual planetary systems are referred to here as a population.}}. 
        In Sect. \ref{sec:ngpps} the \textit{New Generation Planetary Population Synthesis} (NGPPS) used in this work is presented.
        
        Since nature's underlying exoplanetary population is only partially accessible via the transit method, observations cannot be compared directly with the output of population synthesis. To facilitate this comparison, the detection biases of an observational survey must be placed on the synthetic populations. In this work the detection biases of the Kepler survey are applied on the output of NGPPS by simulating the relevant parts of the Kepler pipeline and Kepler's Robovetter \citep{Twicken2016,Twicken2018,Thompson2018}. Section \ref{sec:kobe} introduces  a new computer code, \texttt{Kepler Observes Bern Exoplanets} (\kobe/), which mimics the Kepler transit survey for synthetic planetary systems. \kobe/ computes populations of \editle{synthetic planets which survive the transit detection bias}, like Kepler's planetary candidates. The theory can now be compared with Kepler's findings, as is done in this and forthcoming papers. 
        
        \kobe/ multi-planetary systems (\kmps/) are introduced and compared with the observations in Sect. \ref{sec:kmps}. In Sect. \ref{sec:peasinapod} the peas in a pod trends are formally stated and the architecture of synthetic systems is examined.  In Sect. \ref{sec:roleofbiases} the role of adding detection biases is elucidated. Theoretical scenarios that lead to the peas in a pod trends are discussed in Sect. \ref{sec:theoreticalscenarois}. Section \ref{sec:conclusionandfuture} concludes this paper by summarizing the main results and discussing possible explorations in future works. Appendix \ref{sec:mutual_hill} explains the correlation between the average size of planets and their mutual separation. 

        The aim of this paper is threefold: investigating the peas in \editminor{a} pod trends in the architecture of theoretical planetary systems and compare them with observations, understanding the role of geometrical limitations and detection biases on the observed trends, and exploring the physical mechanisms that could explain the peas in a pod correlations. \editbold{We find that the peas in a pod trends are present in   the theoretically simulated planets (from the Bern Model) and in  the planets that are theoretically observed (via KOBE). The strength of the architecture trends found in CKSM observations (\W18) and \kmps/ are similar. While the limitations and detection biases of the transit method influence the observed architecture, they do not explain the trends. Our work suggests that if nature's `true' exoplanetary population shows the peas in a pod trends, then observation biases from a transit survey can lead to the architecture trends found by \W18. In this manner our work adds support to the hypothesis of an astrophysical origin for the peas in a pod trends.}
        

\section{Generation III \textit{Bern Model}}
\label{sec:bernmodel}
        \begin{figure*}
                \centering
                \includegraphics[width=17cm]{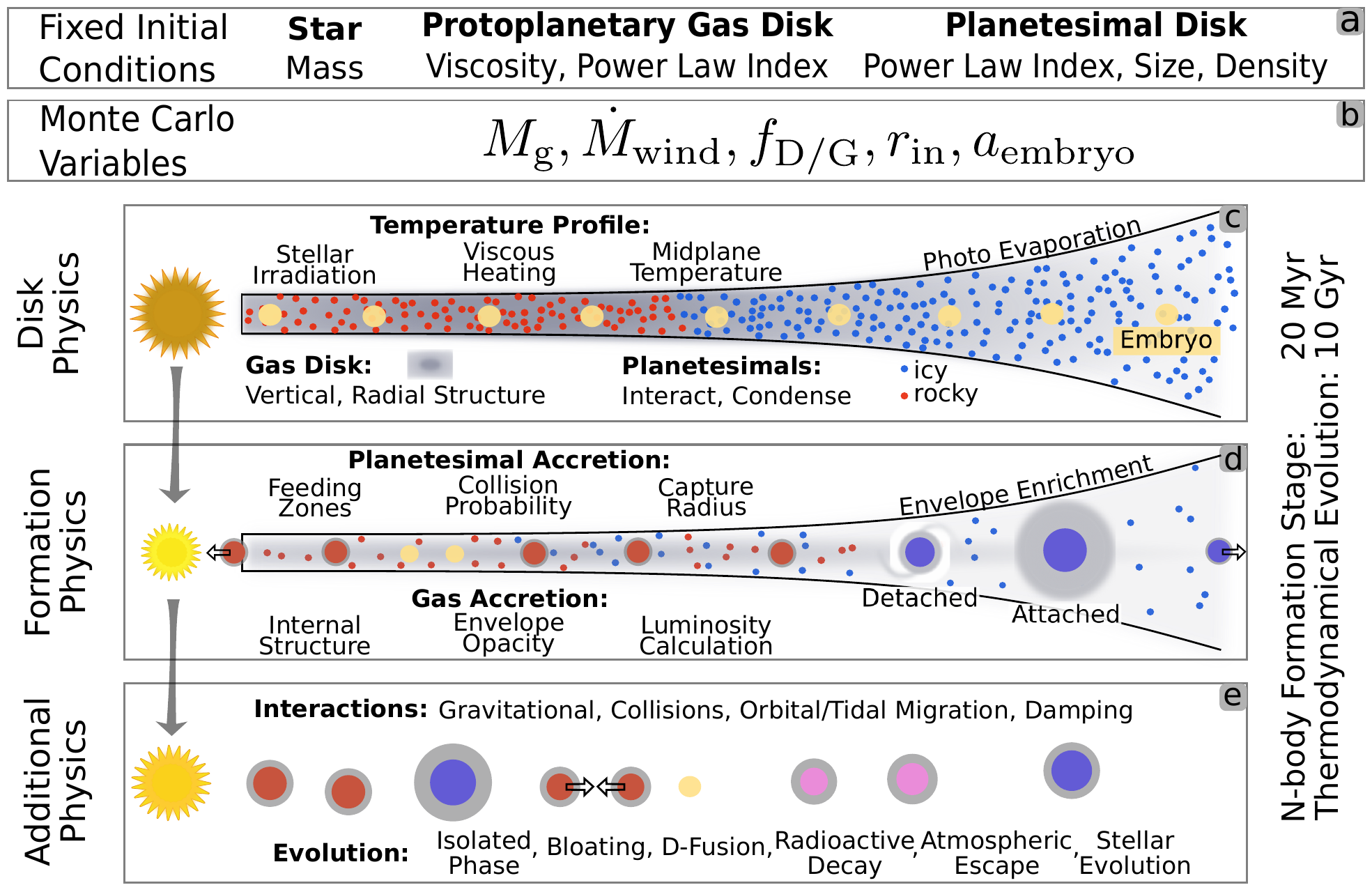}
                \caption{\textit{The Bern Model}: Schematic diagram (not drawn to scale) illustrating the breadth of physical processes incorporated in the Generation III Bern Model of Planet Formation and Evolution. Panels (a) and (b) show the fixed and varying initial conditions, respectively. The physical processes  relevant to the protoplanetary disk are indicated in panel (c). \editle{It represents} a snapshot of the starting point of the model: several protoplanetary embryos are embedded in a disk of gas and planetesimals. The processes that govern planet formation and evolution are displayed in panel (d). In panel (e) additional physics incorporated in the model is shown. The arrows indicate the evolution of a fixed mass star. Most of the depicted physical processes occur simultaneously and not all processes are shown. See text for summary and \paperone/ for details.}
                \label{fig:bernmodel}
        \end{figure*}


        The \textit{Bern Model} is a global model of planet formation and evolution based on the core-accretion paradigm \citep [see] []{Pollack1996, Alibert2004, Alibert2005}\footnote{Readers who are well versed \editle{with} the Bern Model and its updates \editle{may} skip to Sect. \ref{sec:kobe} where \kobe/ is introduced. Other readers \editle{may} use this introductory section as a starting point for key concepts and relevant literature for planet formation in general, and the Bern Model in particular.}\footnote{A global model, which comprises of theoretical models for individual physical processes linked together coherently, calculates the final planetary system based on a set of initial conditions.}. \editmajor{From its initial inception in \cite{Alibert2005}, the model has undergone several updates {\citep{Mordasini2008,Mordasini2009,Alibert2011,Mordasini2012(models),Mordasini2012(MR),Mordasini2012(correlations),Alibert2013,Fortier2013,Marboeuf2014, Thiabaud2014,Dittkrist2014,Mordasini2015,Mordasini2017}}}. The generation III \textit{Bern Model} used in this work is presented in detail in \cite{Emsenhuber2020A, Emsenhuber2020B} (hereafter \paperone/ and \papertwo/, respectively) \citep[for reviews, see][]{Benz2014, Mordasini2018}. For completeness, a summary of the major physical processes included in the model is given below. Figure \ref{fig:bernmodel} shows a schematic diagram of the key physical processes included in the model. 
        
        \subsection{Before planet formation begins}
        The gravitational collapse of cold diffuse molecular clouds leads to the formation of (multiple) stars and circumstellar disks. Conservation of angular momentum implies that gravitationally bound material will flatten into a protoplanetary disk. Dust and gas from the cloud falls onto the protostar and its disk for about $\SI{e5}{\year}$ \citep{Nakamoto1994, Baillie2019}. The surrounding envelope is cleared by this time,  either due to star--disk accretion or dispersion via jets and outflows,  and the thermal emission from this system resembles that of a classical T Tauri Star \citep{Tychoniec2018}. Although debated, dust grains (\SI{e-6}{\meter}) grow quickly by sticky collisions or gravitational instabilities into \editminor{kilometre-sized} planetesimals \citep{Youdin2008,Johansen2007,Williams2011}. Planetesimals, interacting gravitationally as a swarm, undergo rapid runaway growth wherein larger planetesimals grow faster than smaller ones \citep{Kokubo1998}. When runaway growth is no longer possible, either due to significant velocity disruptions or lack of material to accrete, oligarchic growth begins. The resulting lunar mass bodies, called protoplanetary embryos, emerge rapidly $\sim$ \SI{e4}{\year} \citep{Kokubo2002}. This stage marks the starting point for the Bern Model, and is sketched in panel (c) of Fig. \ref{fig:bernmodel}. 
        
        The model studies the \editminor{subsequent} growth of protoplanetary embryos that are embedded in a disk of planetesimals and gas. Multiple physical processes, interactions, and phenomena  simultaneously occur in this star-disk-embryo system, resulting in many kinds of planets and system architectures. \editmajor{The implementation of stellar and protoplanetary disk evolution is presented in Appendix \ref{bernmodelappendix_beforeplanetformation}.} 
        
        \subsection{Planet formation}
        \label{subsec:planetformation}
        In core-accretion models, planet formation 
        occurs \editminor{in two major steps}. Initially all embryos grow by accreting planetesimals at a rate of $\sim\SI{e-5}{\mearth\per\year}$, \editminor{ while the rate of gas accretion is very low} \citep{Alibert2005, Pollack1996}. Eventually, the protoplanetary gas becomes gravitationally bound to these growing planetary cores. If \editminor{the mass of a} core crosses a certain critical mass threshold ($\sim \SI{10}{\mearth}$) while the nebular gas is still present, it can undergo runaway gas accretion and becomes a giant planet (over a few million years). In contrast, planetary cores failing to cross the mass threshold do not undergo runaway gas accretion. Accreting solids from their feeding zones, these cores undergo collisions with other cores and  result in a diverse range of planets (over $\approx$ 10-100 Myr) (see panel (d) of Fig. \ref{fig:bernmodel}). \editmajor{The implementation of solid and gaseous accretion is described in Appendix \ref{bernmodelappendix_planetformation}.}

        \subsection{Additional physics}
        
        The Bern Model considers several additional physical mechanisms (see  panel (e) of Fig. \ref{fig:bernmodel}).
        
        \textit{Orbital Migration: }Gravitational interactions between the planet and the disk lead to the orbital migration of planets and the damping of eccentricity and inclination. The exchange of angular momentum via torques usually results in an inward migration of planets. Low-mass planets undergo type I migration, which is implemented following the approach of \cite{Coleman2014} and \cite{Paardekooper2011}. Massive planets can open a gap in the gas disk and undergo  type II migration, which is implemented following \cite{Dittkrist2014}. In type II migration some planets can migrate outwards. Planets inside the gap, if detached, continue to accrete until the disk disappears \citep{Kley2006}.
        
        \textit{\editminor{N-body} interactions: } Gravitational interactions between the star and multiple planets are included through the \editminor{\textit{N}-body} code \texttt{mercury} \citep{Chambers1999}. This formation stage tracks the dynamical evolution of planetary orbits, resonances, and collisional growth of planets \citep{Alibert2013}. Orbital migration and damping are coherently included in the \editminor{\textit{N}-body}. The \editminor{\textit{N}-body} computations are performed for 20 \editminor{million years} from the start of the model. 
        
        \textit{After \editminor{\textit{N}-body}: } The model calculates the internal structure of all planets for 10 \editminor{billion years}, after which calculations are stopped. This stage includes effects like atmospheric escape \citep{Jin2014}, bloating \citep{Sarkis2020}, and tidal migration. 
                
        \subsection{What is meant by the radius of a planet?}
        \label{subsec:whatisradius}
        
        In \editminor{the} Bern Model, all planets have a spherically symmetric structure composed of several layers of accreted material. These layer are the iron core, silicate (perovskite $\text{MgSiO}_3$) mantle, and water ice (if accreted) for the planetary core, and a H--He gaseous envelope (if accreted). For planets without any gaseous envelope the radius is obtained by solving the core internal structure (see \paperone/). In this study core radius and radius are used interchangeably for such planets.

        Planets with gaseous envelopes, however, do not offer a well-defined surface. The radius of such planets depends on the wavelength at which transits are measured \citep{Heng2019}. To facilitate comparison with transit observations, in this work the concept of transit radius is used. Transit radius is the radial distance from the centre of a planet, where the optical depth for a visible ray of light grazing the planet's terminator (chord optical depth) is $2/3$ \citep{Burrows2007, Guillot2010}. In this study transit radius and radius are   used interchangeably for such planets.

        \section{New Generation Planetary Population Synthesis}
        \label{sec:ngpps}
        
        Population synthesis provides a way to compare theory with observations of exoplanets and their architecture at the population level \citep{Ida2004,Mordasini2009}. The framework of population synthesis rests on one key assumption:   the rich diversity in nature's exoplanetary population emerges due to a  variety of possible initial conditions\editminor{ and \textit{N}-body interactions} \citep{Benz2014, Mordasini2018}. 
        Thus, multiple runs of a global model (while varying the initial conditions for disk and star, \editminor{and including \textit{N}-body interactions}) can produce theoretical exoplanetary populations possessing some of the observed diversity. Panels (a) and (b) of  Fig. \ref{fig:bernmodel} show the fixed and varying initial conditions.
        
        The New Generation Planetary Population Synthesis (NGPPS) \editminor{consists of synthetic planetary systems computed from the generation III} Bern Model (see \papertwo/).
        The Bern Model simulates planet formation and evolution by following the simultaneous growth of multiple planetary embryos embedded in a protoplanetary disk. However, since the number of embryos in a disk is unknown, it is treated as a free parameter. In this work three nominal models are studied with 20, 50, and 100 embryos. Each model is used to simulate 1\,000 planetary systems, wherein different initial conditions are assigned to each system in a Monte Carlo manner\footnote{\editbold{The number of simulated systems (1\,000) is chosen to be of the same order of magnitude as observed exoplanetary systems: 355 in CKSM (\W18), 822 in HARPS-GTO \citep{Mayor2011}, and $\sim 3\,000$ overall \citep{Schneider2011}.}}. The following  Monte Carlo variables are used (for details see \papertwo/):
        
        \begin{itemize}
                \item \textbf{Initial mass:  Protoplanetary gas disk, $\mgasdisk$} 
                
                The \editminor{initial distribution of gas disk mass}, $\mgasdisk$, follows the mass distribution of Class I disks reported by \cite{Tychoniec2018}. The values range from \SIrange{0.004}{0.16}{\msun}. This governs the initial spatial distribution and surface density profile of the disk via eq. \ref{eq:mgasdisk_influences}.
                
                \item \textbf{Disk lifetime: Photo-evaporation rate, $\mwind$}
                
                Varying $\mwind$ allows the model to have disks with different lifetimes. Disk lifetimes closely follow the observed disk lifetimes (see details in \papertwo/).  
                
                \item \textbf{Stellar metallicity:  Dust-to-gas ratio, $\fpg$}
                
                The initial mass of the solids in the disk is a fraction of the initial mass of the gas disk $\mgasdisk$ \editminor{(dust-to-gas ratio,} $\fpg$). Varying this ratio allows the model to capture the observed variation in stellar metallicities. This assumes the  relation
                \begin{equation}
                \label{eq:fpgmetalicity}
                10^{[\text{Fe/H}]} = \frac{\fpg}{\fpgsun} \ \text{, } \fpgsun = 0.0149 \text{ \citep{Lodders2003}.}
                \end{equation}
                The distribution of metallicities is in the range \numrange{-0.6}{0.5} and follows that of \cite{Santos2005}. Additionally, it is assumed that all of the dust in the solid disk is converted to planetesimals\footnote{In an alternative approach, some of the solid disk mass could be partitioned into pebbles. For a comparison of planet formation via pebble accretion and planetesimal accretion, see \cite{Brugger2018, Brugger2020}.}.
                
                \item \textbf{Inner edge of disk, $r_\mathrm{in}$}
                
                Regions of the disk that are close to the star interact with the stellar magnetic field resulting in stellar accretion, ejection, outflows, among other phenomena. The inner edge of the disk is taken at the co-rotation distance from the star, which is  the distance where the Keplerian orbital period matches the rotation period of the star. The stellar rotation periods are sampled from observations \citep{Venuti2017}. The distribution has a mean value of \SI{4.7}{\day} (corresponding to \SI{0.055}{au}), while the lower end is truncated at \SI{0.77}{\day}. 
                
                \item \textbf{Initial location of planetary embryo, $a_\text{embryo}$}
                
                Planetary embryos are initialized with a mass of $\SI{e-2}{\mearth}$. \editmajor{The initial location of embryos follows a distribution that has a uniform probability in the logarithm of distance between $r_\mathrm{in}$ and $\SI{40}{\au}$}. It is ensured that all embryos are at least \SI{10}{\rhill} apart from each other, resulting from their runaway growth \citep{Kokubo1998, Kokubo2002}. 
                
        \end{itemize}
        
\editmajor{The characteristics of all NGPPS planetary systems are strongly distorted by failed embryos due to their tremendous numbers. As a working definition, planetary embryos that fail to grow more than ten times from their initial masses are considered failed embryos. To simplify the discussion that follows, failed embryos are removed from the underlying population by removing objects with mass below $\SI{0.1}{\mearth}$ \footnote{Due to their small size, these objects are virtually undetectable via the transit method and do not affect the results of this paper.}. In addition, for simplicity, only the results of the 100-embryo population are presented (except in Sect. \ref{subsec:dynamicalinteractions} where the 20- and 50-embryo populations are also discussed).}
        
\editbold{Synthesizing thousands of planetary systems using the Bern Model, from a human perspective of current standards, is a theoretically complicated and numerically expensive endeavour;  however, it is  only a simplified approximation of our current understanding for how nature forms planets and planetary systems. The simplifications, choices, and assumptions made in the model may have a strong impact on the outcome of this study. Some of the major caveats are mentioned here, and the details can be found in \paperone/ (for the model) and \papertwo/ (for the population synthesis). The model assumes planets form via core accretion and ignores other formation pathways like disk instability. The protoplanetary disk and the internal structure of planets are solved via 1D models which may not capture the nuances of 3D effects. The model assumes that the time required for forming protoplanetary embryos is negligible compared to the evolution timescale of the gaseous disk, and that all embryos undergo the oligarchic growth regime. The dust-to-gas ratio of the disk is assumed to be the same as that of the star, and all the dust in the disk is assumed to aggregate into planetesimals (rocky or icy) of fixed size and fixed densities. The \textit{N}-body interactions are tracked for only 20 Myr which may not be enough to capture dynamical effects, collisions, or instabilities beyond 2 au. Merger collisions and stripping of planetary envelopes during giant impacts are treated in a simplified manner. Since the geological evolution of planets is ignored, no secondary Earth-like atmosphere is possible in the current model. Despite these and many other assumptions, the model is remarkably successful in capturing a variety of physics (see Fig. \ref{fig:bernmodel}) and produces diverse planets and planetary systems which bear impressive resemblance to those found in nature. }  
                
        \section{Kepler Observes Bern Exoplanets (\kobe/)}
        \label{sec:kobe}
\begin{figure*}
	\centering
	\includegraphics[width=8cm]{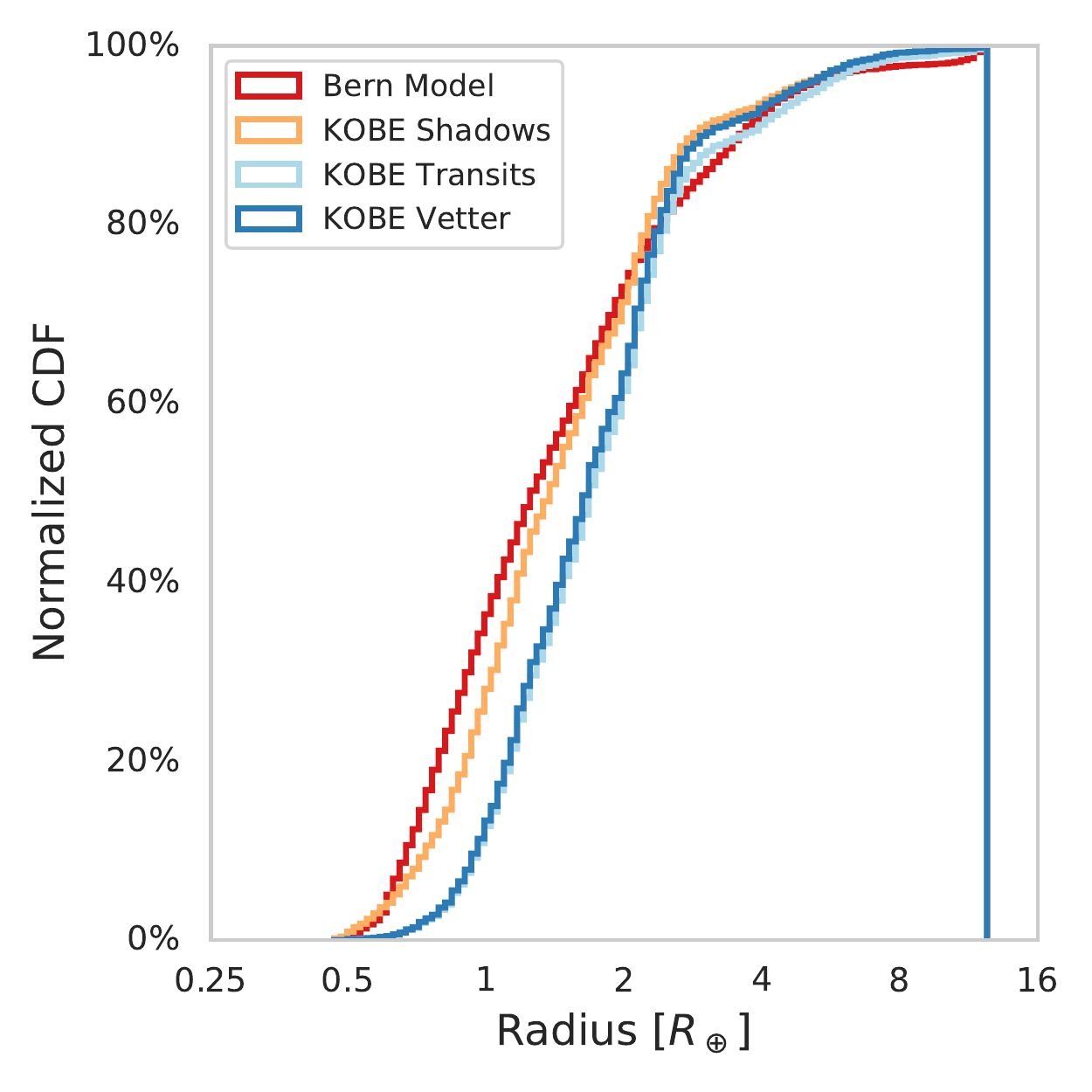}
	\includegraphics[width=8cm]{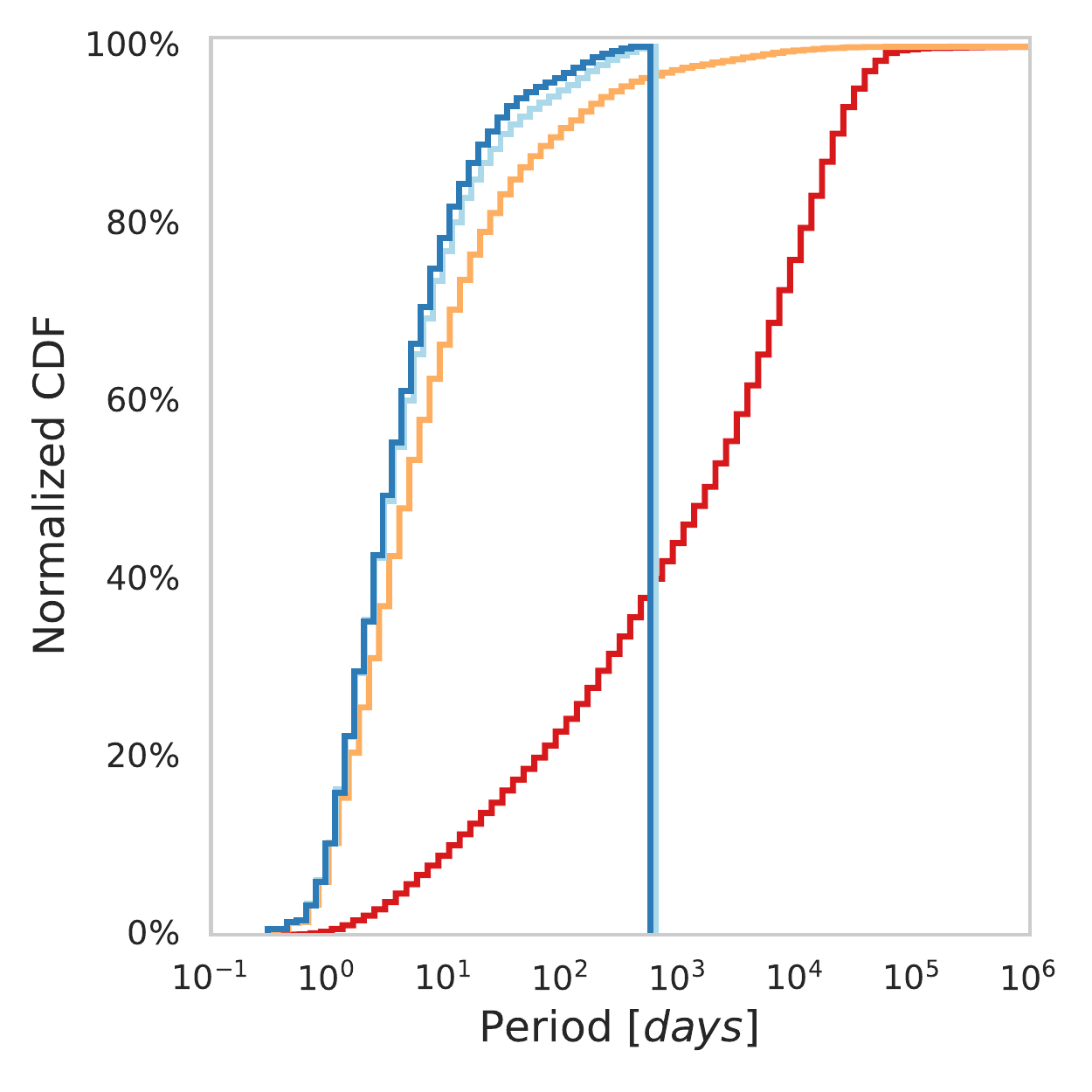}
	\caption{This figure shows how a planetary population is affected as it goes through the various modules of \kobe/. It shows the normalized cumulative distributions for radius (left) and period (right) for the 100 embryo population. The solid red curve represents the underlying population as calculated by the Bern Model. The orange curve is the output of \kobeshadows/. \kobetransits/'s \periodictce/ catalogue is shown in light-blue, and the catalogue of planetary candidates, as vetted by \kobevetter/ is shown in blue. }
	\label{fig:kobeexample}
\end{figure*}        

        \texttt{Kepler Observes Bern Exoplanets} (\kobe/) is a new computer code that simulates transit surveys of exoplanets\footnote{The current version of \kobe/ is designed to simulate NASA's \textit{Kepler} space telescope. However, \kobe/ is not limited to the Kepler survey and can be easily tweaked to simulate other transit surveys like TESS and PLATO \citep{Ricker2014, Rauer2014}.}. Suppose a population of synthetic planets (as in the  Bern Model NGPPS) is hypothetically observed by Kepler's transit survey. The aim of \kobe/ in this scenario is to identify those synthetic planets that would have been detected by the Kepler pipeline. 
        
        Calculations in \kobe/ are organized in three sequential modules. \kobeshadows/, the first module, is tasked with finding transiting planets from a synthetic population of planets. 
        \editminor{     This module produces the \kobeshadows/ catalogue, which consists of systems with at least one transiting planet.  All of the planets in this catalogue will transit, but not all of them will be detected.} The next module, \kobetransits/, computes transit related parameters for transiting planets. Planets that produce a  S/N are detected.
        \editminor{Planets that transit at least two times and have S/N $\ge 7.1$ constitute the \kobe/-periodic threshold crossing event (\periodictce/) catalogue \footnote{Following \W18, the minimum number of transits is fixed at two.}.} The last module, \kobevetter/, applies the completeness and reliability of the Kepler pipeline by emulating Kepler's Robovetter \citep{Twicken2016, Twicken2018,Thompson2018}. Transiting planets that are \editle{identified} as planetary candidates by \kobevetter/ make up the \kobe/ catalogue. The synthetic population in this catalogue is comparable to the exoplanet population discovered by Kepler. Later sections of this paper, analyse the architecture of planetary systems in the \kobe/ catalogue and compare it to observations. In a forthcoming paper the \kobe/ catalogue will be compared with other findings of Kepler.
        
        These three modules encapsulate the three different kinds of biases and limitations of a transit survey. \kobeshadows/ accounts for the geometrical limitation of the transit method. A planet can only transit when its orbit is aligned with respect to an observer. \kobetransits/ applies the detection biases coming from physical limitations; large planets in tight orbits around a quiet star are strongly favoured. Finally, \kobevetter/ imprints the completeness and reliability of a transit survey. \editmajor{In Appendix \ref{kobeappendix} the three modules are described in detail.}

        To understand \kobe/'s effect, Fig. \ref{fig:kobeexample} presents the 100-embryo underlying population (in red) \editminor{as it goes through each} stage of calculation in \kobe/. The shadow catalogue is dominated by planets that have high transit probability (eq. \ref{eq:transitprobability}), which is decided mostly by the star-planet distance and to a minor extent by the planet's size. Therefore, the shadow catalogue closely follows the underlying population in radius, but not in period. The excess of planets in the shadow catalogue around \SI{3}{\rearth} comes from a cluster of planets in the underlying population with high transit probability due to their low periods, $P<\SI{10}{\day}$. As shown in the period distribution, planets with $P<\SI{10}{\day}$ make up $70\%$ of the shadow catalogue, while they account for only $10\%$ of the underlying population. 
        
        The \periodictce/ catalogue strongly favours large planets at shorter orbital distances. Therefore, in the radius distribution the tail of small  planets in the \periodictce/ catalogue is shifted to right. About $30\%$ of the planets in the shadow catalogue have $\rplanet < \SI{1}{\rearth}$, whereas only $10\%$ of the \periodictce/ planets have $\rplanet < \SI{1}{\rearth}$. Requiring a minimum of two transits implies that the maximum period of a planet in the \periodictce/ catalogue will be $P_\mathrm{max} = (3.5\times365.25)/2 \approx \SI{640}{\day}$. This explains the sharp drop at 640 d in the period distribution of the \periodictce/ catalogue. The \kobevetter/ catalogue closely resembles the \periodictce/ catalogue. Differences arise when the completeness, as emulated by \kobevetter/, is considerably low. As seen in Fig. \ref{fig:cdppvetter}, this occurs for planets with large radii or large periods. 
        
        \section{KMPS: KOBE Multi-Planetary Systems}
        \label{sec:kmps}
        
        \begin{figure*}[]
                \centering
                \includegraphics[width=8cm]{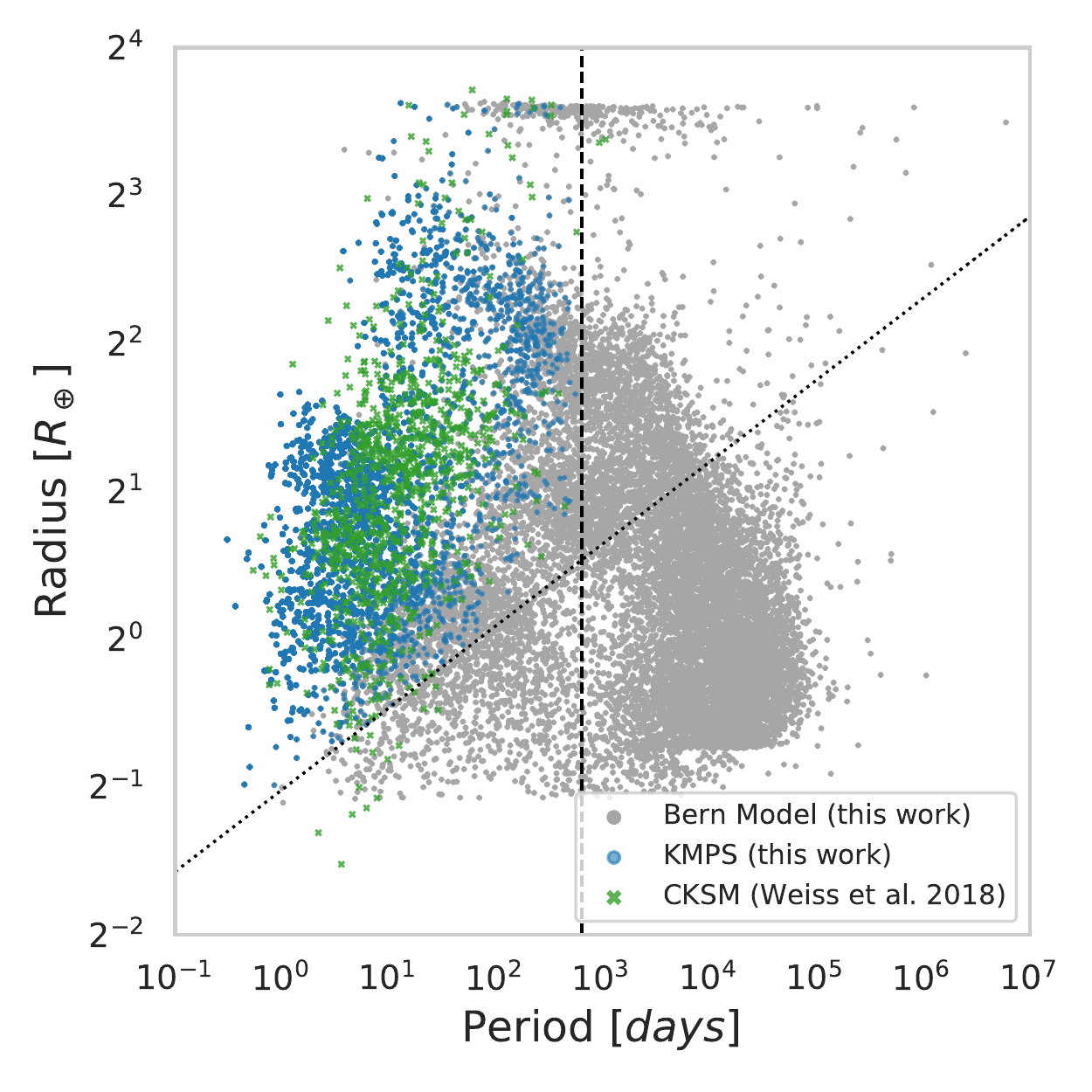}
                \includegraphics[width=8cm]{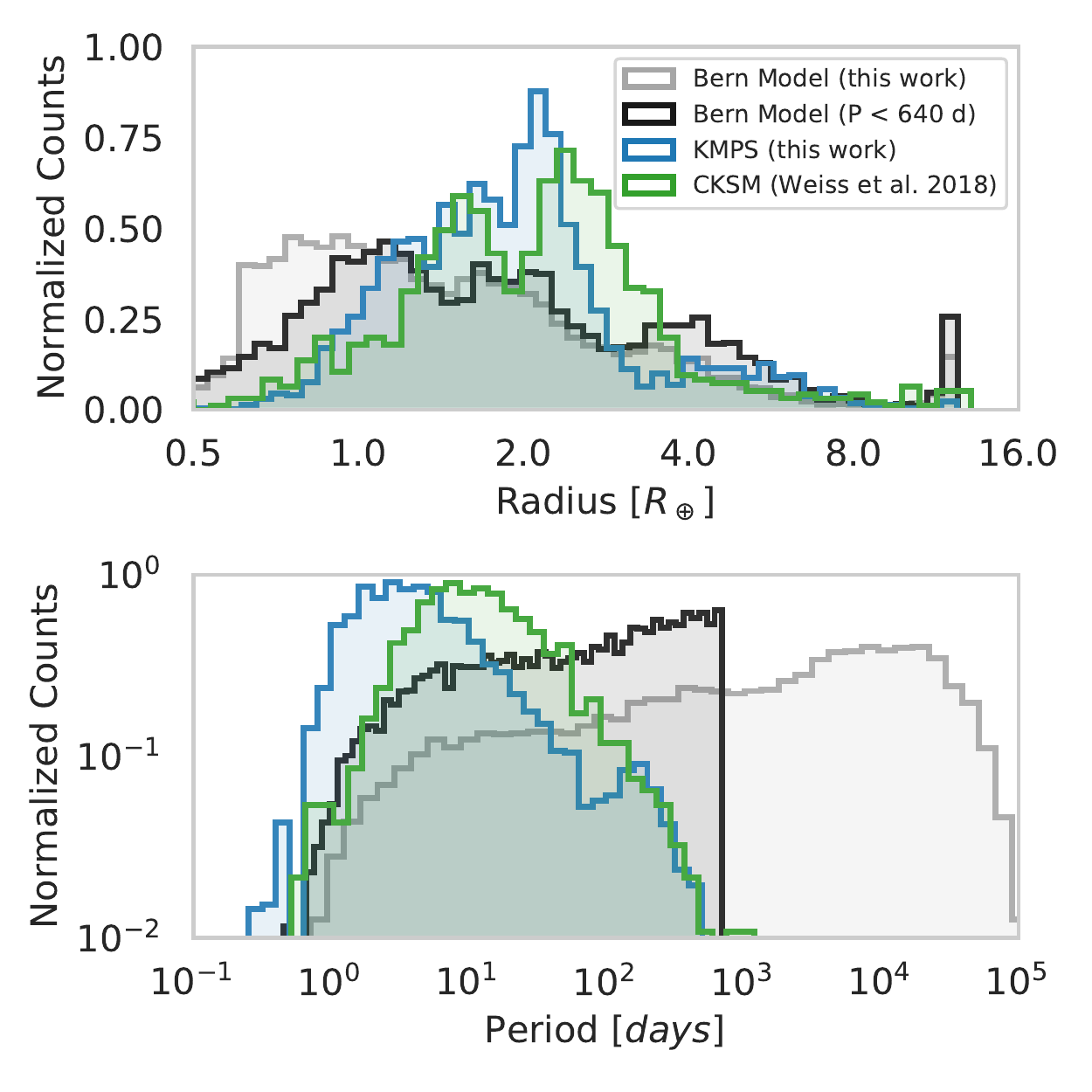}
                \caption{Comparison of planetary populations. Theoretical (blue) represents planets in  \kobe/ multi-planetary systems (\kmps/) and observations (green) are the CKS multi-planetary systems (CKSM).  
                Left: Scatter plot with planetary radius on the y-axis and period on the x-axis. The dashed line shows the maximum period of a planet that  can be found by \kobe/. The dotted line shows the minimum radius of a planet around a $\SI{1}{\rsun}$ star for producing a transit S/N of 10. The underlying theoretical population is shown in grey. 
                Right: Comparison of radius (top) and period (bottom) distributions. The radius valley can be clearly seen in the planets found by \kobe/ and the California-Kepler survey.
                }
                \label{fig:kmps_cksm_distributions}
        \end{figure*}

        In \W18 selection cuts were placed to obtain a `high-purity' population of planetary systems, the CKS-Multis (CKSM). The \kobe/ catalogue described in the last section undergoes similar cuts. Planets with a high impact parameter, $b>0.9$, are removed. Planets with multi-transit S/N $<10$ are also removed \editbold{(defined in eq. \ref{eq:transitsnr})}. Finally, planetary systems with only one remaining planet are removed. This creates a catalogue of \kobe/'s multi-planetary systems (\kmps/), which have the same characteristics as the CKSM catalogue coming from observations. Figures \ref{fig:kmps_cksm_distributions}  and \ref{fig:multiplicity} present a comparison of the theoretically observed planetary population (\kmps/ in blue) with observations (CKSM in green) of exoplanets. Overall, the two catalogues have remarkable similarities and understandable differences. The underlying population (Bern Model in grey) is also shown.
        
        The scatter plot in Fig. \ref{fig:kmps_cksm_distributions} (top) shows the radius of a planet as a function of its orbital period. It shows that the \kmps/ and CKSM planetary populations cover similar parameter space. A comparison with the grey points gives an impression of the planets that are missed by the transit method or removed by the selection cuts. In terms of period, the \kmps/ catalogue is bound by a vertical dashed line. This line marks the maximum period of a planet that can be found by \kobe/. This comes from the requirement of at least two transits ($\ntransit$ is fixed as $\tkepler/P$). There are two planets in the CKSM catalogue that have periods larger than \kobe/'s maximum detectable period, Kepler objects of interest (KOIs) K00435.02 and K00490.02. For a given period, the dotted line \editminor{approximates} the minimum size of a planet that will produce a transit S/N of 10 around a $1~\rsun$ star. There are some CKSM planets below this dotted line. These planets are orbiting a star of $\rstar < 1 \rsun$. 

        \subsection{Radius and period distributions}    
        
	  For radius (Fig. \ref{fig:kmps_cksm_distributions}) the \kmps/ and CKSM populations are quite similar. The \kmps/ radius distribution shows the \editminor{cumulative effects of both \kobe/ and the selection cuts  placed on the underlying population.} \editmajor{This distribution shows a bimodal nature with the two modes being around $1.4 - 1.7~\rearth$ and $2-3~\rearth$.} \editmajor{The observed CKSM radius distribution also shows this feature. This is the well-known radius gap seen around $2 ~\rearth$ \citep{Fulton2017,Julia2020}.} The CKSM population has  more planets with sizes between $3-4~ \rearth$ than the \kmps/ population. This can be attributed to a dearth of $3-4~ \rearth$  planets  with $P < \SI{100}{\day}$ in the underlying population. This is also reflected in the sharp drop seen in the period distribution of \kmps/ planets with $P \approx \SI{100}{\day}$. \editmajor{The radius distribution of the underlying \editbold{populations}, however, does not show a radius gap. This is because the underlying population is dominated by small planets  at a large distance from the host-star. It is the cumulative effect of applying transit probability (via \kobeshadows/) and the detection biases of the transit method (via \kobetransits/ and \kobevetter/) that removes these small and distant planets. This allows the radius valley in the theoretical population to be clearly seen.}
        
        The period distributions of the \kmps/ and the CKSM populations have similar slopes after their respective peaks. This is a reflection of the role played by transit probability (which decreases as $P^{-3/2}$). In the \kmps/ population the period distribution peaks at about \SI{3}{\day}, while the CKSM distribution peaks around \SI{5}{\day}.  
        
        \subsection{Multiplicity distribution}
        
        \begin{figure}
        \centering
        \resizebox{\hsize}{!}{\includegraphics{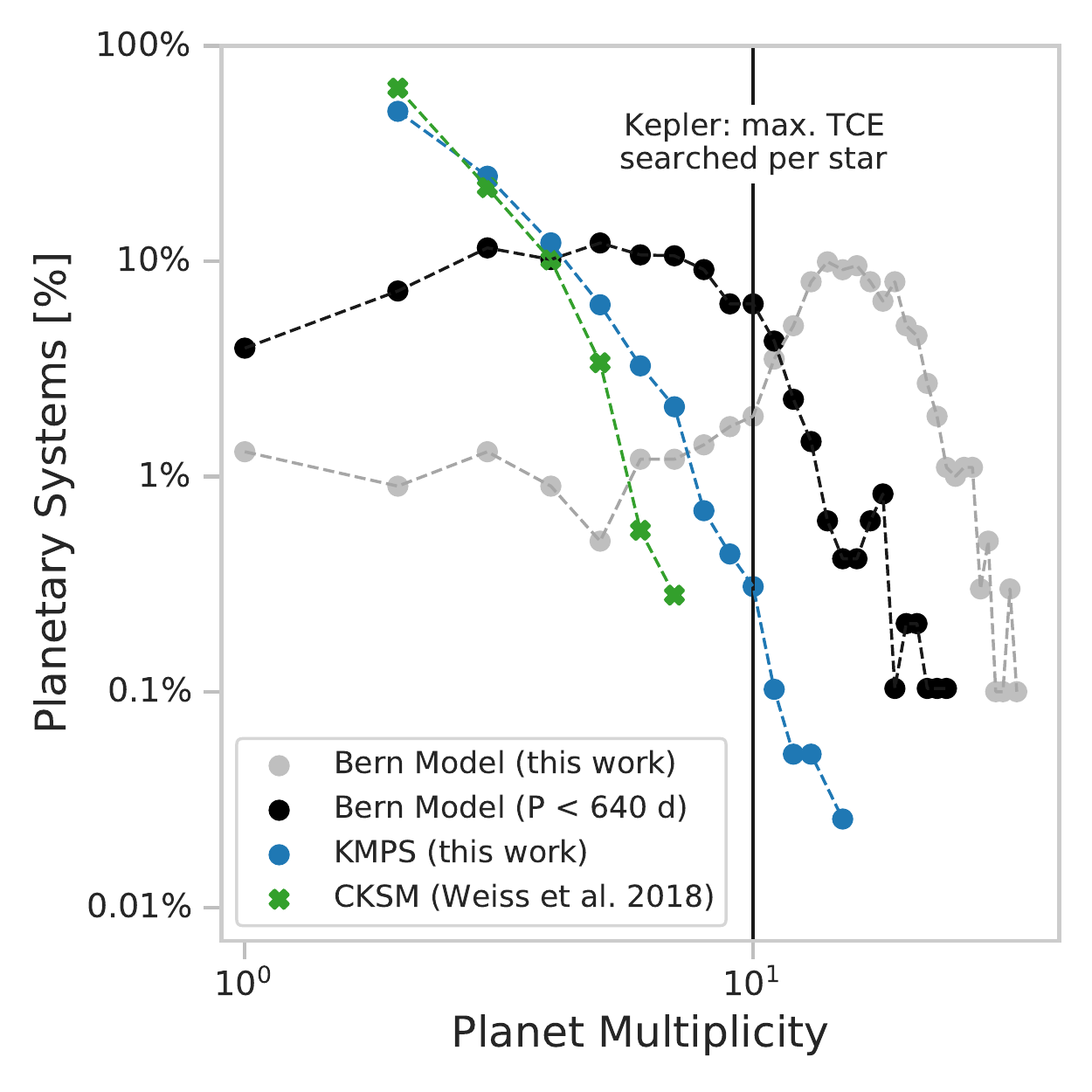}}
        \caption{Distribution of planetary multiplicity: Bern Model in grey, Bern Model detectable planets (P < 640 d) in  black, \kmps/ in blue, and observed CKSM in green. The solid black line indicates the maximum number of TCE searches for a star performed by the Kepler pipeline.}
        \label{fig:multiplicity}
        \end{figure}    
        
        The geometrical limitations of the transit method severely impacts the observed multiplicity of planets in a system. Multiplicity, for an observer, results from the overlap of transit shadow band of multiple planets at the observer's location (see Fig. \ref{fig:kobeshadows}). \editminor{Low mutual inclination between multiple planets leads to a large overlap in the transit shadow bands. This results in a higher probability that observers will find multiple transiting planets.}  The mutual inclination of planets is governed, in part, by the dynamical history of the system. Figure \ref{fig:multiplicity} shows the multiplicity distribution. The theoretical \kmps/ population shows a noteworthy similarity with the observed CKSM population. The vertical solid black line indicates the maximum number of TCEs per star that was searched by the Kepler pipeline \citep{Twicken2016}. The multiplicity distributions of  the \kmps/ and CKSM populations show large differences with the underlying synthetic \editbold{populations}. This is directly attributed to the geometrical limitations inherent in the transit method.
        
        About $60\%$ of the systems in the CKSM catalogue have only two planets. The percentage of systems with higher multiplicity drops sharply, with less than $1\%$ of CKSM systems having six or seven planets. The \kmps/ catalogue closely follows the CKSM catalogue in the multiplicity distribution. \kmps/ systems with two planets are less frequent than CKSM systems   (about $50\%$). However, for three or more planets the \kmps/ catalogue has more systems than the CKSM catalogue. This may arise due to the low mutual inclination between the planets formed in the Bern Model \citep{Mulders2019}. 
        
        \section{Peas in a pod}
        \label{sec:peasinapod}
        
        The so-called peas in a pod trends in the architecture of planetary systems refers to correlations observed in the properties of adjacent exoplanets. The following  statements define the peas in a pod trends in the architecture of multi-planetary systems:
        \begin{itemize}
                \item \textbf{Size: } Planets within a system tend to be either \editle{similar or ordered in size}. Here, similarity implies that for two adjacent planets $\rout/\rin \approx 1$. Two adjacent planets are said to be ordered in size when the outer planet is larger than the inner planet, $\rout/\rin > 1$.
                
                \item \textbf{Mass: } Planets within a system tend to be either \editle{similar or ordered in mass}. Here, similarity and ordering are defined using a planet's mass.
                
                \item \textbf{Spacing: } For systems with three or more planets the spacing between a pair of adjacent planets is similar to the spacing between the next pair of adjacent planets. This is quantified by the ratio of period ratios for adjacent pairs of planets, $(P_{j+2}/P_{j+1})/(P_{j+1}/P_{j}) \approx 1$. The index $j$ identifies different planets,  where $j=1$ is the innermost planet. 
                
                \item \textbf{Packing: } Small planets are found to be closely packed together, while large planets tend to have large orbital spacing. This is quantified by comparing the average radii of adjacent planets $(\rin + \rout)/2$, with their period ratio $\pout/\pin$.
        \end{itemize}
        
        These statements take the results from \W18 and \cite{Millholland2017} into account. These trends were examined in \W18 by measuring the strength of correlations using the Pearson R correlation test. \editbold{Since the aim of this paper is to examine the architecture trends, the same correlation test is used throughout this paper to compare the correlation between synthetic planetary systems and observations\footnote{In addition to the Pearson R correlation test, the Spearman R correlation test was also performed for all correlations. The Spearman R tests for correlations (specifically monotonicity) in the rank of members of two datasets instead of their actual value.}. We note that correlation coefficients can only measure the strength of correlation in one dataset and  that they cannot measure the similarity between two underlying datasets (see \citealt{Bashi2021} for an inter catalogue similarity metric). In addition, since correlation coefficients require large datasets, they cannot be used to measure architecture trends for a single system.}
        
\begin{figure}
                \centering
                \resizebox{\hsize}{!}{\includegraphics{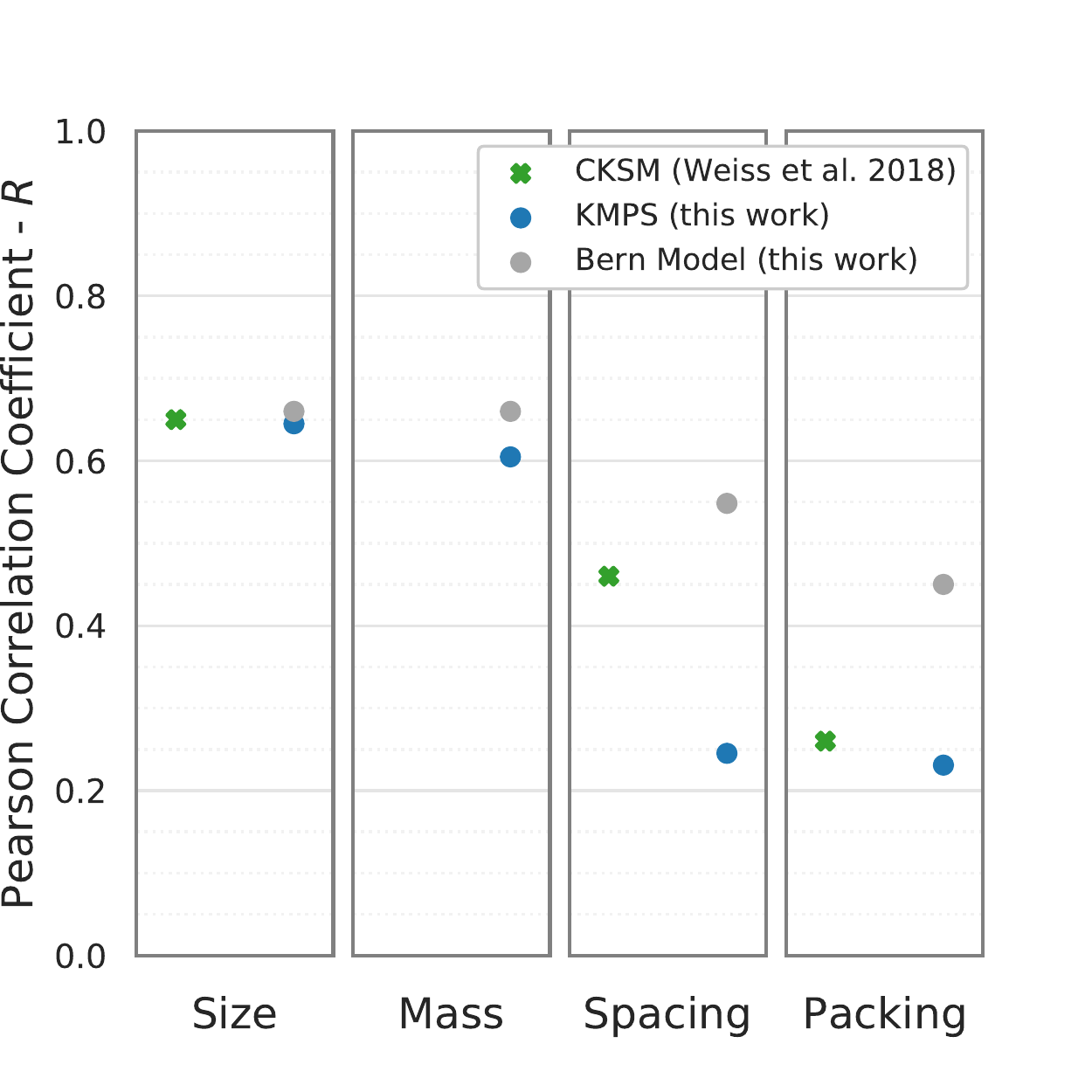}}
                \caption{Peas in a pod. Comparison of the correlations present in the theoretically observed \kmps/ catalogue (theory) and the CKSM catalogue (observations). Also shown are the correlations present in the underlying Bern Model population (grey).}
                \label{fig:kmps_pip_comparison}
        \end{figure}

        Figure \ref{fig:kmps_pip_comparison} presents a comparison of the correlation coefficient found in the observed exoplanetary systems (CKSM) and the  \editminor{observable synthetic} population (\kmps/). There is a remarkable agreement in the correlations of size and packing. Although present in \kmps/, the spacing correlations are not as strong as those found in CKSM. As transit observations do not yield planetary masses, the correlation coefficient for mass is not available for the CKSM systems. Each panel in Fig. \ref{fig:kmps_pip_comparison} is discussed below with additional details. 

        \subsection{Peas in a pod: Radius}
        \label{subsec:pip_size}
        \begin{figure*}
                \centering
                \includegraphics[width=6cm]{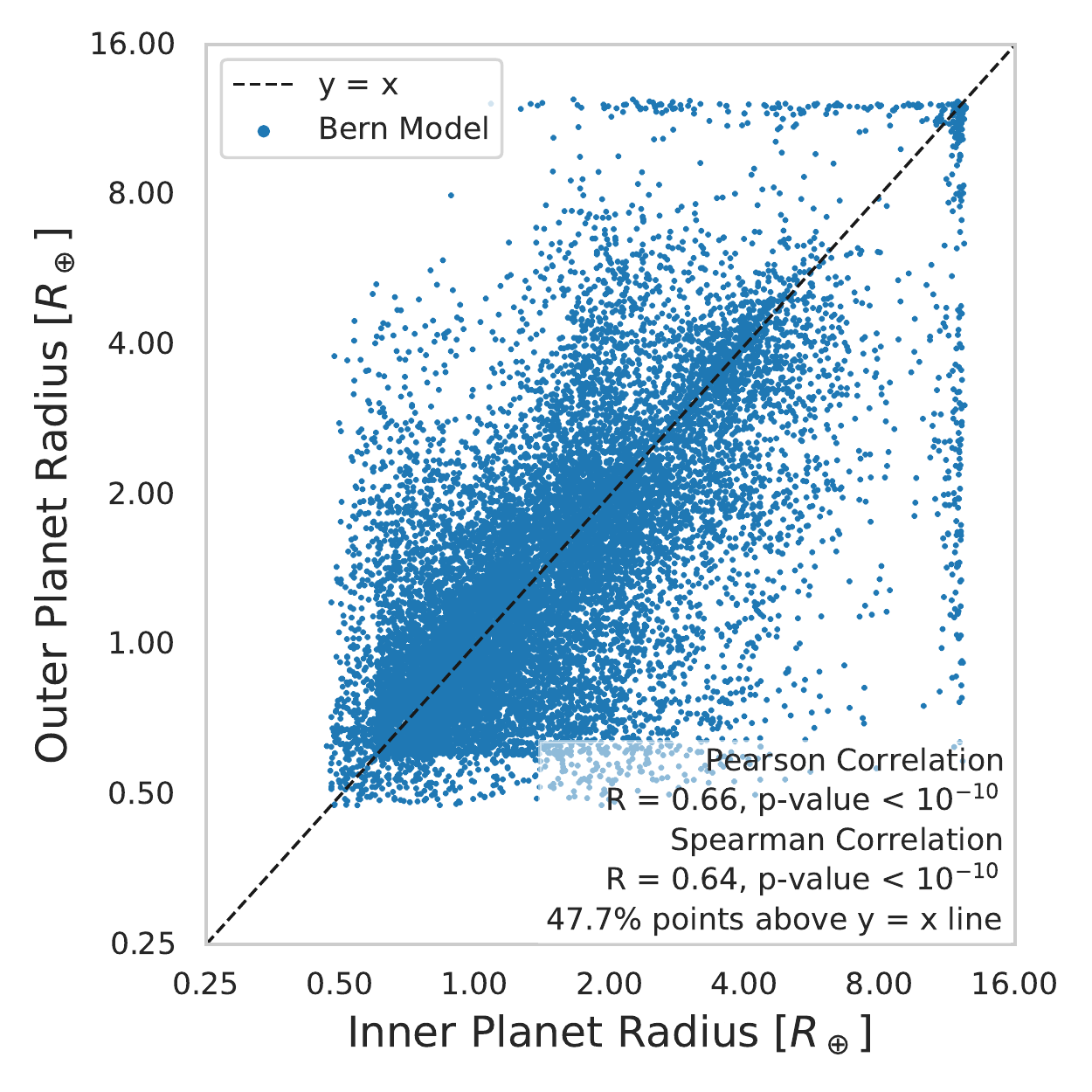}
                \includegraphics[width=6cm]{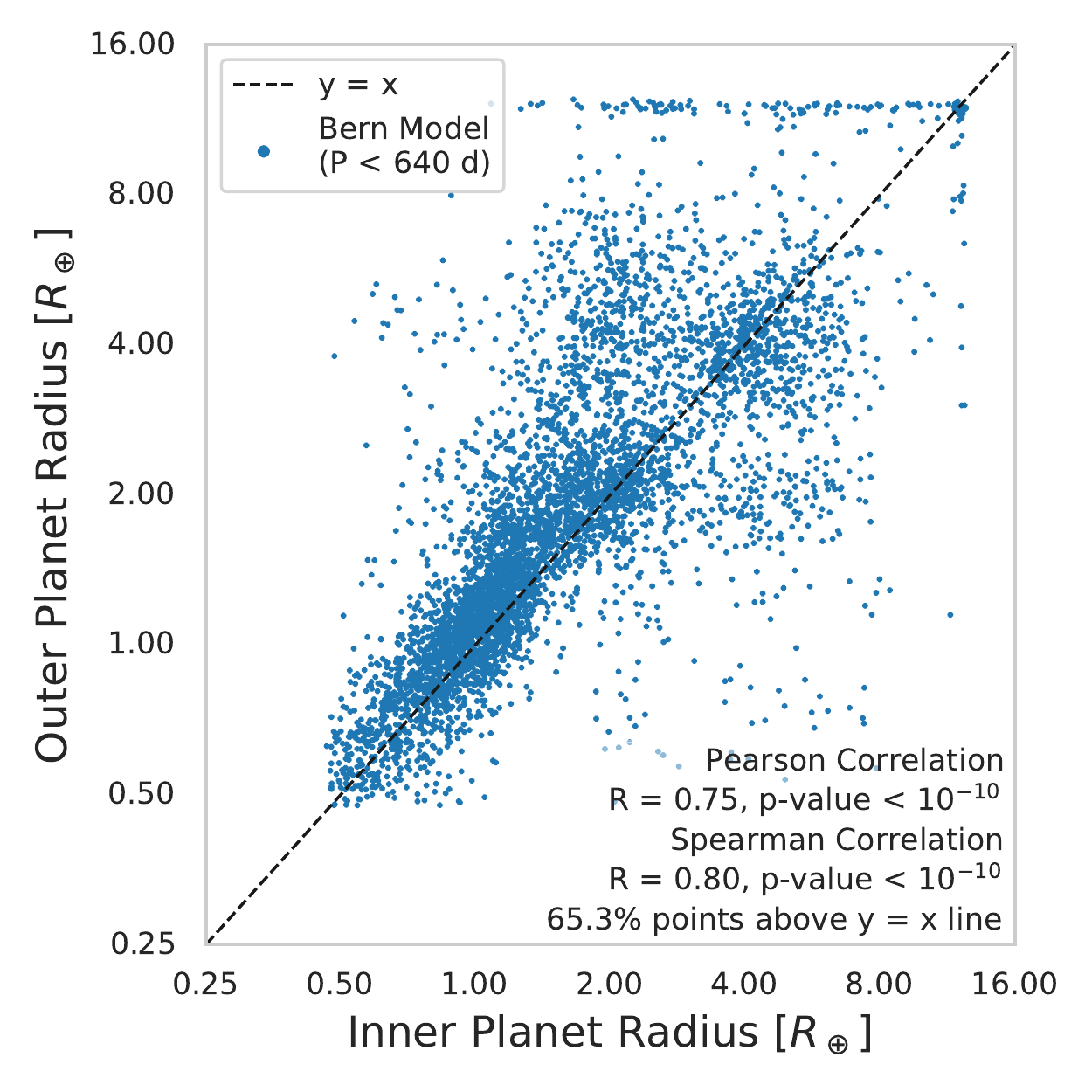}
                \includegraphics[width=6cm]{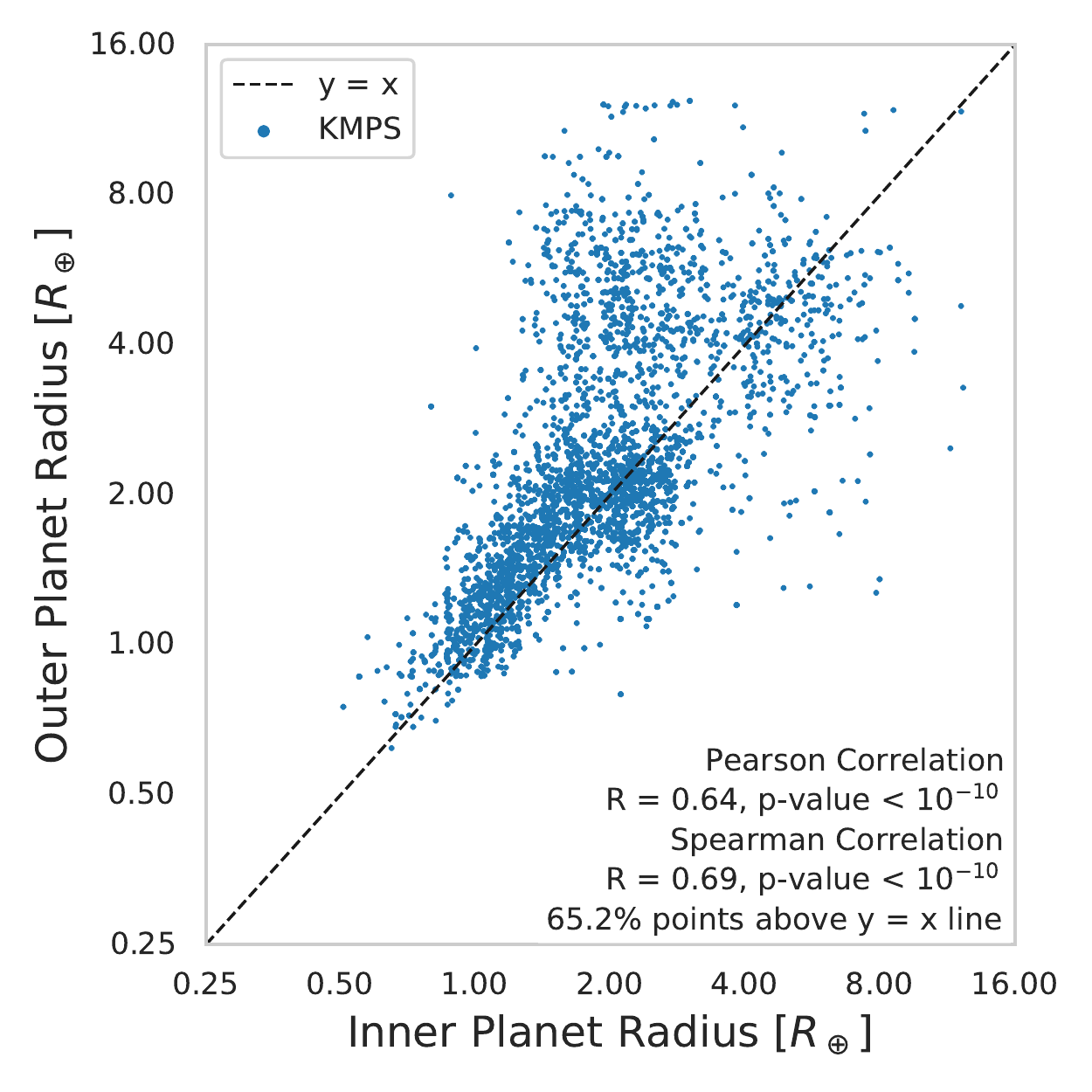}            
                \caption{\textit{Peas in a pod:  Size.} The sizes of adjacent planets are shown for the underlying population (left), for the underlying population of detectable planets (P < 640 d) (middle), and for  theoretical observed planets (right).}
                \label{fig:kmps_pip_radius}
        \end{figure*}

        \begin{figure*}
        \centering
        \includegraphics[width=6cm]{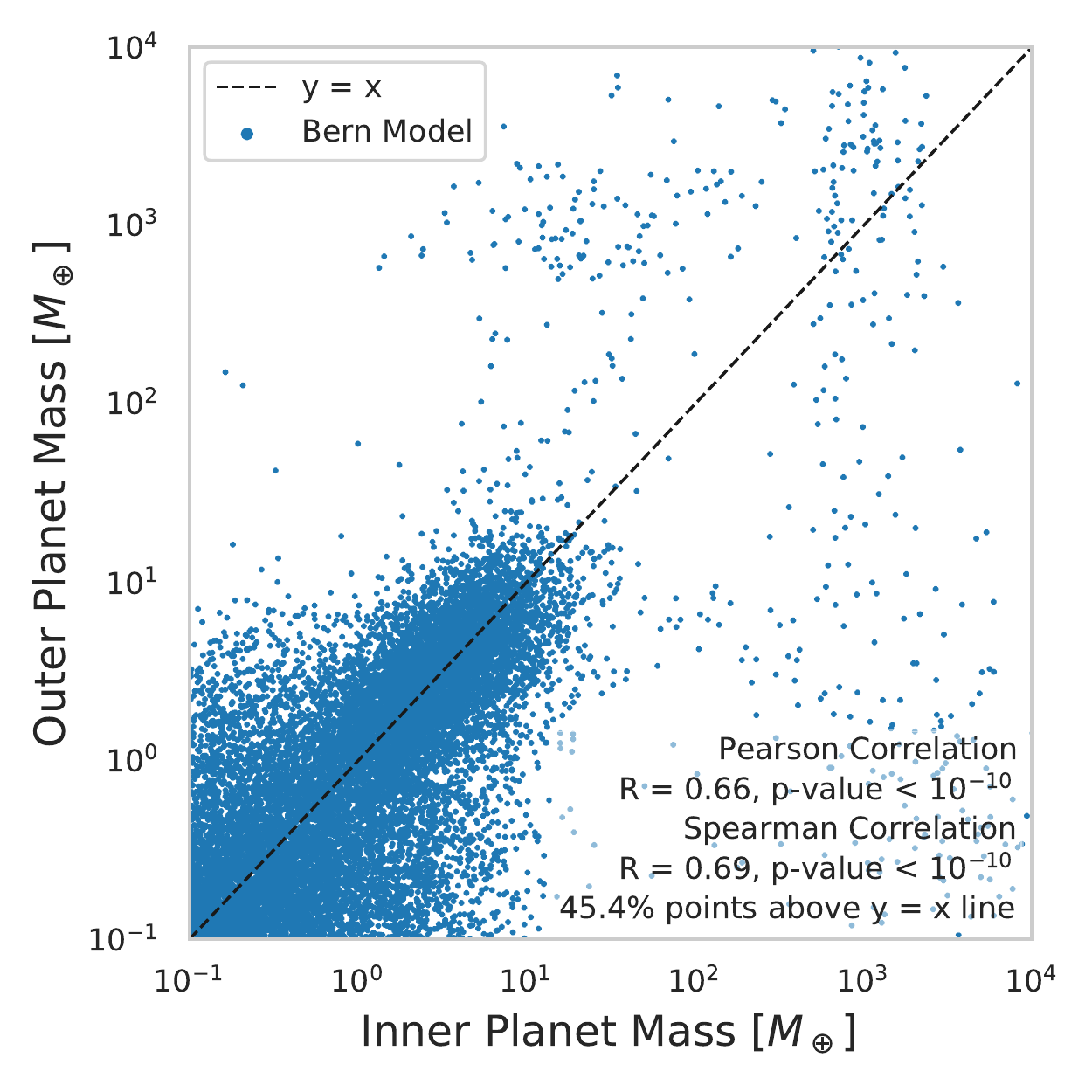}
        \includegraphics[width=6cm]{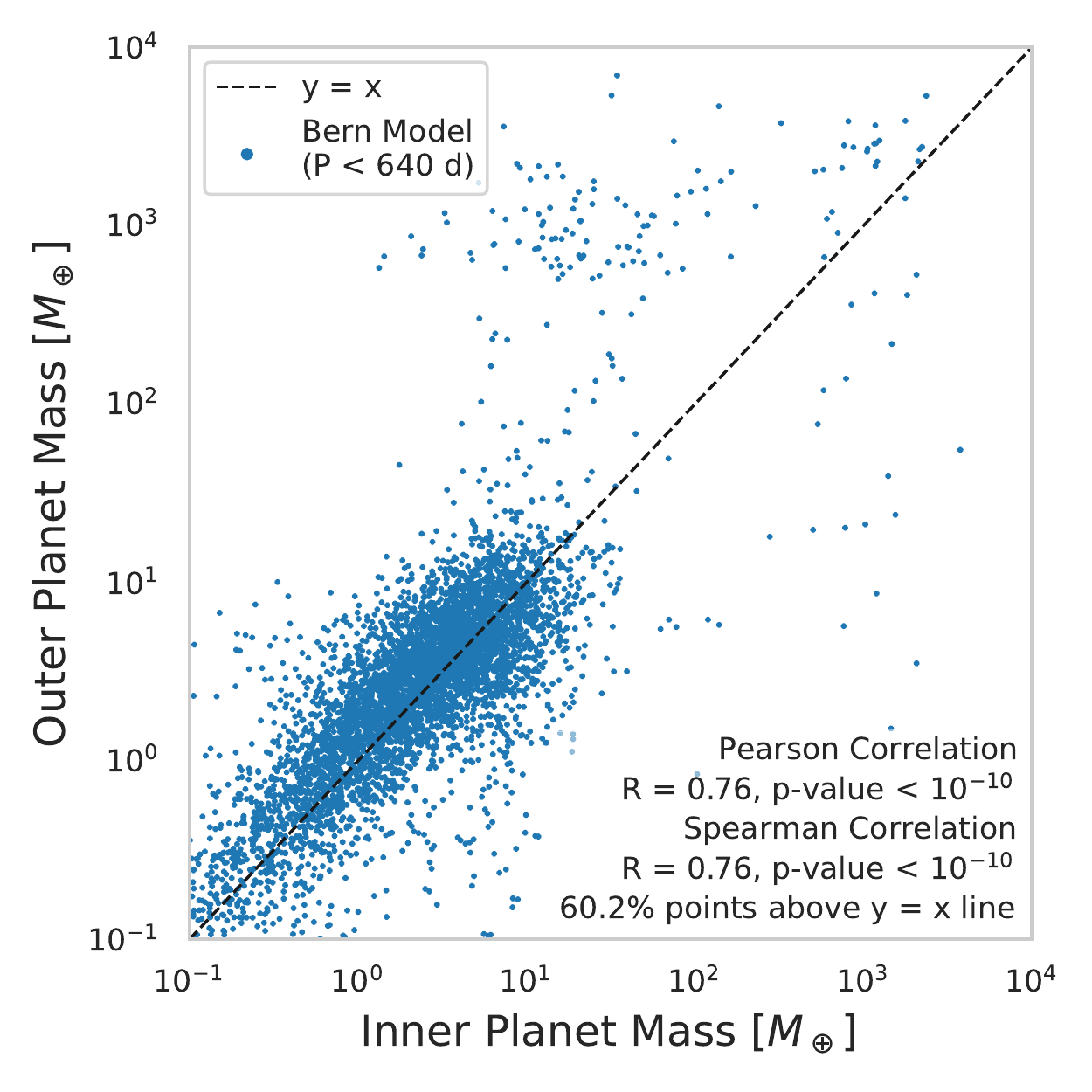}
        \includegraphics[width=6cm]{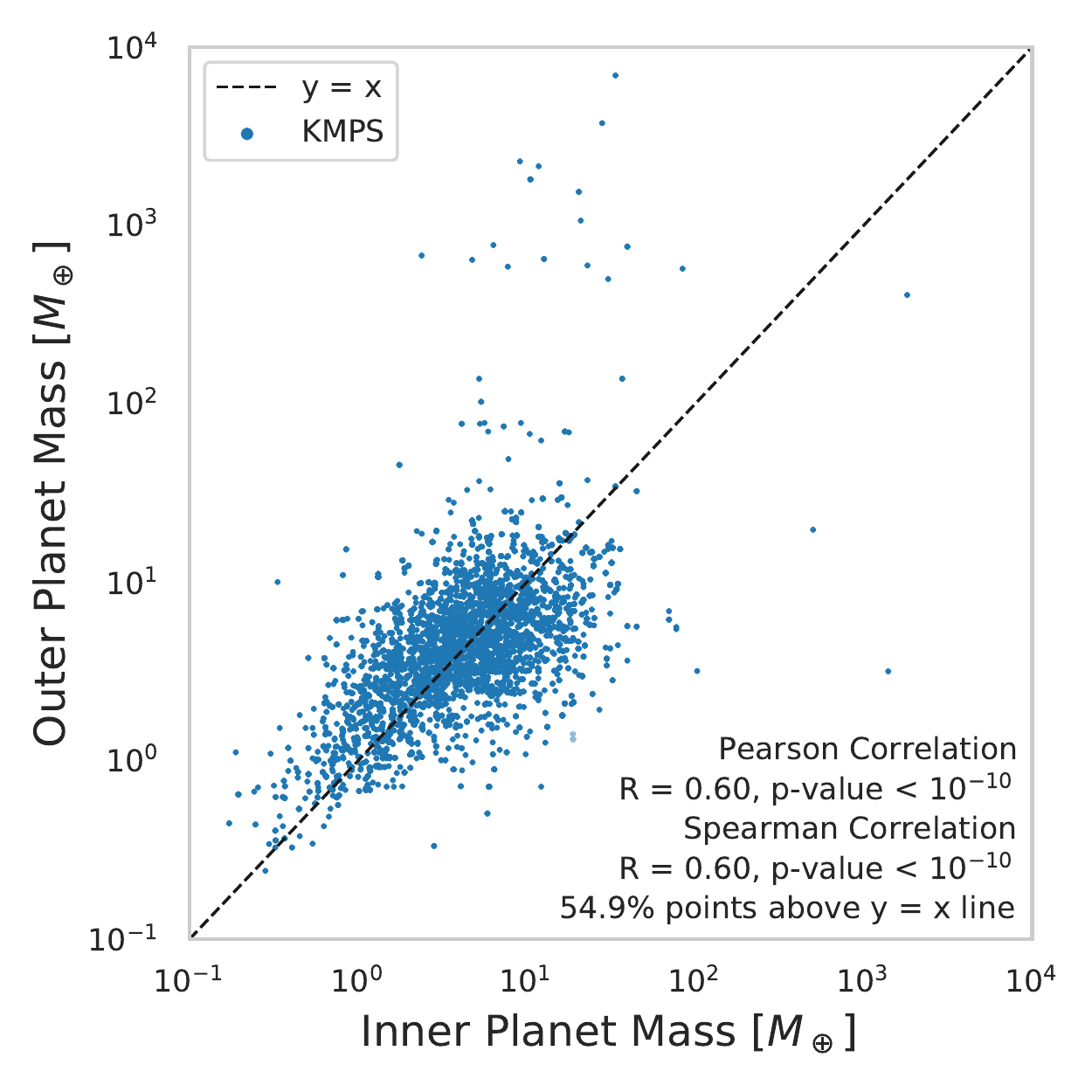}
        \caption{\textit{Peas in a pod:  Mass.} The masses of adjacent planets are shown for the underlying population (left), for the underlying population of detectable planets (P < 640 d) (middle), and for  theoretical observed planets (right).}
        \label{fig:kmps_pip_mass}
        \end{figure*}

        Following \W18, before testing for correlations in size (and also in mass), all adjacent pairs of planets in the \kmps/ population are required to undergo a swapping test. If both planets in a pair produce transit S/N$\ge 10$ (see eq. \ref{eq:transitsnr}) when their orbital positions are swapped, then this pair is included for testing correlations. \editminor{Figure \ref{fig:kmps_pip_radius} shows the radius of an outer planet as a function of the inner planet's radius, for the underlying (left) and the \kmps/ (right) populations.} \editmajor{The middle panel shows the same for planets from the underlying population that could have been detected by  \kobe/ (P < 640 d).}

        For the \kmps/ population there is a strong ($\mathrm{Pearson}~R = 0.64$) and significant correlation present in the size of adjacent planets\footnote{Following \W18, the size correlation coefficient $\mathrm{Pearson}~R$ is calculated on the log of radius. Since the Spearman $R$ is calculated on the rank of the members in a dataset, taking a log produces no difference in the coefficient's value.}. The size correlation between adjacent planets is also seen through the Spearman R test ($\mathrm{Spearman}~R = 0.69$). This implies that adjacent planets in the \kmps/ catalogue have sizes that are similar to their neighbours. The plot also shows that for $65\%$ of adjacent pairs the outer planet is also the largest planet.  This frequency of ordered adjacent pairs in \kmps/ is exactly the same as seen in CKSM (\W18) and similar to the findings of \cite{Ciardi2013}. This implies that the outer planet in an adjacent pair of planets is often the larger planet in the \kmps/ catalogue.
        
        It is interesting to compare the size correlations present in the \kmps/ populations, with the size correlations present in the underlying populations. The underlying population (Fig. \ref{fig:kmps_pip_radius} (left)) shows strong ($\mathrm{Pearson}~R = 0.66, \mathrm{Spearman}~R = 0.64$) and significant (\textit{p}-value$\ll 10^{-10}$) correlation in size of adjacent planets. This is an important result, and it strongly suggests that size correlation between adjacent pairs of planets is already present in the underlying population. However, only $48\%$ of the pairs in the underlying population are ordered. 
        
        \editmajor{Keeping only detectable planets (with P < 640 d) shows the size-correlation present in the underlying population of detectable planets. Compared to the entire underlying population, this population shows a stronger size similarity and size ordering. Removing non-detectable planets tends to remove many small planets that occur frequently in the outer parts of a system. However, adjacent pairs where the outer planet is smaller are removed more often than the adjacent pairs where the outer planet is larger. This shows that the inner region of many planetary systems is populated by size-ordered pairs. This also explains the increase in size correlation seen in this population, which arises from a decrease in adjacent pairs where the outer planet is smaller.}
                
        \editmajor{The role of detection biases becomes clear when the \kmps/ population is compared with the population of detectable planets. Since, small planets are harder to detect via the transit method, many planets with $\rplanet < 1 \rearth$ are not found by \kobe/. This effectively decreases the size-similarity correlation. However, as the \kmps/ population is a subset of the population of detectable planets, it inherits the frequency of size-ordered pairs.}

        Overall, the theoretically observed \kmps/ catalogue shows similarity and ordering in the size of adjacent planets. These trends are in excellent agreement with observations. The theoretical underlying population \editmajor{of detectable planets} also shows these correlations. While the size similarity and size ordering are affected by the detection biases,  these correlations do not originate from detection biases of the transit method. This suggests that the correlations seen in observations may have an astrophysical origin. 
        
        \subsection{Peas in a pod: Mass}
        \label{subsec:pip_mass}
        
        
        Figure \ref{fig:kmps_pip_mass} shows  the masses of the inner and outer planets in an adjacent pair. For the \kmps/ population a swapping test as described in Sect. \ref{subsec:pip_size} was implemented There is a strong and significant correlation present between the masses of adjacent pairs of \kmps/ planets. These correlations are also confirmed by the Spearman correlation test. This implies that adjacent planets in the \kmps/ populations have similar masses. Figure \ref{fig:kmps_pip_mass} also shows that about $55\%$ of \kmps/ adjacent pairs lie above the $y=x$ line (i.e. they are ordered in mass). This means that there are more planetary pairs where the outer planet is also the more massive planet.
        
        Whether this trend is also present in the underlying population is an interesting question. Figure \ref{fig:kmps_pip_mass} (left and middle) shows that the underlying population have an even stronger and significant mass similarity correlation. 
        \editmajor{     In the underlying population, the outer regions of a system are heavily dominated by small planets with $\mplanet < 1\mearth$. When the population of detectable planets is considered, these small planets are noticeably missing (Fig. \ref{fig:kmps_pip_mass} (middle)). In addition, mass-ordered adjacent pairs are more common in the inner region of many planetary systems. Thus, as noted in the last section, considering only detectable planets has two important consequences: increase in mass similarity correlation and increase in frequency of mass-ordered planetary pairs. This suggests that detectable planets in the Bern Model tend to have masses similar to their adjacent neighbour or that the outer planet in an adjacent pair is often the more massive planet.}

        Overall, adjacent planets in the \kmps/ catalogue show mass similarity and ordering.   Since mass similarity and ordering are already present in the underlying population of detectable planets, these correlations do not emerge from the detection biases of the transit method. This implies that the peas in a pod mass similarity and mass ordering trend is probably astrophysical in origin. \editmajor{However, detection biases seem to diminish the strength of these correlations (see Sect. \ref{sec:roleofbiases}).}

        The patterns seen in the mass trend are strikingly similar to the size trends discussed above. This suggests that the two trends may not be independent of each other. This is understandable since the size of a planet strongly depends on its mass. Planetary mass is evaluated directly from formation physics, whereas the planetary radius has to be evaluated from additional considerations. 
        
\begin{figure}
                \centering
                \resizebox{\hsize}{!}{\includegraphics{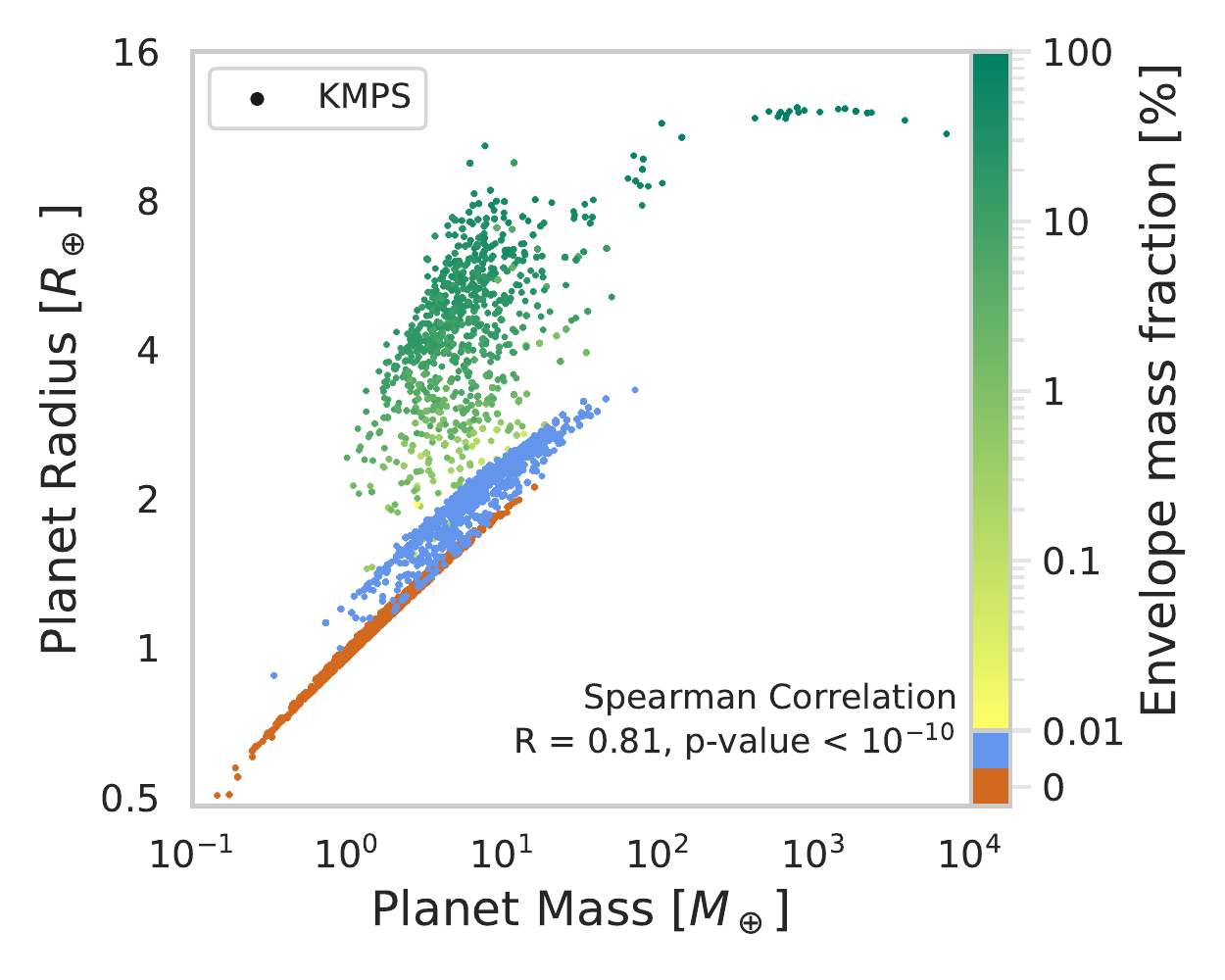}}
                \caption{\textit{Mass-radius relationship.} Planetary radii are  plotted as a function of planetary masses for the planets in the \kmps/ 100-embryo population. For planets with non-zero H--He envelopes, the colour denotes their envelope mass fraction, $\frac{\menv}{\mplanet}$. Planets without envelopes and without volatiles in their cores are  in brown, while planets without envelopes that have volatiles in their cores are  in blue.}
                \label{fig:kmps_pip_m_r}
        \end{figure}

        Figure \ref{fig:kmps_pip_m_r} shows a mass-radius diagram of all planets in the \kmps/ 100 embryo population. Planets with a large envelope mass fraction are composed mainly of H--He gases that they accreted during their formation. On the other hand, planets with a low envelope mass fraction are mostly dominated by their cores and have a small H--He gas envelope. The plot shows that Jupiter-sized planets have high envelope mass fractions, while planets with sizes $< \SI{4}{\rearth}$ are mostly core-dominated. {The high value of the Spearman correlation coefficient ($R = 0.81$) indicates that the radius as a function of planetary mass is a highly monotonic function\footnote{The Spearman correlation coefficient is unity ($R=1$) for a strictly monotonic function.}}. This implies that for the \kmps/ planets an increase in planetary mass is very likely to result in an increase in the planetary radius as well. 
        
        These factors suggest that the trends of planetary masses are probably more fundamental in the system architecture. The trend seen in the size of adjacent planets is likely to be a derivative trend from the mass correlation. 

        \subsection{Peas in a pod: Spacing}
        \label{subsec:pip_spacing}
        \begin{figure*}
                \centering
                \includegraphics[width=6cm]{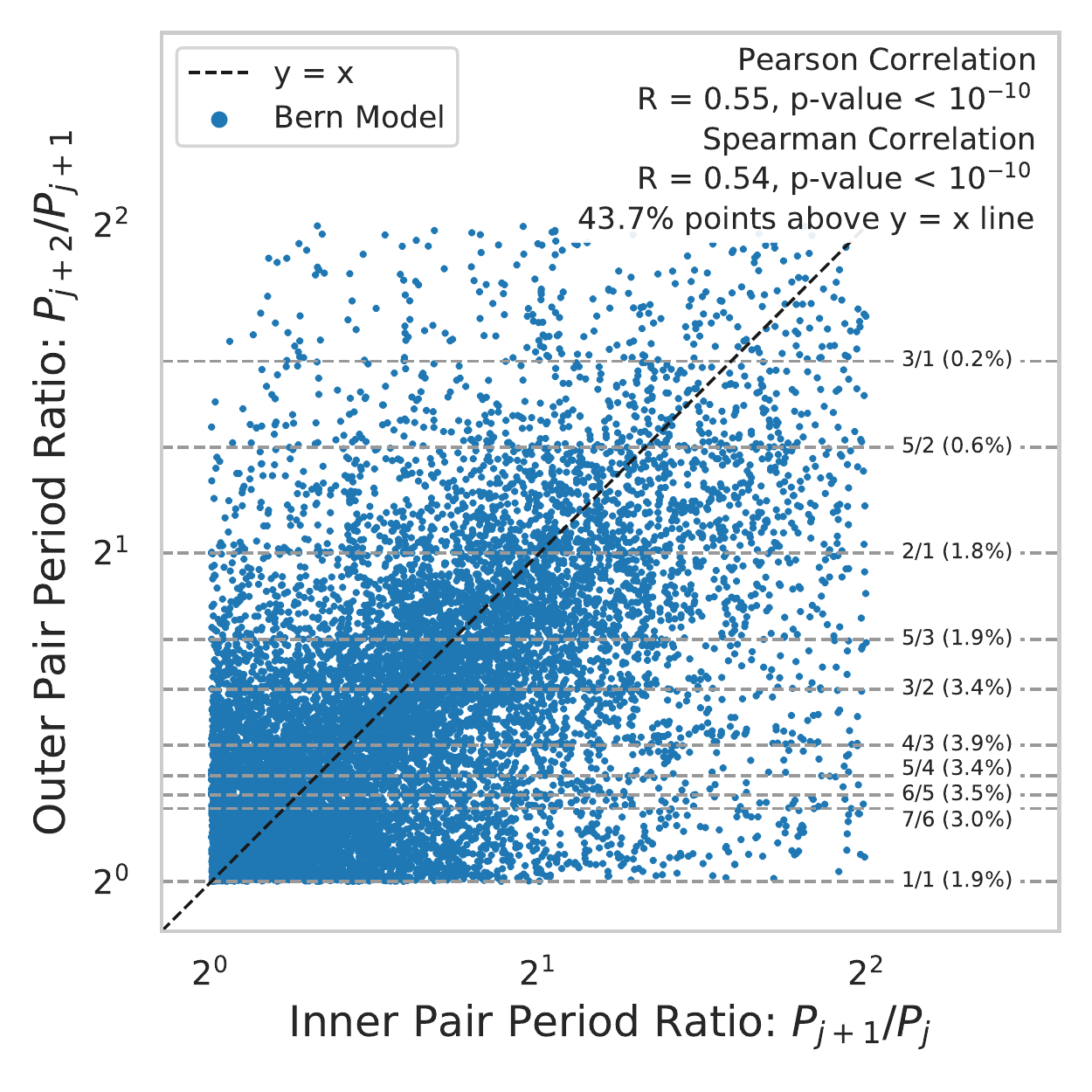}                    
                \includegraphics[width=6cm]{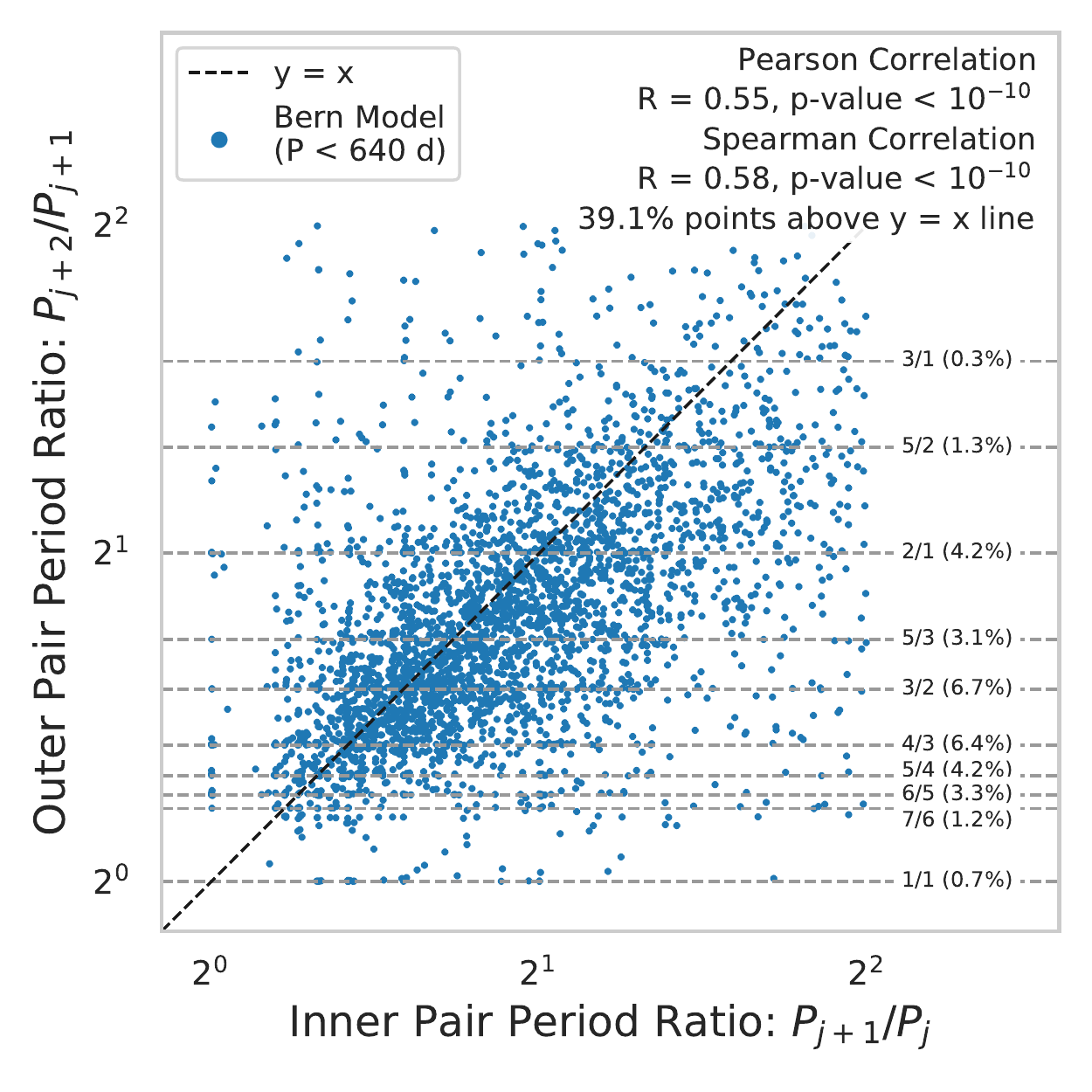}       
                \includegraphics[width=6cm]{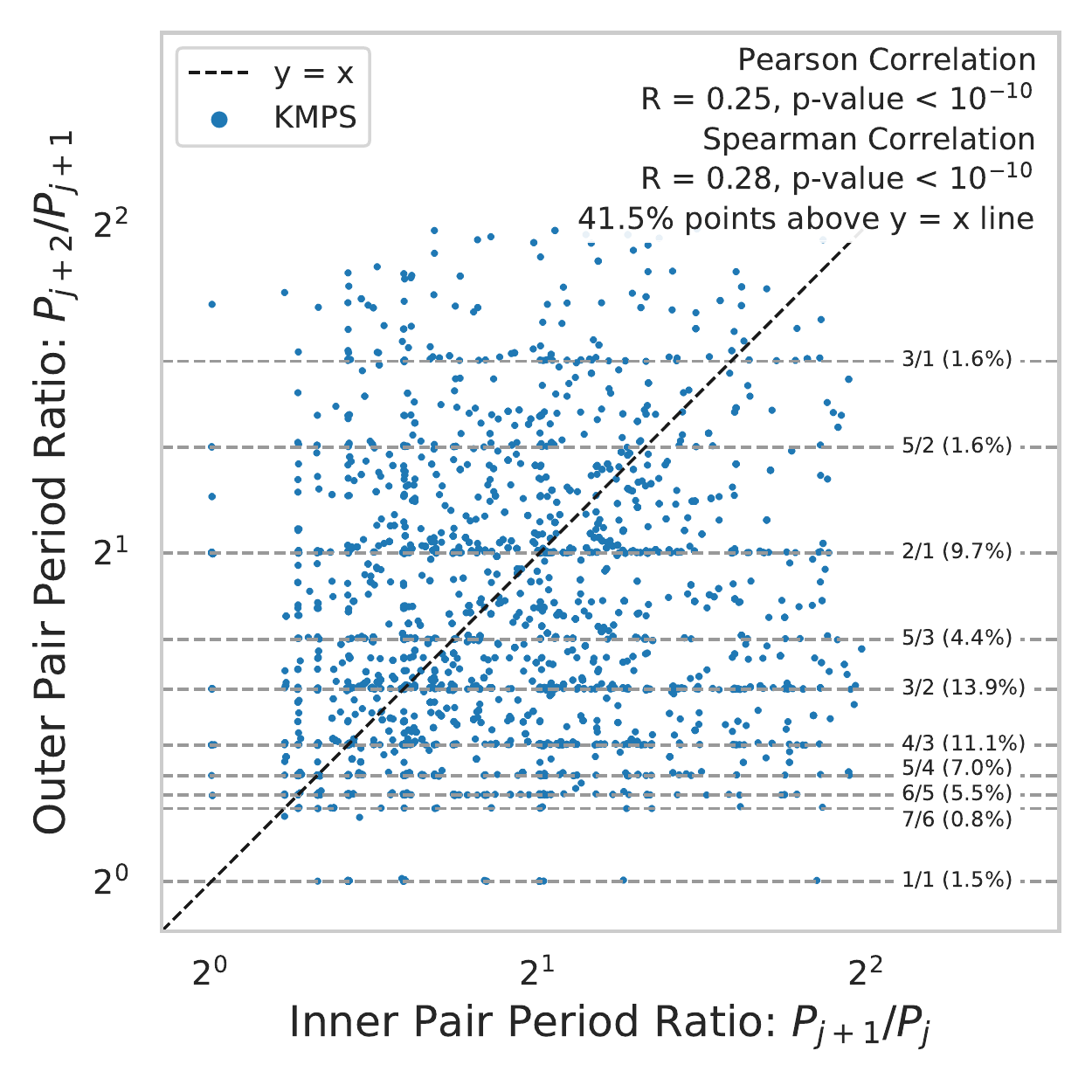}
                \caption{\textit{Peas in a pod: Spacing.} The plots shows the period ratio of the outer pair of planets as a function of the period ratio of  the inner pair for the underlying population (left), for the underlying population of detectable planets (middle), and for the \kmps/ systems (right). The dashed horizontal lines mark some of the important \editbold{commensurabilities}. The number inside the parenthesis is the percentage of outer planetary pairs that have a period ratio within $1\%$ of the indicated \editbold{commensurability}.}
                \label{fig:kmps_pip_spacing}
        \end{figure*}

        To investigate the correlation in spacing between adjacent pairs of planets (for systems with three or more planets) the ratio of periods are used. Figure \ref{fig:kmps_pip_spacing} shows the period ratio for an outer pair of adjacent planets $P_{j+2}/P_{j+1}$ as a function of the period ratio of an adjacent inner pair of planets $P_{j+1}/P_{j}$. Following \W18, the period ratios are limited to 4\footnote{\editbold{See \cite{Zhu2019} and \cite{Weiss2019} for a discussion and justification of this choice.}}.

        The correlation tests reveals that there is a positive correlation for spacing in the \kmps/ catalogue ($R = 0.25$). The observed CKSM catalogue showed even stronger spacing correlation with $R = 0.46$ (\W18). The underlying population shows a much stronger ($R = 0.55$) and significant correlation. This implies that for the theoretically observed and underlying population, the period ratio of one pair of planets is correlated with the period ratio of the next pair of planets. However, this trend is notably diminished when the underlying population is analysed by \kobe/ (discussed further in Sect. \ref{subsec:role_biases_spacking}). 
                
        The plots in Fig. \ref{fig:kmps_pip_spacing} shows that many pairs of planets are found in orbital mean motion \editbold{commensurability}. The dashed horizontal lines are shown to guide the eye for some of the important \editbold{commensurabilities}. The number in brackets is the percentage of outer planetary pairs that have a period ratio within $1\%$ of the indicated \editbold{commensurability}. For example, in the \kmps/ 100-embryo population, about $14\%$ of outer planetary pairs are in the $3/2$ orbital \editbold{commensurability},  and $11\%$ and $10\%$ of planetary pairs are close to the $4/3$ and $2/1$ \editbold{commensurability}, respectively. With a period ratio of $1/1$ there are also some cases of co-orbital \editbold{commensurabilities} \citep{Leleu2019stability}.
                
        The spacing correlation increases sharply as the number of embryos increases in the underlying populations (not shown). The introduction of more embryos in a system has several consequences. Most importantly, it increases the dynamical interactions between growing embryos and planets causing more merger collisions and ejection of planets. In addition to creating new planetary neighbours, these scenarios also lead to a dynamical clearing of space. For example, if three consecutive planets in a system have periods 1, 10, and 100 d, respectively, then all adjacent pairs have a period ratio of 10. The ejection of the middle planet creates new adjacent pairs with a period ratio of 100. If multiple planets, within a system, are clearing space through dynamical interactions, then this provides a mechanism for \editmajor{adjacent planets with correlated spacing to emerge}. In Sect. \ref{subsec:dynamicalinteractions}, the effects of dynamical interactions are analysed further. 
                
        Figure \ref{fig:kmps_pip_spacing} also shows that the frequency of spacing ordered adjacent planetary pairs (i.e.  where the period ratio of the outer pair is \editle{larger} than the inner pair) is always less than $50\%$. This suggests that it is more common to have larger spacing between the inner pair of planets for any three consecutive planets. This frequency decreases with increasing the number of protoplanetary embryos (not shown). This also suggests that increasing dynamical interactions plays a role in allowing adjacent planetary pairs with larger spacings to emerge. 
        
        The frequency of ordered adjacent pairs falls sharply for the population of detectable planets. This indicates that for three consecutive planets an inner pair that has a larger spacing than an outer adjacent pair is much more common in the inner region of a system. This could, potentially, be a result of limited \editminor{\textit{N}-body} calculation time. In NGPPS the \editminor{\textit{N}-body} calculations are done until 20 Myr. This means for a planet located at 1 au or 365 d that  the \editminor{\textit{N}-body} tracks its evolution for 20 M orbits. For planets that are further out their orbital \editle{evolution is tracked for lesser number of orbits}, which could thereby influence the results. 
        
        In the context of spacing between adjacent planets, another possibility to explore is the role played by the initial location of embryos (described in Sect. \ref{sec:ngpps}). A simple calculation allows us to derive the \editbold{expected value of} initial period ratio of embryos by converting uniform log spacing in semi-major axis $a_\text{embryo}$ to periods: 
        \begin{equation}        
        \log \ \frac{P_{j+1}}{P_{j}} = \frac{3}{2} \frac{1}{n_\text{emb}} \log \ \left(\  \frac{a_\text{embryo}^\text{outer}}{a_\text{embryo}^\text{inner}} \ \right)
        .\end{equation}  
        Here, $n_\text{emb}$ is the \editminor{total number of embryos initially} placed in a simulation and the factor $3/2$ comes from the application of Kepler's third law, and  $a^\text{emb}_\text{outer} = \SI{40}{au}$ is the maximum distance from the star at which an embryo can be placed. For the inner edge the mean value of $r_\mathrm{in} = \SI{0.055}{au}$ can be used. This provides an approximate value of the \editbold{average} initial period ratio of embryos. The value of this is \editminor{$1.6, 1.2, \text{and}\ 1.1$} for the populations with $20, 50, \text{and}\ 100$ embryos respectively. This implies that all planetary embryos start with \editbold{period ratios close to 1. While the initial period ratios of embryos is close to unity, the location of an embryo is assigned randomly (see Sect. \ref{sec:ngpps}). This would result in the absence of the spacing correlation at early times (see Fig. \ref{fig:pip_time}). It is clear from the plot that there is little trace of these initial values at 4 Gyr.} 
                
        Overall, all theoretical populations show a positive spacing correlation in agreement with observations. The spacing between one pair of planets is similar to the spacing between the next pair of planets. The large correlations present in the underlying population suggest that this trend is probably astrophysical in origin. Geometrical limitations and detection biases have a noticeable influence on the spacing correlation (see Sect. \ref{sec:roleofbiases}). For synthetic populations the period ratio of an inner pair of planets is often larger than the period ratio of the next outer pair. 
        
        \subsection{Peas in a pod: Packing}
        \label{subsec:pip_packing}

        \begin{figure*}
                \centering
                \includegraphics[height=5.6cm]{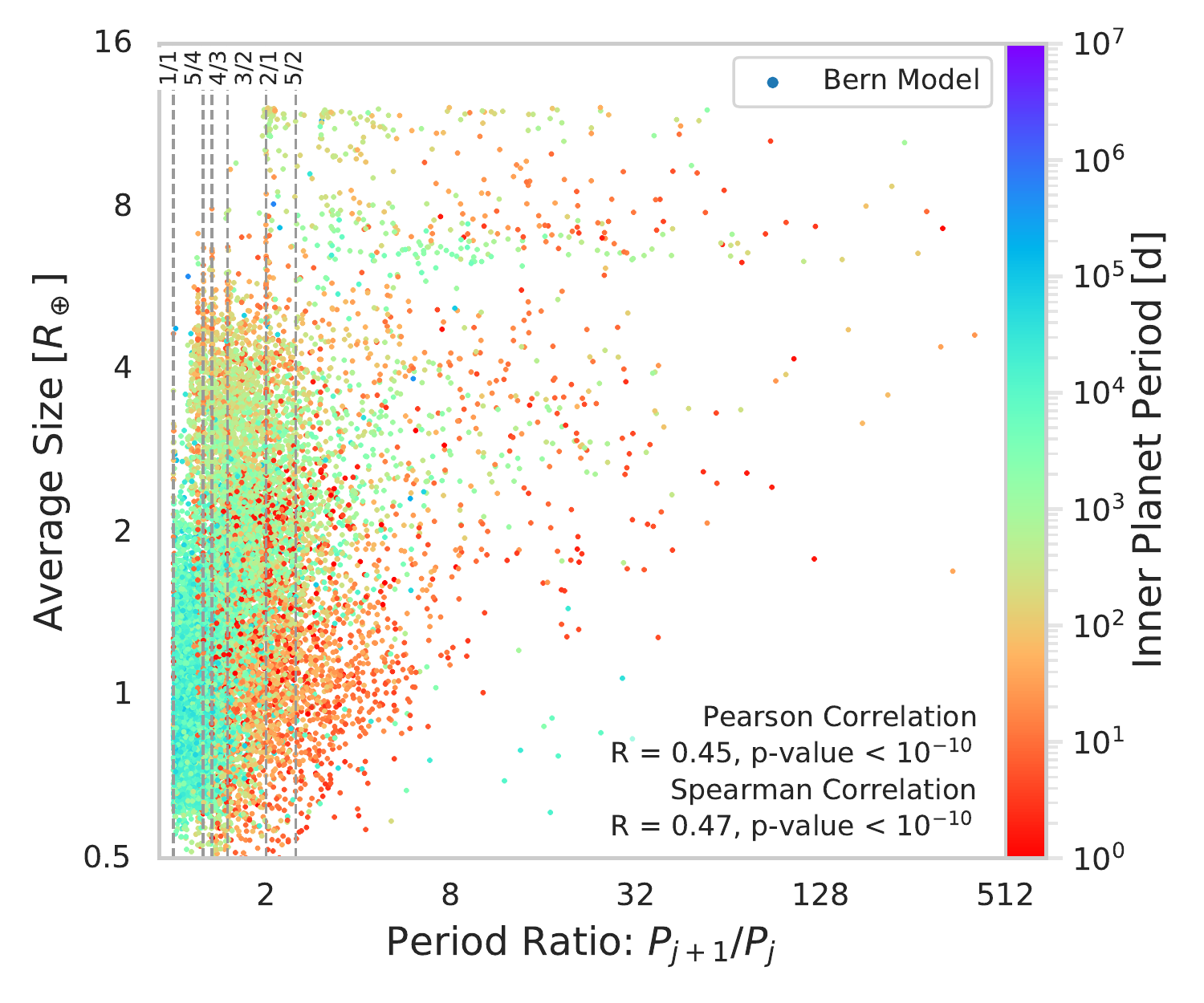}              
                \includegraphics[height=5.6cm]{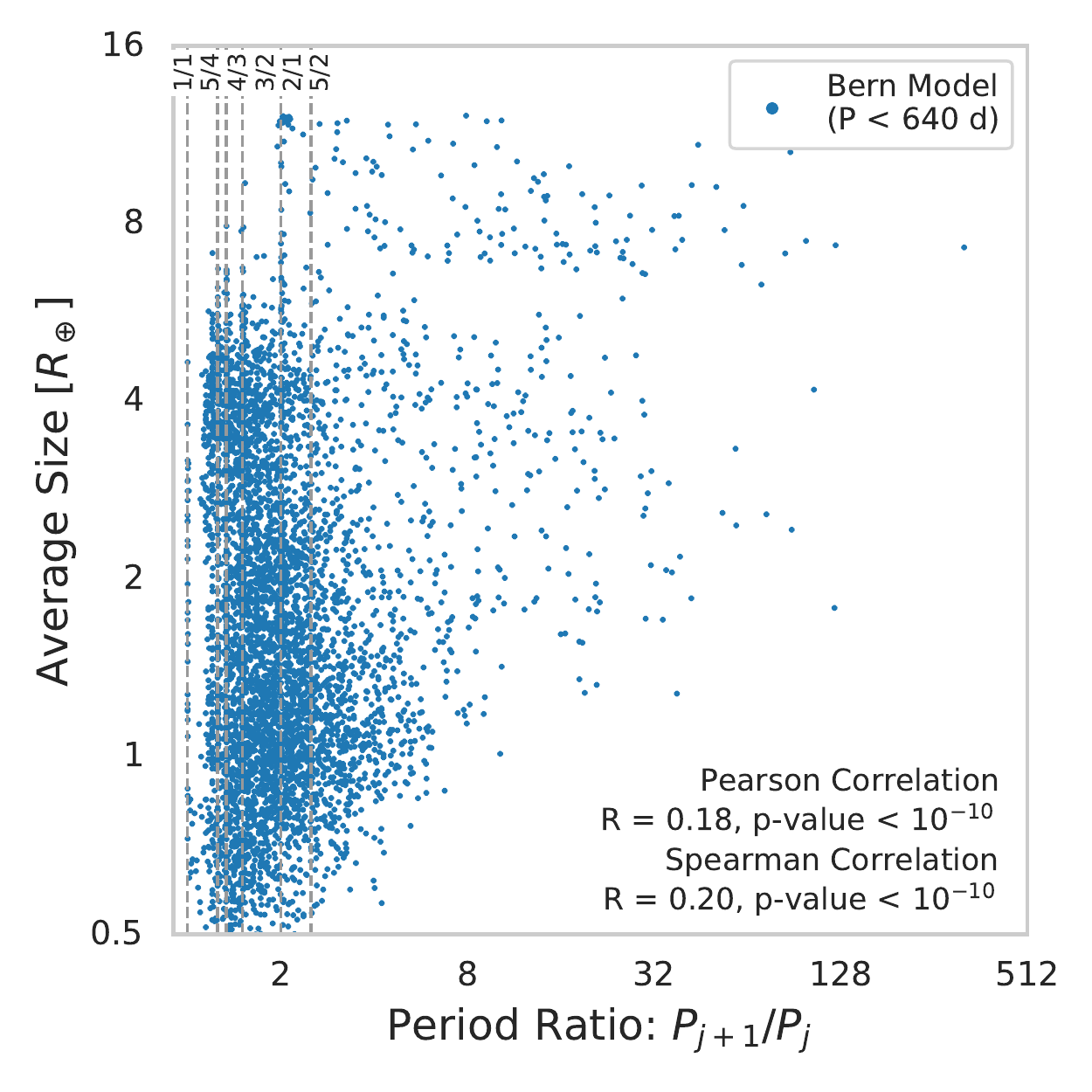}         
                \includegraphics[height=5.6cm]{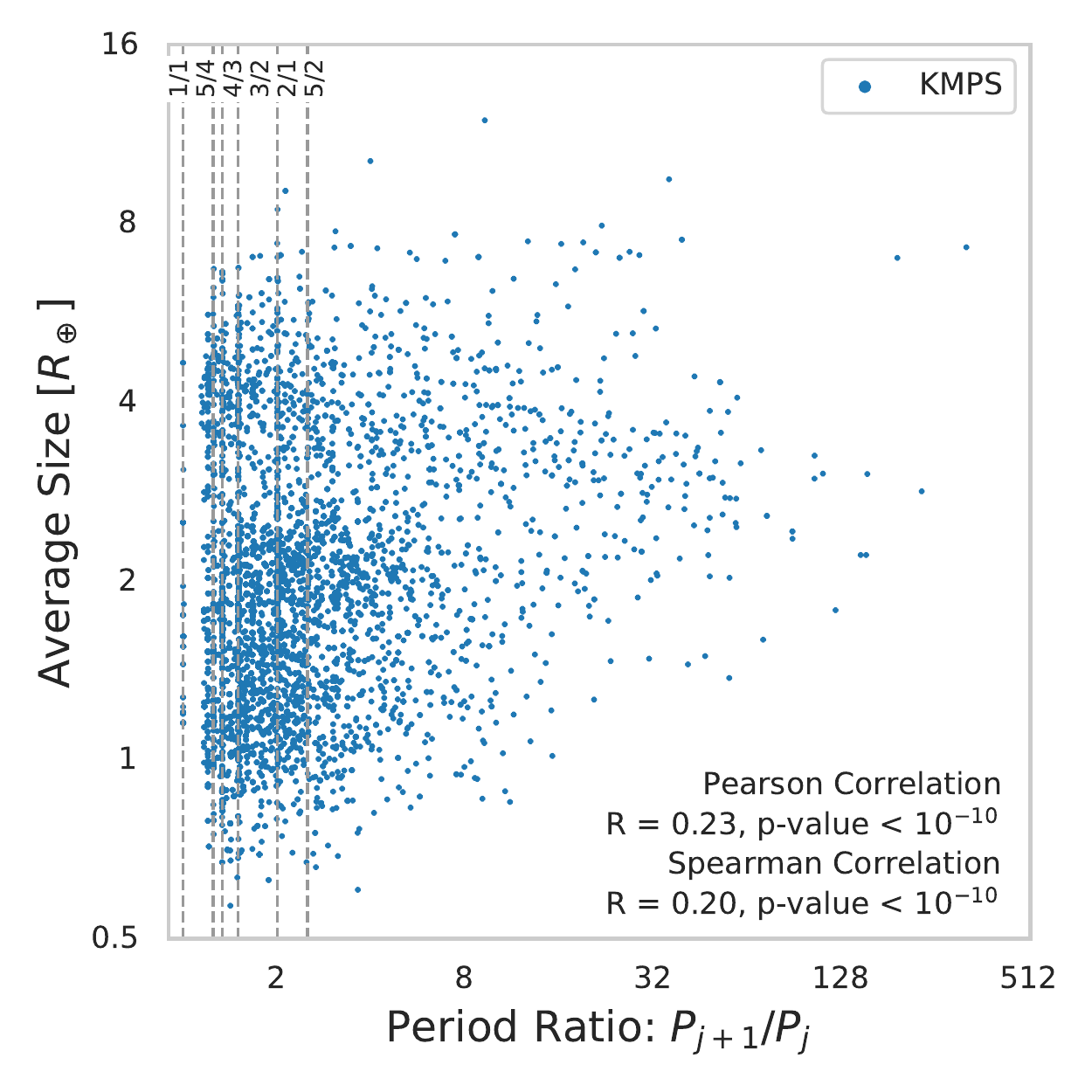}
                \\
                \includegraphics[height=5.6cm]{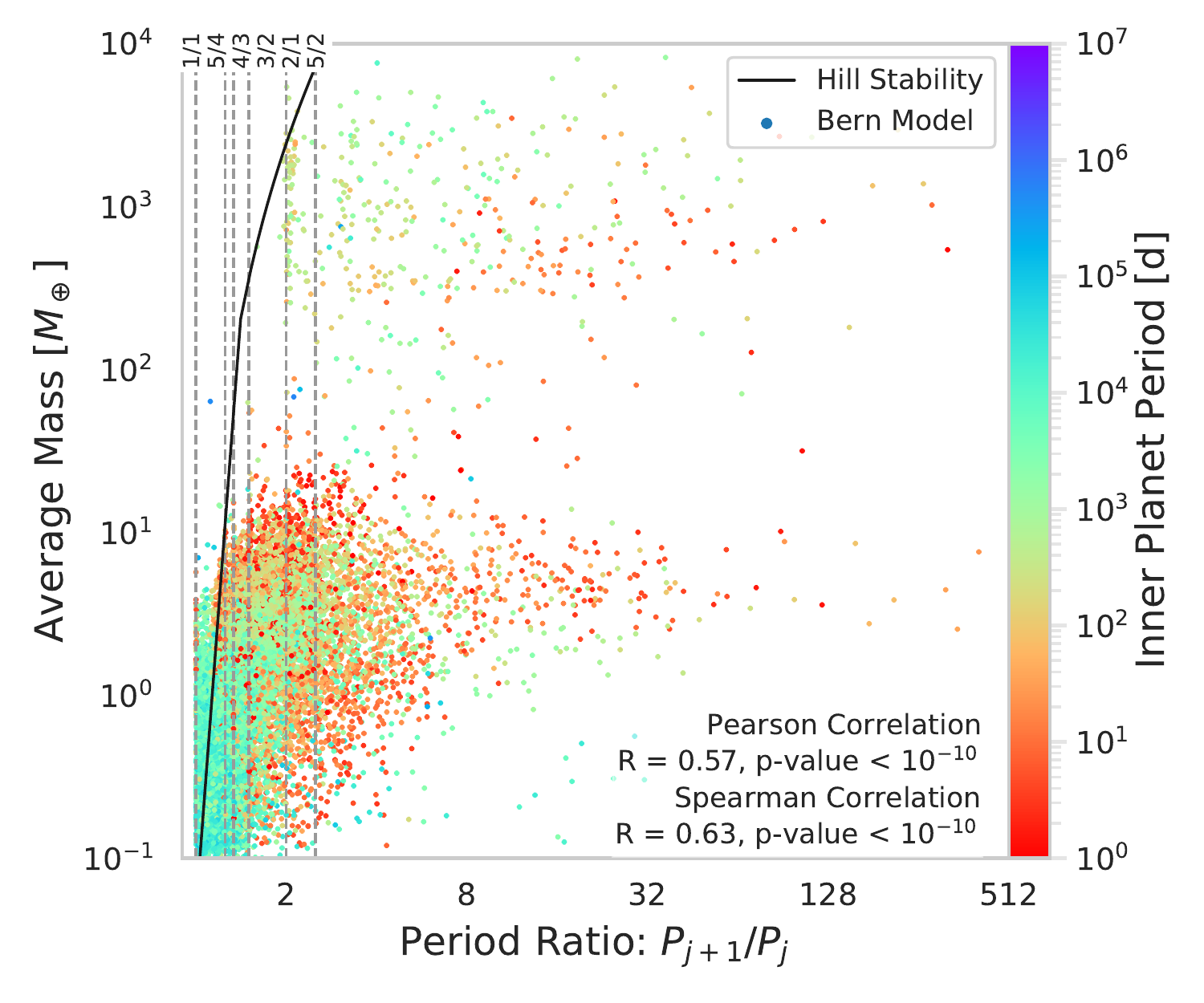}
                \includegraphics[height=5.6cm]{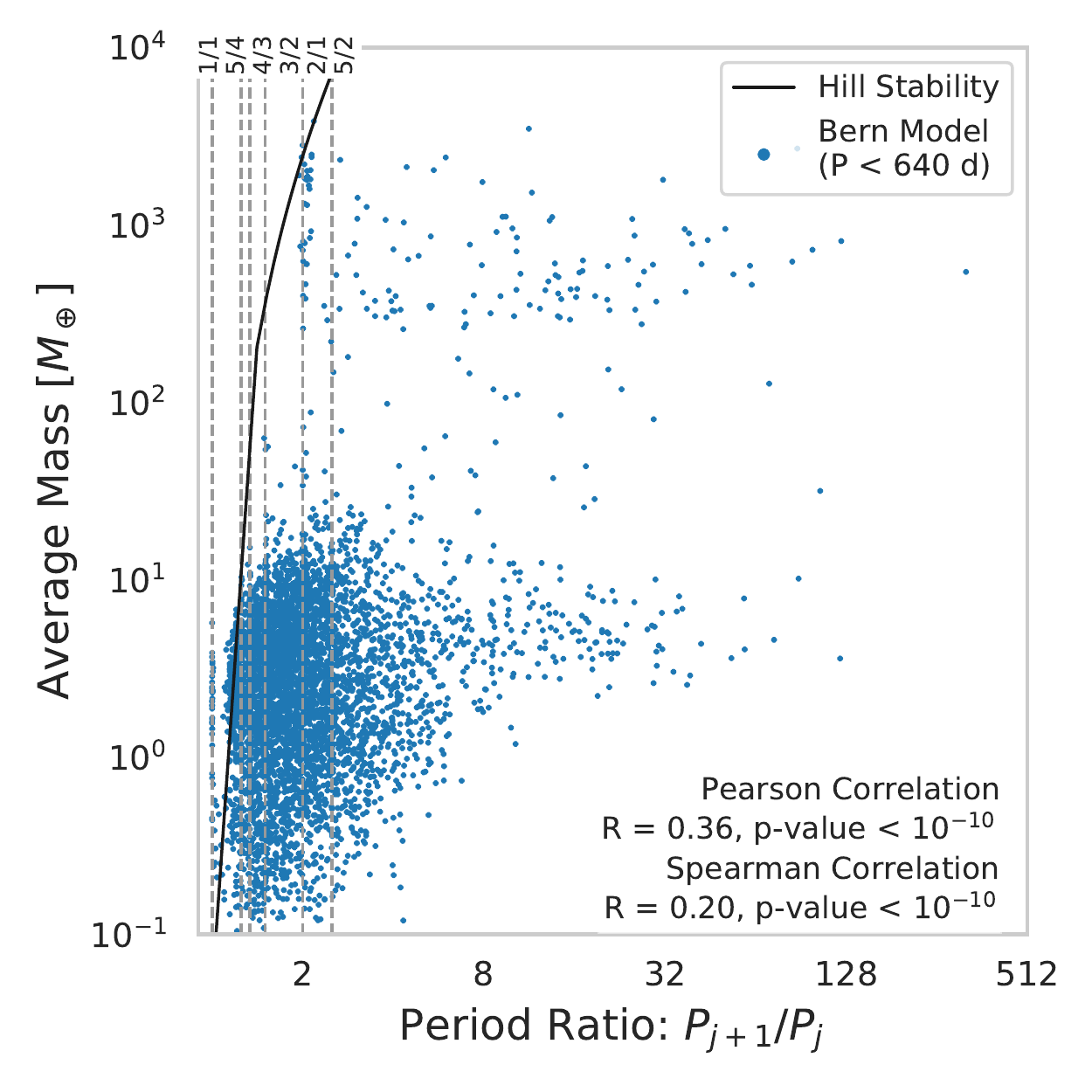}
                \includegraphics[height=5.6cm]{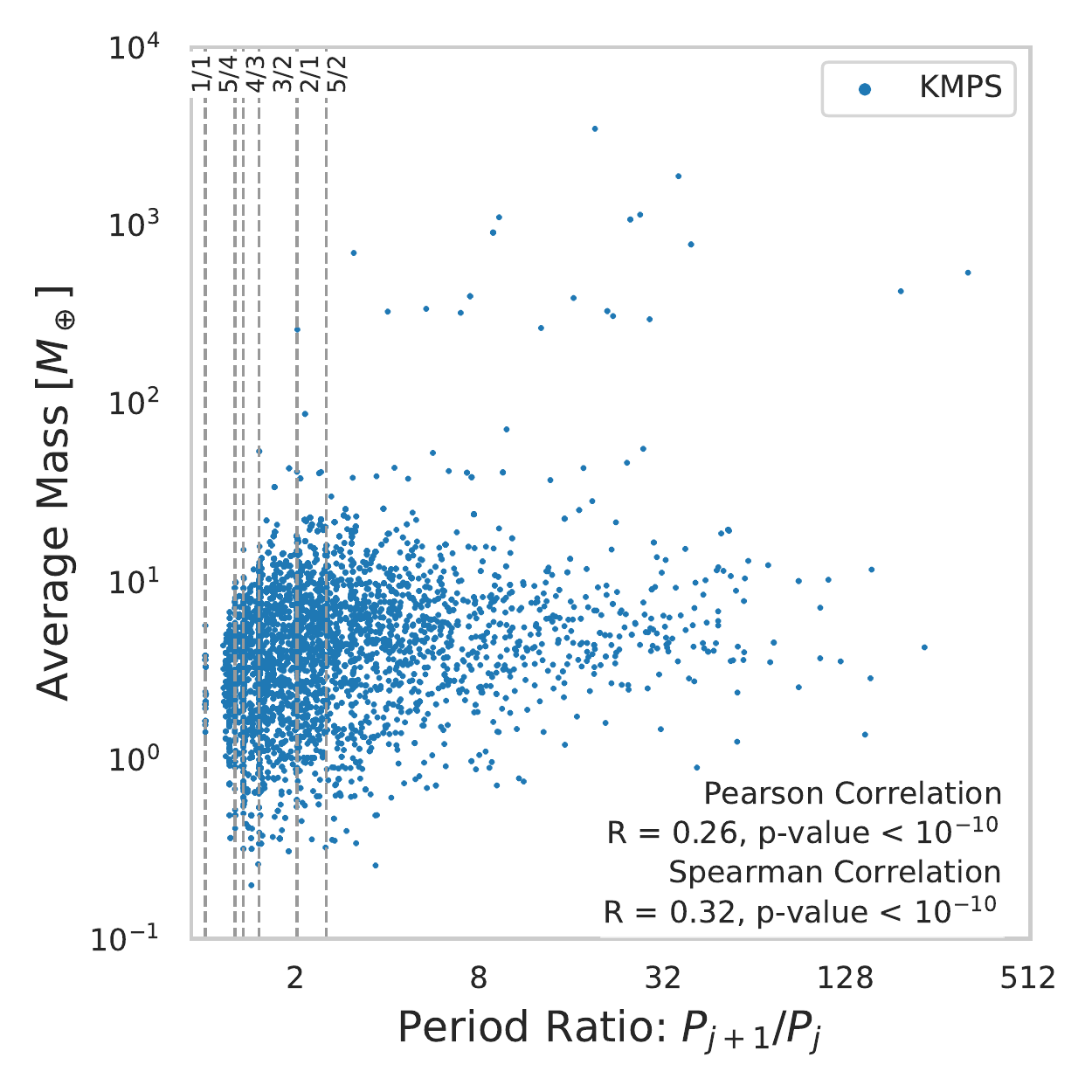}           
                \caption{\textit{Peas in a pod: Packing.} The average sizes (top) and average masses (bottom) of adjacent planets are shown as a function of their orbital period ratios $P_{j+1}/P_j$ for the underlying population (left), underlying population of detectable planets (middle), and \kmps/ planets (right). For the underlying population, the position of the inner planet in each pair is in a different colour, showing that the trend is due to those planetary pairs where the inner planet is close to the host star. The black curve shows the Hill stability criterion from \cite{Deck2013}. Adjacent planetary pairs on the right side of this curve are Hill stable. Points on the left side are Hill unstable and will probably be removed with longer N-body calculations.}
                \label{fig:kmps_pip_packing}
        \end{figure*}
     
        \cite{Weiss2018} have found that smaller planets tend to have small spacing while larger planets are likely to have large spacing. There is a correlation between the average size of an adjacent planetary pair with their period ratio. The Pearson correlation coefficient for the packing trend in the CKSM catalogue is $ R = 0.26$. 
        
        It is suggested in Sect. \ref{subsec:pip_mass} that the correlations seen in sizes  probably arise from underlying correlations present in planetary masses. The correlations in planetary masses are probably more fundamental than those of planetary radii. A further test of this idea could be if the packing correlations were to also exist in planetary masses. This has not been reported in the literature before. Figure \ref{fig:kmps_pip_packing} shows the average size (top) and average mass (bottom) of adjacent pairs of planets as a function of their period ratios $P_{j+1}/P_j$.
        
        For the \kmps/ population the Pearson correlation coefficient for the packing trend (with average sizes) is $R = 0.23$, which is in good agreement with observations. The plot shows that for planetary pairs of average size $1~\rearth$, the spacing is generally lower than pairs of average size $2~\rearth$. The correlation stems from the lack of planetary pairs with small average sizes and large spacing between them. Figure \ref{fig:kmps_pip_packing} (top left) shows that this correlation is even stronger in the underlying population. Here the correlation coefficient is $R=0.45$. The plot shows that while there is a cluster of points with low period ratios ($P_{j+1}/P_j < 2$) extending from average sizes $0.5$ to $5~ \rearth$, the correlation seems to emerge from the lack of small  planetary pairs with large spacing. For example, there is only one pair of adjacent planets in the underlying population with average size between $1-2~\rearth$ and period ratio between $128 - 512 $. For the same period ratio bin, there are two pairs of planets with average sizes between $2-4~\rearth$, while there are eight pairs of planets with average sizes between $4-8~\rearth$. 
        
        Figure \ref{fig:kmps_pip_packing} (bottom) shows that the average mass of planetary pairs is indeed correlated with their spacing. The correlation of period ratios is stronger with average mass than with average size. The correlation coefficient is $R=0.26$ for the \kmps/ population and increases to $R=0.57$ for the underlying population. These plots show features that are similar in quality to the plots with average sizes. 
        For all populations there are no planetary pairs with average mass $> 1~000 \mearth$ and spacing \editminor{$P_{j+1}/P_j < 3/2$}. \editmajor{This can be explained by invoking stability arguments. \cite{Deck2013} studied long-term stability of planetary systems and provided stability criteria. The Hill stability criteria (eq. 59 from their paper), relating masses and locations of two planets, is plotted in Fig. \ref{fig:kmps_pip_packing}. A pair of planets are Hill stable if they are on the right side of the black curve.}
        
        To further understand the packing trend, the location of the inner planet (in an adjacent pair) is shown in colour for the underlying population. The coloured plot shows several interesting features. This trend is mostly driven by planetary pairs where the inner planet is located close to the host star ($P \lesssim 10 \ \si{\day}$). For these pairs of planets the spacing seems to increase with their average size and mass. 
        
        As mentioned in Sect. \ref{subsec:pip_spacing}, dynamical interactions can lead to merger collisions and ejection of planets. This results in dynamical clearing of space between planets. Large planets may undergo several collisions leading to the ejection or accretion of several planets. This may allow them to have wider orbital spacing. On the contrary, small planets may not  have undergone several collisions, thereby remaining in compact configurations. This could explain how small planets tend to have smaller orbital spacing and large planets have wider orbital spacing. Due to limited \editminor{\textit{N}-body} integration time, the inner region (< \SI{1}{au}/\SI{365}{\day}) of a planetary system experiences many more dynamical interactions than the outer region. This explains the small contribution towards the packing trend from planets that are in the outer region (green points). This scenario is discussed further in Sect. \ref{subsec:dynamicalinteractions}.
        
        Overall, the findings of this section indicate that the average mass (and therefore the average size) of a planetary pair is correlated with their spacing. Planets with smaller masses are packed closely together, while massive planets seem to have larger orbital spacing between them. As these correlations are also present in the underlying population, it hints towards an astrophysical origin of this trend. In \W18 this trend was further examined through the mutual separation ($\Delta$) of adjacent planet in units of mutual Hill radius. Their findings can be explained by detection biases, and is discussed in Appendix \ref{sec:mutual_hill}.
        
        \section{Role of detection biases in peas in a pod trend}
        \label{sec:roleofbiases}
        
        \begin{figure*}
                \centering
                \includegraphics[width=8cm]{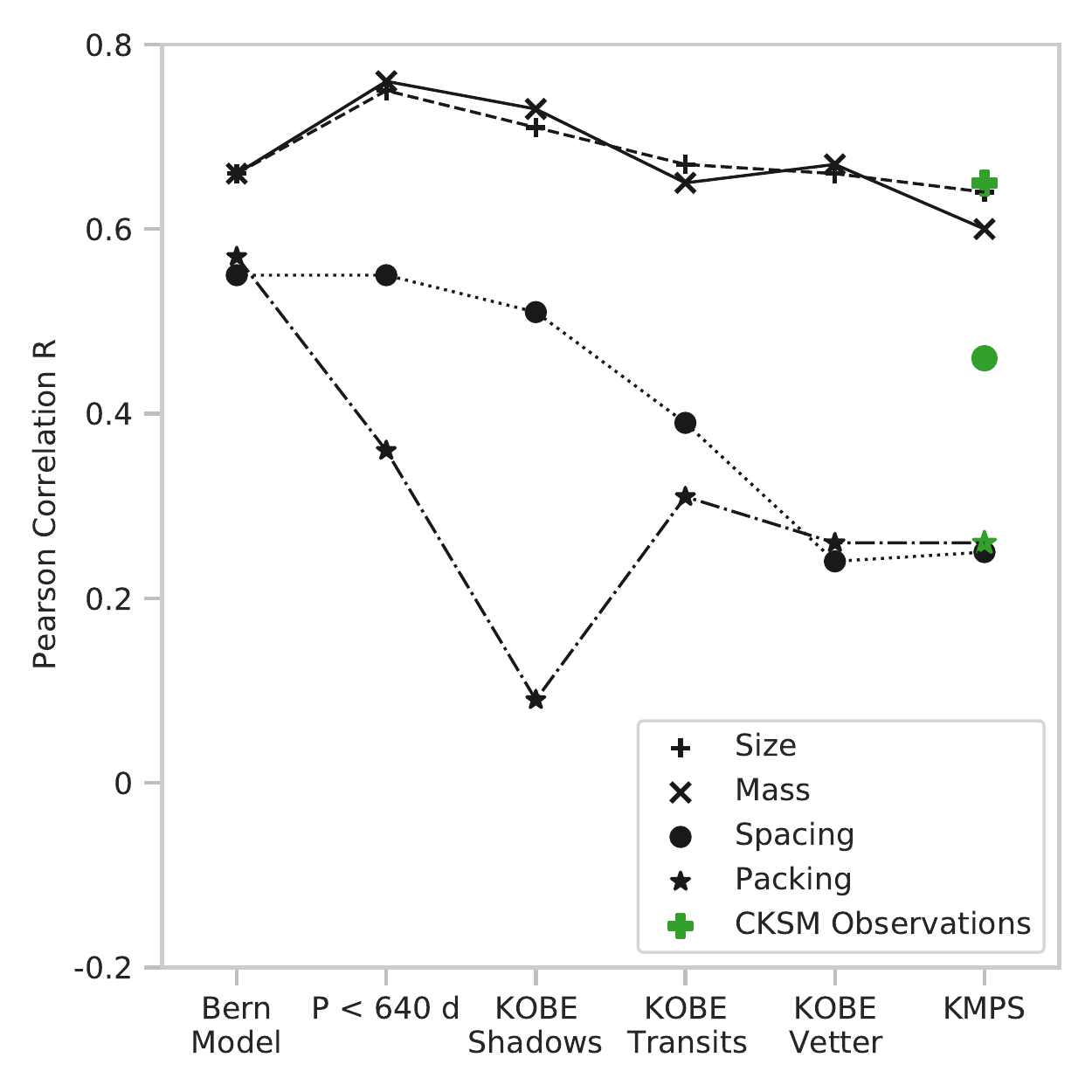}
                \includegraphics[width=8cm]{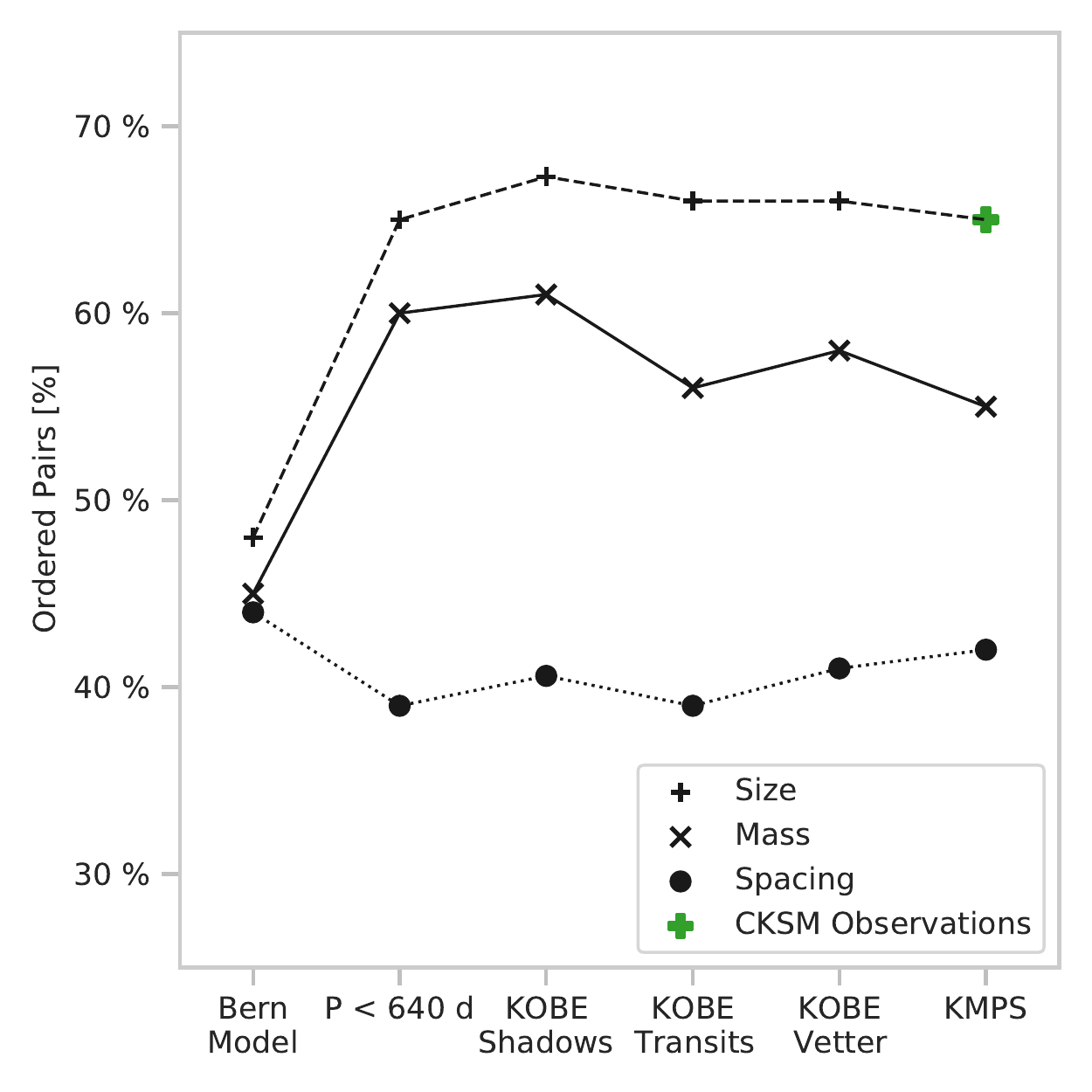}
                \caption{Influence of the geometrical limitations of the transit method (\kobeshadows/), the transit detection biases (\kobetransits/), and the completeness of the Kepler survey (\kobevetter/) on the peas in a pod trend. The plot shows how the correlation coefficients (left) and the frequency of ordered pairs (right) varies in the underlying Bern Model population, the underlying population of detectable planets (P < 640 d), and the theoretically observed \kmps/ population \editbold{(for the size--mass trends in the \kmps/ catalogue adjacent planetary pairs have undergone a swapping test, as mentioned in Sect. \ref{subsec:pip_size})}. Observations from the CKSM exoplanetary catalogue are shown in green.}
                \label{fig:role_kobe}
        \end{figure*}
        
        Population synthesis based on planet formation models provides a natural playground for testing \editbold{the role of detection biases of the transit method in the peas in a pod} trends. \editminor{The Bern Model} consists of theoretical description for many physical phenomena that are active during planet formation. Supplying them with randomized initial conditions and \editminor{\textit{N}-body} calculations, the NGPPS provides a theoretical \editminor{version} of nature's underlying population. The \kmps/ catalogue, from \kobe/, stands on the same footing as observations (CKSM). This work thus  allows both the theoretically observed exoplanetary populations (from \kobe/) and the theoretical underlying populations (from NGPPS) to be investigated for the peas in a pod trend.
        
        To understand how the geometrical limitations and the detection biases of the transit method affect the peas in a pod trends, the correlation test was performed after each stage of calculations in \kobe/. Figure \ref{fig:role_kobe} (left) shows the Pearson correlation coefficient for the similarity trends in size, mass, spacing, and packing. Figure \ref{fig:role_kobe} (right) shows the percentage of ordered pairs for size, mass, and spacing \footnote{An ordered pair is one in which the outer planet (from the star) has the larger value for a given quantity (e.g. radius, mass).}. Observations from CKSM are in green.

        The similarity and the differences of these trends can be understood via the following statement.The chances of detecting a transiting exoplanet depend strongly on its location (star--planet distance and orbital period) and weakly on its size (radius). Specifically, the size dependence is from $R_\mathrm{planet}/\rstar$ \editminor{(see eq. \ref{eq:transitprobability}), which varies from $10^{-3}$ for sub-}Earth-size planets to $10^{-1}$ for Jupiter-size planets around a Sun-like star. This suggests that the effect of geometrical limitations and detection biases will be much more severe on orbital periods than on planetary sizes. This is easily seen from the plots in Fig. \ref{fig:kobeexample} and Fig. \ref{fig:role_kobe}.
        
        \subsection{Peas in a pod: Mass and size}
        
        One striking feature in Fig. \ref{fig:role_kobe} is that the size trend closely follows the mass trend. The small variations between the two trends  probably arise from the scatter seen in the mass-radius diagram (see Fig. \ref{fig:kmps_pip_m_r}). The underlying population shows strong mass (and thereby size) similarity and ordering correlations. This strongly indicates that the peas in a pod mass (and thereby size) trend  arises from planet formation. 
        
        \editmajor{The geometrical limitations and detection biases of the transit method tend to decrease the strength of the similarity correlations.} The vetting procedure, in \kobevetter/, seems to have little effect on the mass (size) similarity correlations. Although the completeness of Kepler's Robovetter drops sharply with radius (see Fig. \ref{fig:cdppvetter}), the frequency of large planets in the \kobevetter/ catalogue is also low: about $70\%$ of planets have $\rplanet \leq 3~\rearth$. Finally. the size correlations seen in the \kmps/ catalogue match  the observations very closely. 
        
        However, \kobe/ \editmajor{has little influence on} the mass--size ordering trend. For the underlying population of detectable planets, the frequency of mass--size ordered pairs is close to $60\%$. This means that there is a higher chance for an outer planet in a pair to be heavier and/or larger. The frequency of size-ordered pairs in \kmps/ matches  CKSM observations very closely.

        Since the size-similarity and ordering trend in the \kmps/ populations very closely  matches the  observations, one could extrapolate this to learn about the  nature of the  underlying exoplanetary population. These results suggest that the size--mass similarity and ordering correlations found in observations are probably astrophysical and are not severely affected by detection biases.         
        
        \subsection{Peas in a pod: Spacing and packing}
        \label{subsec:role_biases_spacking}
        
        The underlying populations shows strong spacing (for systems with three or more planets, period ratios limited to 4) and packing trends. Both of these trends involve period ratios of adjacent planets, already hinting that these trends will be strongly influence by \kobe/. \editminor{One way in which \kobe/ influences the spacing and packing trends is due to missing planets.}
        
        \kobeshadows/ finds transiting planets that  have a fortuitous alignment with an observer. Transiting planets found by \kobeshadows/ are not necessarily consecutive. In several cases many intermediate planets are missed, resulting in a strong affect on period ratios. However, the effect of missing planets may be more adverse on the packing trend than on the spacing trend. Consider a hypothetical system with five planets at periods of 1, 10, 100, $1\,000$, and $10\,000$ d. The period ratio for all four adjacent pairs is 10, and the ratio of period ratios for any three consecutive planets is 1. If the planets with periods of 10 and $1\,000$ d do not transit for an observer, then the period ratios of the two transiting adjacent pairs jumps to 100. However, for the three transiting planets the ratio of their period ratios is still 1. If the two transiting adjacent planets have small average sizes, then the jump in the period ratio will weaken the packing correlation. This example demonstrates the adverse effect of missing planets on the packing trend\footnote{Each stage of \kobe/'s calculation introduces some randomness (location of observers or vetting planetary candidates). Even if all planets in a system were  in the same orbital planet, the randomness inherent in \kobe/'s calculation \editmajor{or the variation in their sizes} could lead to random missing planets.}. This explains the diminishing of the spacing trend and the absence of packing correlation from the 100-embryo population in the \kobeshadows/ catalogue.
        
        \kobetransits/ requires that all transiting planets have at least two transits, which implies that only planets with $P < 640 \si{\day}$ can be included. This means that only the inner region of a planetary system is now considered. This helps in removing pairs with abnormally high period ratios caused by missing planets. This may explain how the packing trend is restored in the catalogue from \kobetransits/. The spacing trend is reduced further by \kobetransits/ and \kobevetter/. These modules provide imprints of the physical detection biases and completeness profile of the Kepler pipeline. 
        
        The role of adding biases on the spacing ordering trend can be seen in Fig. \ref{fig:role_kobe} (right). For the underlying population the frequency of ordered pairs is less than $50\%$ (for three consecutive planets there are more inner pairs with larger spacing than their next outer pair). There seems to be little influence of \kobe/, and thereby detection biases, on the frequency of ordered pairs. 
        
        Overall, the underlying populations show strong spacing and packing trends. Geometrical limitations and detection biases of the transit method are responsible for reducing the strength of these correlations.

        \section{Discussion:  Theoretical scenarios}
        \label{sec:theoreticalscenarois}
        
        The results of Sect. \ref{sec:peasinapod} indicate that the peas in a pod trend is present in the synthetic planetary systems from the Bern Model. This section is dedicated to the discussion of some theoretical scenarios that offer partial explanations for these trends. 
        
        \subsection{The evolution of peas in a pod}
        \label{subsec:pip_time}
        \begin{figure}
                \centering
                \resizebox{\hsize}{!}{\includegraphics{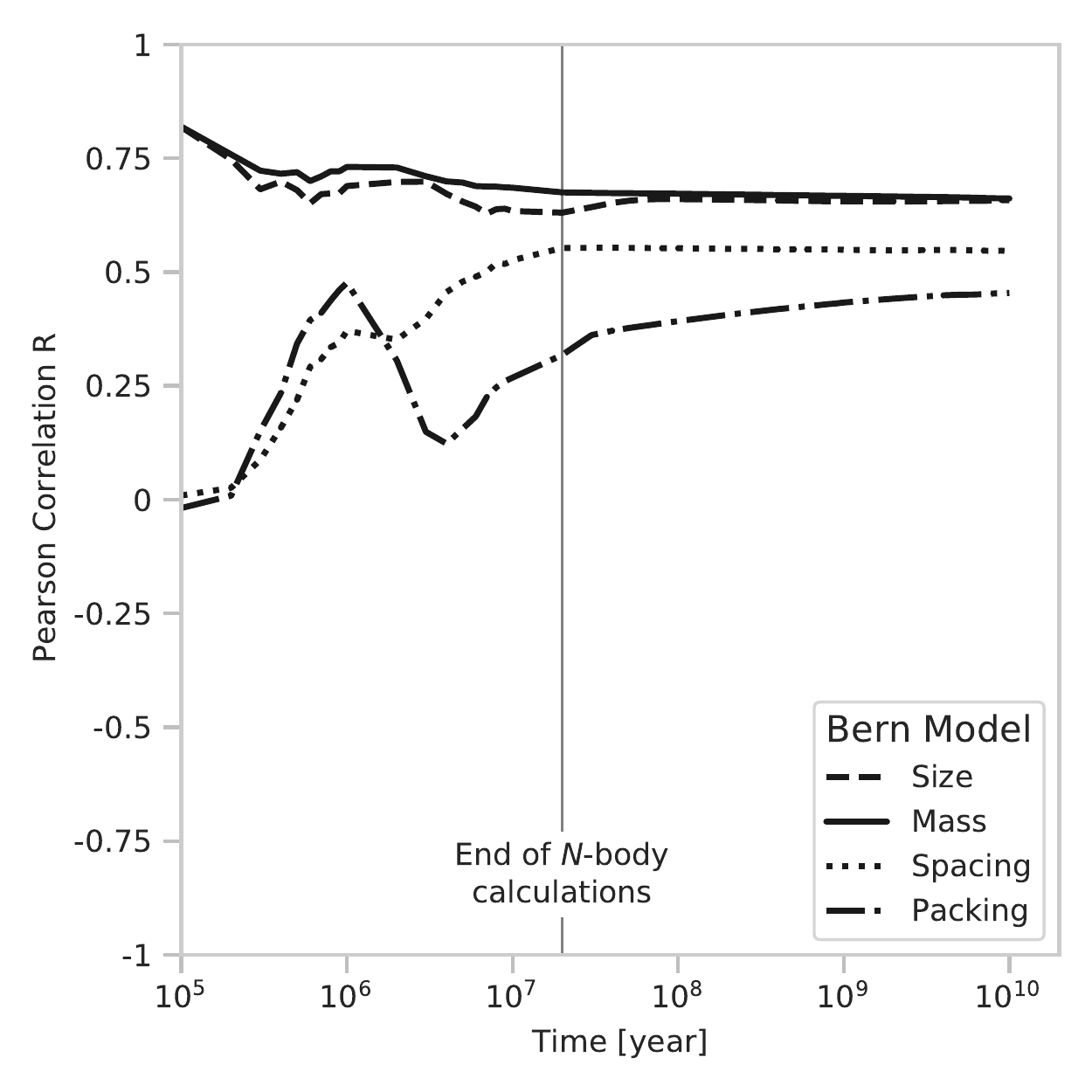}}
                \caption{Evolution of the peas in a pod trends. The vertical solid line represents the end of \editminor{\textit{N}-body} calculations.}
                \label{fig:pip_time}
        \end{figure}

        \begin{figure*}
        \centering
        \includegraphics[width=8cm]{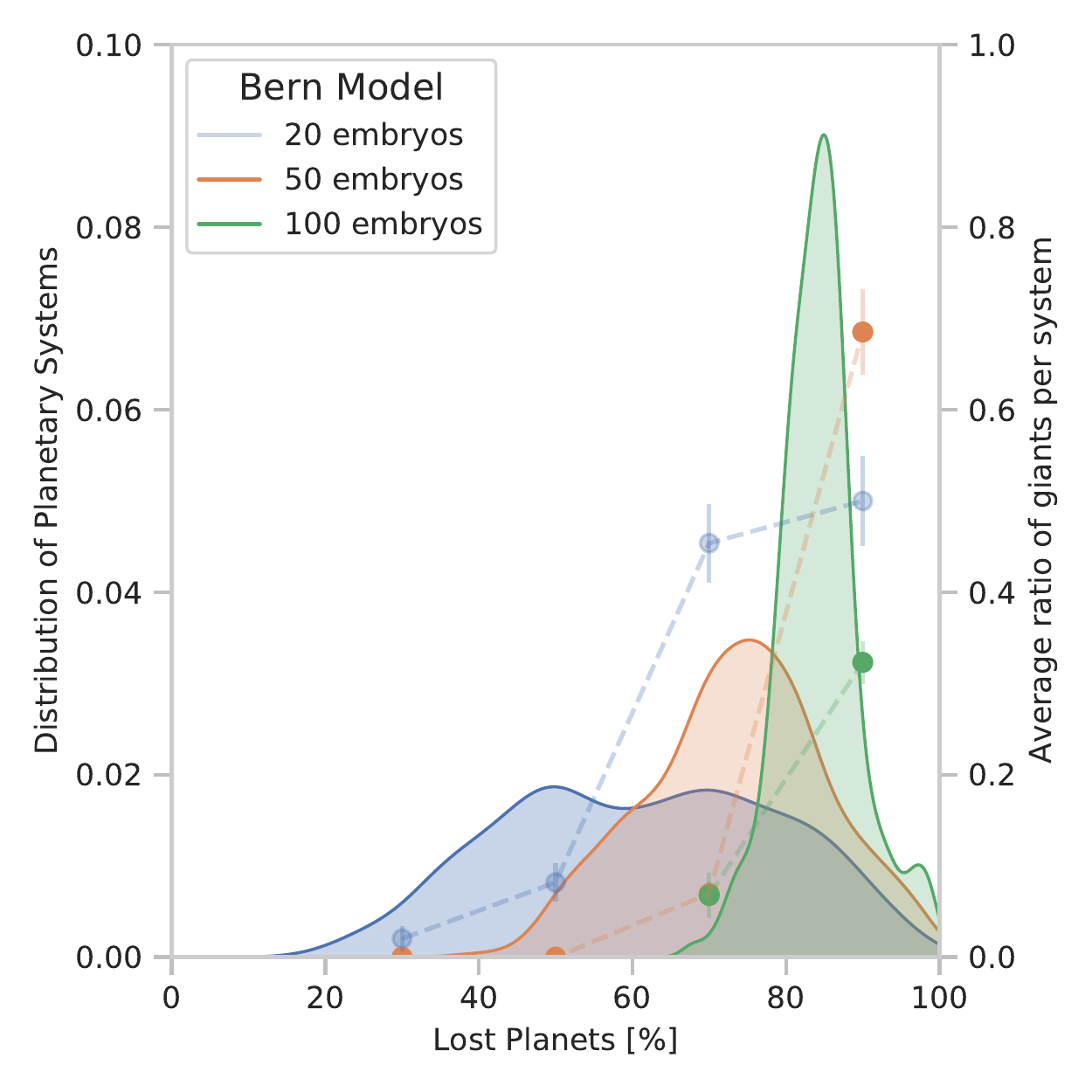}
        \includegraphics[width=8cm]{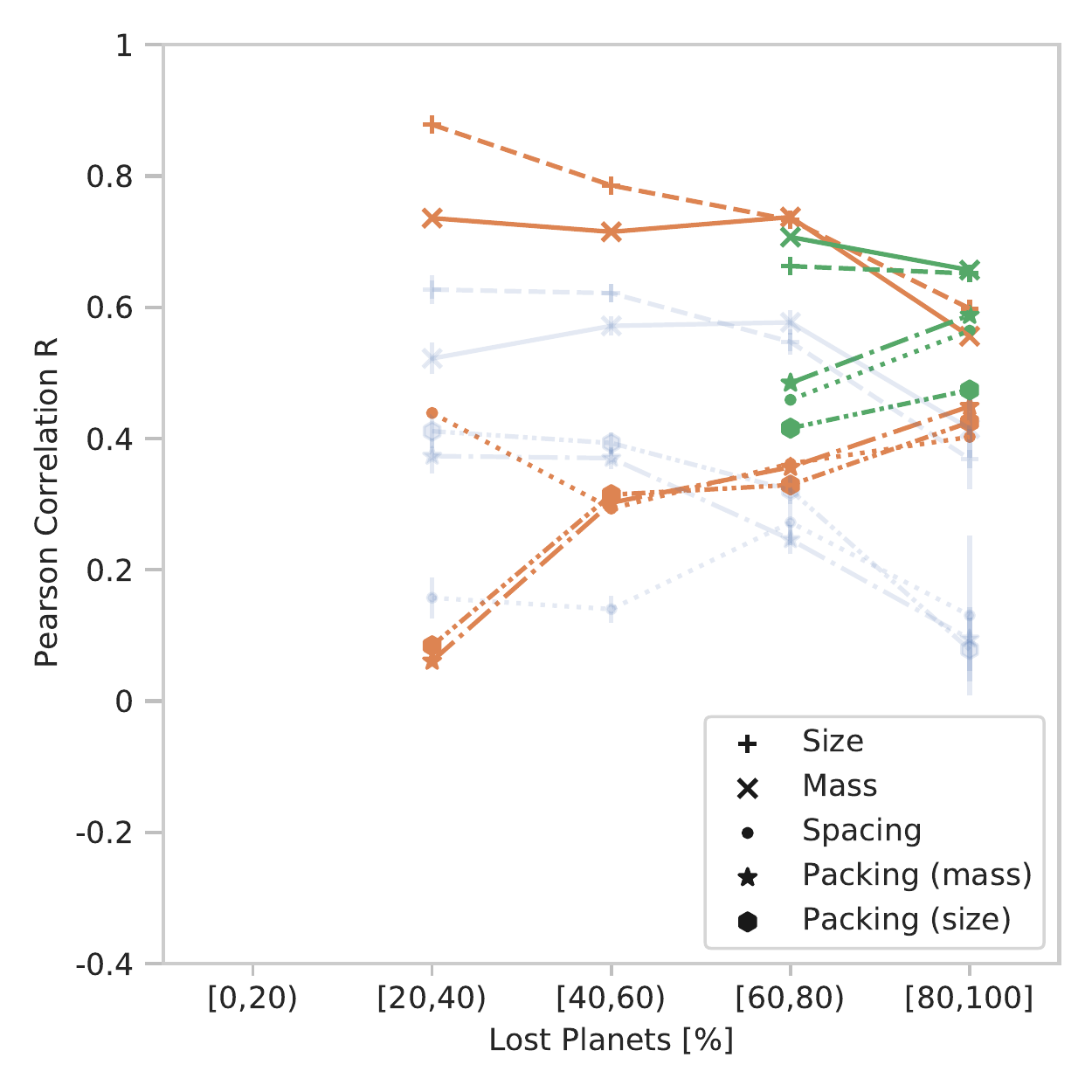}
        \caption{Role of dynamical interactions on the peas in a pod trends. Left: Distribution of lost planets [$\%$] by systems in the 20-, 50-, and 100-embryo NGPPS populations. The fraction of planets lost by a system can be used as a proxy for the cumulative dynamical interactions experienced by a system. The dashed lines show the ratio of giant planets per system averaged over bins.  
        Right: Correlation coefficient for the peas in a pod trends  for each bin. The error bars correspond to the standard error of the correlation coefficient \citep{Zar2014}. Increasing dynamical interactions results in the strengthening of the spacing and packing trend.}
        \label{fig:lost_planets}
        \end{figure*}
        
        One way to understand how the peas in a pod trends emerge is by investigating when the trends emerge. To this end, the correlation tests for all trends were performed for all underlying populations at all time steps. Figure \ref{fig:pip_time} shows the evolution of the correlation coefficients for the underlying 100-embryo population.
        Since most of the variations happen during the \editminor{\textit{N}-body} calculations, this suggests that dynamical interactions during the formation stage play a key role in shaping these trends.
        
        The plot shows that the underlying population shows a very strong correlation for the size and mass trends, already at the beginning of the calculations. This suggests that the peas in a pod mass (and thereby size) similarity trends are present at very early times. This high correlation can be attributed to two factors: oligarchic growth of protoplanetary embryos and uniform accretion of solids by protoplanets at early times (see Sect. \ref{subsec:planetformation}). The Bern Model starts with lunar mass embryos that are separated by at least $10~\rhill$. Runaway growth of planetesimals leads to protoplanets, which eventually grow oligarchically. The oligarchic growth stage results in mass ratios approaching unity \citep{Kokubo1998}. In this way the seeds for the peas in a pod mass trend (and therefore also the size trend) are already planted
        \footnote{\editbold{Oligarchic embryos offer only a partial explanation of the early mass--size similarity. In other words, oligarchic embryos do not explain why adjacent planets have similar masses on Myr timescales. To check this, a 50-embryo population was simulated where the embryos had initial masses between $0.0001 \mearth$ and  $ 0.01\mearth$. We find that this synthetic population shows similar (but  slightly weaker) mass similarity correlations as the population with fixed initial embryo mass (at early and late times). In addition, the final mass of a planet seems to have little dependence on the initial embryo mass. If fixed initial embryo mass could explain mass similarity on Myr timescales, then we would have seen weak or no mass similarity correlation in the population where embryo masses were varied,  which does not seem to  be the case here. This indicates that additional physics involved in planet formation is essential for obtaining a detailed understanding of the peas in a pod mass similarity trend.}}. In the Bern Model, protoplanetary embryos accrete solids from the planetesimal disk at a rate given by eq. \ref{eq:coreaccretion}. This core accretion rate prominently depends on the location and mass of the embryo as well as the surface density of the disk, $\sigmamean$. Since these factors (surface density of solid disk and location and mass of embryos) do not undergo any drastic changes at early times, the accretion of solids by neighbouring protoplanets is uniform. Thus, uniformly growing oligarchic embryos may explain the high mass--size correlation seen at $t = 10^5~\si{\year}$ in Fig. \ref{fig:pip_time}. 
        
        Between $10^5$ and  $10^6 ~\si{\year}$ the correlation coefficient for mass (and thereby size) drops. This could be attributed to the differences in the rate of solid accretion by cores of different types of planets. The cores of giant planets have to reach a critical mass ($\mcore \approx 10 - 20 \mearth$) before the gas disk dissipates \citep{Pollack1996, Alibert2005}. On the other hand, planetary cores which will fail to reach this critical mass (for runaway gas accretion), are known to have longer formation times (\paperone/). When adjacent planetary cores in a system  grow at different rates, the correlation between their masses and sizes may decrease. The size correlation seems to trace the mass correlation (with some scatter). That the size trend follows the mass trend is not surprising, since planetary sizes are calculated from their masses (via internal structure calculations). 
        
        Between $10^6$ and  $2\times 10^7~\si{\year}$ the correlation coefficient for mass decreases slightly. Most giant planets  have acquired their final masses in the first few million years. Other planets, however, continue to grow by solid accretion, gas accretion (before the gas disk dissipates), and merger collisions. This implies that adjacent neighbours may be growing at different rates depending on their local environment. Different growth rates imply that the mass correlation will decrease. The local environment around planets growing in the same disk does not suffer any drastic changes. This may explain why the mass correlation also does not show any drastic changes. Additionally, the dissipation of the gas disk has a strong effect on planetary radii since planets contract rapidly after disk dispersal. This may contribute to the decreasing size correlation in this time period. 
        
        
        The spacing and packing trends start with almost no correlation and undergo interesting variations before ending with their final value.   Initially, adjacent planets have uncorrelated small period ratios and small sizes. This may explain the absence of these trends at early times. Some physical processes that affect the location of a planet are orbital migration, resonance capture, and ejection or collision of planets. When a planet is lost (via ejection or collision), it clears up space allowing new adjacent pairs to emerge with wider orbital spacing. This dynamical sculpting may explain how planets within a system evolve towards similar spacing. 
        Large planets may undergo several collisions that lead to the ejection of several planets allowing them to have wider spacings. This offers a possible explanation for the emergence of the packing trend. After a few million years most systems  have lost their gas disks, which leads to rapid contraction of planetary radii. This is responsible for the sharp drop in the packing trend (in the range $t = 10^6 - 2\times 10^7~\si{\year}$). As planets continue to grow via merger collisions, the packing trend re-emerges slowly.  
        
        The spacing and packing trends are seen to have several common behaviours. They are both absent at early times, arise from dynamical interactions, and are strongly influenced by the detection biases. It is a possibility that these two trends are not independent of each other. In fact there is a simple scenario that could unify them. The peas in a pod spacing trend could be a reflection of the  mass similarity and the packing trends. Since adjacent planets are more likely to have similar masses, and the orbital spacing between planetary bodies is related to their masses (heavy planets have wider orbital spacing, while small planets tend to have smaller orbital spacing), the spacing trend can emerge. Planets with large or small masses have neighbours with similar masses, and this leads them to also have period ratios that are  similar. Further tests are required to confirm this scenario. 
        
		Overall, this section presents two important findings. First, the similarity in mass--size trends are already present at early times. These are, \editbold{perhaps,} due to oligarchic growth of protoplanetary embryos and uniform growth of these protoplanets at early times. Uniformly growing neighbouring planets will continue to show size--mass similarity. Different growth rates amongst adjacent planets during the formation stage tends to decrease the mass--size trends. Second, the spacing and packing trends are absent at early times. Dynamical interactions (especially merger collisions) tend to increase spacing and packing correlations.

        \subsection{Role of dynamical interactions}
        \label{subsec:dynamicalinteractions}

        Dynamical interactions can often lead to the ejection of planets and merger collisions. This would lead to a decrease in the number of planets in a system. Systems that have had more dynamical interactions will have lost more planets than systems with less dynamical interactions. Since the number of embryos ($n_\mathrm{emb}$) that a theoretical system begins with is fixed for each synthetic population, the percentage of lost planets can be used as a diagnostic for its dynamical history. With $n_\mathrm{mul}$ as the multiplicity of systems at 4 Gyrs, the percentage of planets lost by a system is 
        \begin{equation}
        \mathrm{Lost~planets~[\%]} = \frac{n_\mathrm{emb} - n_\mathrm{mul}}{n_\mathrm{emb}} \times 100.
        \end{equation}
        
        Figure \ref{fig:lost_planets} (left) shows the distribution of lost planets in the underlying NGPPS populations. This plot shows that the distribution shifts to the right, as the number of embryos increases from 20 to 50, and from 50 to 100. This demonstrates that adding more embryos in a system tends to increases their dynamical interactions, which in turn forces these systems to lose more planets. This verifies that lost planets can be used as a proxy for the dynamical interactions experienced by a system. 
        
        Now the planetary systems are divided into five sub-populations depending on the percentage of planets they lose: [0,20), [20-40), [40-60), [60-80), and [80-100]. The ratio of giant planets to the total number of planets in each system is calculated\footnote{Following \papertwo/, giant planets are defined as planets with $\mplanet >= 300 \mearth$.}. This ratio is then averaged over each sub-population and is shown in Fig. \ref{fig:lost_planets} (left). There is a clear increase in the ratio of giant planets in a system with increasing dynamical interactions. This shows that systems with more giant planets have more cumulative dynamical interactions. 
        
        For each peas in a pod trend (size, mass, spacing, and packing), the correlation coefficient is measured across each sub-population. This is shown in Fig. \ref{fig:lost_planets} (right). 
        As the percentage of lost planets increases, the spacing and packing correlations also increase (for the 50- and 100-embryo populations). This strongly suggests that increasing dynamical interactions results in strengthening of the spacing and the packing trends. This adds support to the finding of the last section that dynamical interactions amongst growing planets leads to the spacing and packing trend. For the 20-embryo population the result of the Pearson correlation test becomes unreliable due to low multiplicities.
        Going from left to right, the size-mass correlations show little variations at first. However, the size--mass correlations drop sharply in the last two bins. This drop may arise from the presence of giant planets in these sub-populations, indicating an anti-correlation between mass similarity and presence of giant planets.

        \section{Summary, conclusions, and future work}
        \label{sec:conclusionandfuture}
        
        In this paper the peas in a pod trends in the architecture of planetary systems was studied. Using the Bern Model, thousands of synthetic planetary systems were simulated. To compare this population of theoretical systems with observations, a new computer code, \kobe/, was developed and was introduced in this paper.  
        
        \kobe/ closely simulates the geometrical limitations of the transit method and the detection biases of the Kepler transit survey. \kobeshadows/ finds transiting planets via their transit shadow bands, thereby imprinting the transit probability and including the geometrical limitations of the transit method. By selecting only high S/N transiting planets, as calculated by \kobetransits/, the detection biases of the Kepler mission are simulated. Finally, \kobevetter/ rejects some of the planets as false positives, emulating the completeness and reliability of the Kepler Robovetter. Transiting planets that are dispositioned as planetary candidates make up the \kobe/ catalogue.
        
        Additional selection cuts are placed on the \kobe/ catalogue to generate the \kobe/ multi-planetary systems population  (\kmps/). This population is compared with the multi-planetary systems catalogue of the California-Kepler Survey (CKSM from \W18). The \kmps/ and CKSM planetary populations showed similar radius and period distributions. The peas in a pod trend was investigated for several populations. The main conclusions of this paper are:
        \begin{enumerate}
                \item 
                The peas in a pod size and mass similarity trends are present in the theoretically observed (\kmps/) and the theoretical underlying (Bern Model) populations. This means that adjacent planets within a \editbold{synthetic} system tend to have similar sizes and masses. The strength of the size trend in the \kmps/ population is in good agreement with observations. Detection biases tend to diminish the strength of these correlations. 
                                
                \item 
                The peas in a pod size and mass ordering trends are present in the theoretically observed (\kmps/) population. The frequency of size-ordered pairs is in good agreement with observations. This trend is also present in the theoretical underlying population of detectable planets (planets with periods of less than 640 days). Thus, in the inner region of a \editbold{synthetic} system there is a higher chance for an outer planet in an adjacent pair to be larger and/or more massive. Detection biases of the transit method have little influence on this trend.  
                
                \item \editbold{The presence of the size and mass similarity and ordering trends in both the theoretical underlying (Bern Model) and the theoretically observed (by \kobe/) populations implies that this trend, also seen in observations, may have an astrophysical origin.}
                
                \item The peas in a pod mass--size trends are present at very early times. The primordial origin of these trends is \editbold{probably} due to oligarchic growth of protoplanetary embryos and the uniform growth of planets at early times. Later stages of planet formation, including dynamical \editminor{\textit{N}-body} effects, allows planets within a system to grow at different rates. This tends to  decrease the strength of these trends. 
                
                \item 
                In the peas in a pod spacing similarity trend, for three consecutive planets in a system the period ratio of the inner pair tends to be similar to the period ratio of the outer pair. This correlation is present in the theoretically observed (\kmps/) and the theoretical underlying (Bern Model) populations.  
                
                \item The strength of this trend is higher in the underlying population. Detection biases are responsible for reducing the strength of these correlations. \editbold{This suggests that the spacing trend, as reported by \W18, probably has an astrophysical origin.}
                
                \item This trend is absent at early times and likely arises from the dynamical interactions taking place during planet formation stage. Merger collisions and ejection of planets are some of the ways through which planets become evenly spaced. Additionally, this trend increases when the number of embryos in a population is increased, further suggesting that dynamical interactions increase this trend.
                
                \item 
                Observations suggest that large planets tend to have wider orbital spacing, while small planets are often packed in compact configurations. This packing trend is also present in theoretically observed (\kmps/) and the theoretical underlying (Bern Model) catalogues. The strength of this trend is in good agreement with observations. 
                
                \item Detection biases and missing intermediate planets have a strong influence on this trend. These effects tend to diminish these correlations. \editbold{However, this trend is likely to have an astrophysical origin since it  is also present in the underlying population.}

                \item This trend is not present at early times and probably arises from \editminor{\textit{N}-body} dynamical interactions. Large planets undergo several merger collisions, thereby clearing more space around them.
                
                \item 
                The peas in a pod size trends are probably derivative of the peas in a pod mass trends. The existence of mass trends is probably an astrophysical phenomenon. It has also been suggested that the mass similarity and packing trend may combine to give rise to the spacing trend.
        \end{enumerate}
        
        The results of this paper imply that physical processes involved in planet formation gives rise to adjacent planets that have similar masses (and therefore sizes), and that  are evenly spaced. Large planets tend to have wider orbital spacing, while smaller planets tend to be packed in compact configurations. Detection biases of the transit method diminish the size--mass similarity trends and influence the spacing and packing trend. We suggest that the peas in a pod similarity and ordering trends seen in observations may have an astrophysical origin. 
        
        One of the shortcoming of this and other studies on the peas in a pod trends is the use of correlation coefficients in measuring architecture trends. Although useful in making population-level studies, the reliable calculation of correlation coefficients requires large datasets which hinder the study of these trends at the system level. One line of future work could be the development of system level architecture metrics (\citealt{Alibert2019,Mishra2019, Gilbert2020}). These metrics could allow the architecture of an individual system, the Solar System for example, to be studied. This could allow the disentanglement of the role played by specific initial conditions from the effects of planet formation processes in engendering these trends. Furthermore, system level studies are required to establish the unification of peas in a pod trends, as mentioned previously. 
        
        In addition, the present study can be improved by studying different stellar types. To facilitate comparison with \W18, several aspects of \kobe/ were restricted in this paper. For example, the calculation of transit S/N in \kobe/ assumes that all planets are in circular obits, the sampling of CDPP used fixed value of  $\ttrial = 6 h$. Future versions of \kobe/ will include the effect of eccentricity on transit S/N, and will use $\ttrial$ values based on calculated transit durations. Additionally, \kobe/ can be further improved by including stellar limb darkening. 
        
        \begin{acknowledgements}
                The authors thanks the anonymous referee for their comments and suggestions.
                This work has been carried out within the framework of the NCCR PlanetS supported by the Swiss National Science Foundation. The authors acknowledge support from the Swiss National Science Foundation under grant BSSGI0\_155816 ``PlanetsInTime''. A.E. acknowledges the support from the University of Arizona. Calculations were performed on the Horus cluster at the University of Bern.          
                \\
                \textit{Data:} The synthetic planetary populations (NGPPS) used in this work are available online at \url{http://dace.unige.ch} under section ``Formation \& evolution''. This research has made use of the NASA Exoplanet Archive, which is operated by the California Institute of Technology, under contract with the National Aeronautics and Space Administration under the Exoplanet Exploration Program: \url{https://exoplanetarchive.ipac.caltech.edu} (DOI: 10.26133/NEA6). The CKS dataset is available at \url{https://california-planet-search.github.io/cks-website/}. 
                \\
                \textit{Software:} \kobe/ (this paper), Python \citep{python3}, NumPy \citep{numpy}, SciPy \citep{scipy}, Seaborn \citep{seaborn}, Pandas \citep{pandas}, Matplotlib \citep{matplotlib}.
                
        \end{acknowledgements}
               
        \bibliographystyle{aa}
        \bibliography{ngpps_pip.bib}

\begin{thebibliography}{125}
\expandafter\ifx\csname natexlab\endcsname\relax\def\natexlab#1{#1}\fi

\bibitem[{{Adams}(2019)}]{Adams2019}
{Adams}, F.~C. 2019, \mnras, 488, 1446

\bibitem[{Adams {et~al.}(2020)Adams, Batygin, Bloch, \& Laughlin}]{Adams2020}
Adams, F.~C., Batygin, K., Bloch, A.~M., \& Laughlin, G. 2020, \mnras, 493,
  5520

\bibitem[{Alibert(2019)}]{Alibert2019}
Alibert, Y. 2019, \aap, 624, A45

\bibitem[{Alibert {et~al.}(2013)Alibert, Carron, Fortier, Pfyffer, Benz,
  Mordasini, \& Swoboda}]{Alibert2013}
Alibert, Y., Carron, F., Fortier, A., {et~al.} 2013, \aap, 558, A109

\bibitem[{Alibert {et~al.}(2004)Alibert, Mordasini, \& Benz}]{Alibert2004}
Alibert, Y., Mordasini, C., \& Benz, W. 2004, \aap, 417, L25

\bibitem[{Alibert {et~al.}(2011)Alibert, Mordasini, \& Benz}]{Alibert2011}
Alibert, Y., Mordasini, C., \& Benz, W. 2011, \aap, 526, 1

\bibitem[{Alibert {et~al.}(2005)Alibert, Mordasini, Benz, \&
  Winisdoerffer}]{Alibert2005}
Alibert, Y., Mordasini, C., Benz, W., \& Winisdoerffer, C. 2005, \aap, 434, 343

\bibitem[{{Armstrong} {et~al.}(2020){Armstrong}, {Lopez}, {Adibekyan}, {Booth},
  {Bryant}, {Collins}, {Deleuil}, {Emsenhuber}, {Huang}, {King}, {Lillo-Box},
  {Lissauer}, {Matthews}, {Mousis}, {Nielsen}, {Osborn}, {Otegi}, {Santos},
  {Sousa}, {Stassun}, {Veras}, {Ziegler}, {Acton}, {Almenara}, {Anderson},
  {Barrado}, {Barros}, {Bayliss}, {Belardi}, {Bouchy}, {Brice{\~n}o}, {Brogi},
  {Brown}, {Burleigh}, {Casewell}, {Chaushev}, {Ciardi}, {Collins},
  {Col{\'o}n}, {Cooke}, {Crossfield}, {D{\'\i}az}, {Delgado Mena}, {Demangeon},
  {Dorn}, {Dumusque}, {Eigm{\"u}ller}, {Fausnaugh}, {Figueira}, {Gan},
  {Gandhi}, {Gill}, {Gonzales}, {Goad}, {G{\"u}nther}, {Helled}, {Hojjatpanah},
  {Howell}, {Jackman}, {Jenkins}, {Jenkins}, {Jensen}, {Kennedy}, {Latham},
  {Law}, {Lendl}, {Lozovsky}, {Mann}, {Moyano}, {McCormac}, {Meru},
  {Mordasini}, {Osborn}, {Pollacco}, {Queloz}, {Raynard}, {Ricker}, {Rowden},
  {Santerne}, {Schlieder}, {Seager}, {Sha}, {Tan}, {Tilbrook}, {Ting}, {Udry},
  {Vanderspek}, {Watson}, {West}, {Wilson}, {Winn}, {Wheatley}, {Villasenor},
  {Vines}, \& {Zhan}}]{Armstrong2020}
{Armstrong}, D.~J., {Lopez}, T.~A., {Adibekyan}, V., {et~al.} 2020, \nat, 583,
  39

\bibitem[{Bailli{\'{e}} {et~al.}(2019)Bailli{\'{e}}, Marques, \&
  Piau}]{Baillie2019}
Bailli{\'{e}}, K., Marques, J., \& Piau, L. 2019, \aap, 624, 1

\bibitem[{Baraffe {et~al.}(2015)Baraffe, Homeier, Allard, \&
  Chabrier}]{Baraffe2015}
Baraffe, I., Homeier, D., Allard, F., \& Chabrier, G. 2015, \aap, 577, A42

\bibitem[{Barnes(2007)}]{Barnes2007}
Barnes, J. 2007, \pasp, 119, 986

\bibitem[{{Bashi} \& {Zucker}(2021)}]{Bashi2021}
{Bashi}, D. \& {Zucker}, S. 2021, arXiv e-prints, arXiv:2106.00688

\bibitem[{Benz {et~al.}(2014)Benz, Ida, Alibert, Lin, \& Mordasini}]{Benz2014}
Benz, W., Ida, S., Alibert, Y., Lin, D., \& Mordasini, C. 2014, in Protostars
  and Planets VI, ed. H.~Beuther, R.~Klessen, C.~Dullemond, \& T.~Henning
  (University of Arizona, Tucson), 691--713

\bibitem[{Borucki(2016)}]{Borucki2016}
Borucki, W.~J. 2016, Reports on Progress in Physics, 79

\bibitem[{Borucki {et~al.}(2010)Borucki, Koch, Basri, Batalha, Brown, Caldwell,
  Caldwell, Christensen-Dalsgaard, Cochran, Devore, Dunham, Dupree, Gautier,
  Geary, Gilliland, Gould, Howell, Jenkins, Kondo, Latham, Marcy, Meibom,
  Kjeldsen, Lissauer, Monet, Morrison, Sasselov, Tarter, Boss, Brownlee, Owen,
  Buzasi, Charbonneau, Doyle, Fortney, Ford, Holman, Seager, Steffen, Welsh,
  Rowe, Anderson, Buchhave, Ciardi, Walkowicz, Sherry, Horch, Isaacson,
  Everett, Fischer, Torres, Johnson, Endl, MacQueen, Bryson, Dotson, Haas,
  Kolodziejczak, {Van Cleve}, Chandrasekaran, Twicken, Quintana, Clarke, Allen,
  Li, Wu, Tenenbaum, Verner, Bruhweiler, Barnes, \& Prsa}]{Borucki2010}
Borucki, W.~J., Koch, D., Basri, G., {et~al.} 2010, Science, 327, 977

\bibitem[{{Borucki} {et~al.}(2011){Borucki}, {Koch}, {Basri}, {Batalha},
  {Brown}, {Bryson}, {Caldwell}, {Christensen-Dalsgaard}, {Cochran}, {DeVore},
  {Dunham}, {Gautier}, {Geary}, {Gilliland}, {Gould}, {Howell}, {Jenkins},
  {Latham}, {Lissauer}, {Marcy}, {Rowe}, {Sasselov}, {Boss}, {Charbonneau},
  {Ciardi}, {Doyle}, {Dupree}, {Ford}, {Fortney}, {Holman}, {Seager},
  {Steffen}, {Tarter}, {Welsh}, {Allen}, {Buchhave}, {Christiansen}, {Clarke},
  {Das}, {D{\'e}sert}, {Endl}, {Fabrycky}, {Fressin}, {Haas}, {Horch},
  {Howard}, {Isaacson}, {Kjeldsen}, {Kolodziejczak}, {Kulesa}, {Li}, {Lucas},
  {Machalek}, {McCarthy}, {MacQueen}, {Meibom}, {Miquel}, {Prsa}, {Quinn},
  {Quintana}, {Ragozzine}, {Sherry}, {Shporer}, {Tenenbaum}, {Torres},
  {Twicken}, {Van Cleve}, {Walkowicz}, {Witteborn}, \& {Still}}]{Borucki2011}
{Borucki}, W.~J., {Koch}, D.~G., {Basri}, G., {et~al.} 2011, \apj, 736, 19

\bibitem[{Borucki \& Summers(1984)}]{Borucki1984}
Borucki, W.~J. \& Summers, A.~L. 1984, Icarus, 58, 121

\bibitem[{Br{\"{u}}gger {et~al.}(2018)Br{\"{u}}gger, Alibert, Ataiee, \&
  Benz}]{Brugger2018}
Br{\"{u}}gger, N., Alibert, Y., Ataiee, S., \& Benz, W. 2018, \aap, 619, A174

\bibitem[{Br{\"{u}}gger {et~al.}(2020)Br{\"{u}}gger, Burn, Coleman, Alibert, \&
  Benz}]{Brugger2020}
Br{\"{u}}gger, N., Burn, R., Coleman, G. A.~L., Alibert, Y., \& Benz, W. 2020,
  \aap, 640, A21

\bibitem[{{Bryson} {et~al.}(2020){Bryson}, {Coughlin}, {Batalha}, {Berger},
  {Huber}, {Burke}, {Dotson}, \& {Mullally}}]{Bryson2019}
{Bryson}, S., {Coughlin}, J., {Batalha}, N.~M., {et~al.} 2020, \aj, 159, 279

\bibitem[{Burrows {et~al.}(2007)Burrows, Hubeny, Budaj, \&
  Hubbard}]{Burrows2007}
Burrows, A., Hubeny, I., Budaj, J., \& Hubbard, W.~B. 2007, \apj, 661, 502

\bibitem[{Chambers {et~al.}(1996)Chambers, Wetherill, \& Boss}]{Chambers1996}
Chambers, J., Wetherill, G., \& Boss, A. 1996, Icarus, 119, 261

\bibitem[{Chambers(1999)}]{Chambers1999}
Chambers, J.~E. 1999, \mnras, 304, 793

\bibitem[{{Chevance} {et~al.}(2021){Chevance}, {Kruijssen}, \&
  {Longmore}}]{Chevance2021}
{Chevance}, M., {Kruijssen}, J.~M.~D., \& {Longmore}, S.~N. 2021, \apjl, 910,
  L19

\bibitem[{Christiansen {et~al.}(2012)Christiansen, Jenkins, Barclay, Burke,
  Caldwell, Clarke, Li, Seader, Smith, Stumpe, Tenenbaum, Thompson, Twicken, \&
  {Van Cleve}}]{Christiansen2012}
Christiansen, J.~L., Jenkins, J.~M., Barclay, T.~S., {et~al.} 2012, \pasp, 124,
  1279

\bibitem[{Ciardi {et~al.}(2013)Ciardi, Fabrycky, Ford, Gautier, Howell,
  Lissauer, Ragozzine, \& Rowe}]{Ciardi2013}
Ciardi, D.~R., Fabrycky, D.~C., Ford, E.~B., {et~al.} 2013, \apj, 763, 41

\bibitem[{Clarke {et~al.}(2001)Clarke, Gendrin, \& Sotomayor}]{Clarke2001}
Clarke, C.~J., Gendrin, A., \& Sotomayor, M. 2001, \mnras, 328, 485

\bibitem[{Coleman \& Nelson(2014)}]{Coleman2014}
Coleman, G.~A. \& Nelson, R.~P. 2014, \mnras, 445, 479

\bibitem[{Coughlin(2017)}]{Coughlin2017}
Coughlin, J.~L. 2017, KSCI-19114-001, 1

\bibitem[{{Deck} {et~al.}(2013){Deck}, {Payne}, \& {Holman}}]{Deck2013}
{Deck}, K.~M., {Payne}, M., \& {Holman}, M.~J. 2013, \apj, 774, 129

\bibitem[{Demory {et~al.}(2016)Demory, Gillon, {De Wit}, Madhusudhan, Bolmont,
  Heng, Kataria, Lewis, Hu, Krick, Stamenkovi{\'{c}}, Benneke, Kane, \&
  Queloz}]{Demory2016}
Demory, B.~O., Gillon, M., {De Wit}, J., {et~al.} 2016, Nature, 532, 207

\bibitem[{Dittkrist {et~al.}(2014)Dittkrist, Mordasini, Klahr, Alibert, \&
  Henning}]{Dittkrist2014}
Dittkrist, K.~M., Mordasini, C., Klahr, H., Alibert, Y., \& Henning, T. 2014,
  \aap, 567 [\eprint[arXiv]{1402.5969}]

\bibitem[{{Emsenhuber} {et~al.}(2020{\natexlab{a}}){Emsenhuber}, {Mordasini},
  {Burn}, {Alibert}, {Benz}, \& {Asphaug}}]{Emsenhuber2020A}
{Emsenhuber}, A., {Mordasini}, C., {Burn}, R., {et~al.} 2020{\natexlab{a}},
  arXiv e-prints, arXiv:2007.05561

\bibitem[{{Emsenhuber} {et~al.}(2020{\natexlab{b}}){Emsenhuber}, {Mordasini},
  {Burn}, {Alibert}, {Benz}, \& {Asphaug}}]{Emsenhuber2020B}
{Emsenhuber}, A., {Mordasini}, C., {Burn}, R., {et~al.} 2020{\natexlab{b}},
  arXiv e-prints, arXiv:2007.05562

\bibitem[{Espinoza {et~al.}(2020)Espinoza, Brahm, Henning, Jord{\'{a}}n, Dorn,
  Rojas, Sarkis, Kossakowski, Schlecker, D{\'{i}}az, Jenkins, Aguilera-Gomez,
  Jenkins, Twicken, Collins, Lissauer, Armstrong, Adibekyan, Barrado, Barros,
  Battley, Bayliss, Bouchy, Bryant, Cooke, Demangeon, Dumusque, Figueira,
  Giles, Lillo-Box, Lovis, Nielsen, Pepe, Pollacco, Santos, Sousa, Udry,
  Wheatley, Turner, Marmier, S{\'{e}}gransan, Ricker, Latham, Seager, Winn,
  Kielkopf, Hart, Wingham, Jensen, He{\l}miniak, Tokovinin, Brice{\~{n}}o,
  Ziegler, Law, Mann, Daylan, Doty, Guerrero, Boyd, Crossfield, Morris, Henze,
  \& Chacon}]{Espinoza2020}
Espinoza, N., Brahm, R., Henning, T., {et~al.} 2020, \mnras, 491, 2982

\bibitem[{{Evans} {et~al.}(2016){Evans}, {Sing}, {Wakeford}, {Nikolov},
  {Ballester}, {Drummond}, {Kataria}, {Gibson}, {Amundsen}, \&
  {Spake}}]{Evans2016}
{Evans}, T.~M., {Sing}, D.~K., {Wakeford}, H.~R., {et~al.} 2016, \apjl, 822, L4

\bibitem[{Fabrycky {et~al.}(2014)Fabrycky, Lissauer, Ragozzine, Rowe, Steffen,
  Agol, Barclay, Batalha, Borucki, Ciardi, Ford, Gautier, Geary, Holman,
  Jenkins, Li, Morehead, Morris, Shporer, Smith, Still, \& {Van
  Cleve}}]{Fabrycky2014}
Fabrycky, D.~C., Lissauer, J.~J., Ragozzine, D., {et~al.} 2014, \apj, 790, 146

\bibitem[{{Flaherty} {et~al.}(2020){Flaherty}, {Hughes}, {Simon}, {Qi}, {Bai},
  {Bulatek}, {Andrews}, {Wilner}, \& {K{\'o}sp{\'a}l}}]{Flaherty2020}
{Flaherty}, K., {Hughes}, A.~M., {Simon}, J.~B., {et~al.} 2020, \apj, 895, 109

\bibitem[{Fortier {et~al.}(2013)Fortier, Alibert, Carron, Benz, \&
  Dittkrist}]{Fortier2013}
Fortier, A., Alibert, Y., Carron, F., Benz, W., \& Dittkrist, K.-M. 2013, \aap,
  549, A44

\bibitem[{Fulton {et~al.}(2017)Fulton, Petigura, Howard, Isaacson, Marcy,
  Cargile, Hebb, Weiss, Johnson, Morton, Sinukoff, Crossfield, \&
  Hirsch}]{Fulton2017}
Fulton, B.~J., Petigura, E.~A., Howard, A.~W., {et~al.} 2017, \aj, 154, 109

\bibitem[{Gilbert \& Fabrycky(2020)}]{Gilbert2020}
Gilbert, G.~J. \& Fabrycky, D.~C. 2020, \aj, 159, 281

\bibitem[{Guillot(2010)}]{Guillot2010}
Guillot, T. 2010, \aap, 520, 1

\bibitem[{Hadden \& Lithwick(2017)}]{Hadden2017}
Hadden, S. \& Lithwick, Y. 2017, \aj, 154, 5

\bibitem[{He {et~al.}(2019)He, Ford, \& Ragozzine}]{He2019}
He, M.~Y., Ford, E.~B., \& Ragozzine, D. 2019, \mnras, 490, 4575

\bibitem[{Heng(2019)}]{Heng2019}
Heng, K. 2019, \mnras, 490, 3378

\bibitem[{Hoeijmakers {et~al.}(2018)Hoeijmakers, Ehrenreich, Heng, Kitzmann,
  Grimm, Allart, Deitrick, Wyttenbach, Oreshenko, Pino, Rimmer, Molinari, \&
  {Di Fabrizio}}]{Hoeijmakers2018}
Hoeijmakers, H.~J., Ehrenreich, D., Heng, K., {et~al.} 2018, Nature, 560, 453

\bibitem[{Hsu {et~al.}(2018)Hsu, Ford, Ragozzine, \& Morehead}]{Hsu2018}
Hsu, D.~C., Ford, E.~B., Ragozzine, D., \& Morehead, R.~C. 2018, \aj, 155, 205

\bibitem[{Hueso \& Guillot(2005)}]{Hueso2005}
Hueso, R. \& Guillot, T. 2005, \aap, 442, 703

\bibitem[{Hunter(2007)}]{matplotlib}
Hunter, J.~D. 2007, Computing in science \& engineering, 9, 90

\bibitem[{Ida \& Lin(2004)}]{Ida2004}
Ida, S. \& Lin, D. N.~C. 2004, \apj, 616, 567

\bibitem[{{Jin} {et~al.}(2014){Jin}, {Mordasini}, {Parmentier}, {van Boekel},
  {Henning}, \& {Ji}}]{Jin2014}
{Jin}, S., {Mordasini}, C., {Parmentier}, V., {et~al.} 2014, \apj, 795, 65

\bibitem[{Johansen {et~al.}(2007)Johansen, Oishi, Low, Klahr, Henning, \&
  Youdin}]{Johansen2007}
Johansen, A., Oishi, J.~S., Low, M. M.~M., {et~al.} 2007, Nature, 448, 1022

\bibitem[{Johnson {et~al.}(2017)Johnson, Petigura, Fulton, Marcy, Howard,
  Isaacson, Hebb, Cargile, Morton, Weiss, Winn, Rogers, Sinukoff, \&
  Hirsch}]{Johnson2017}
Johnson, J.~A., Petigura, E.~A., Fulton, B.~J., {et~al.} 2017, \aj, 154, 108

\bibitem[{King(2009)}]{King2009}
King, A. 2009, As, 500, 53

\bibitem[{Kipping(2018)}]{Kipping2018}
Kipping, D. 2018, \mnras, 473, 784

\bibitem[{Kipping \& Sandford(2016)}]{Kipping2016}
Kipping, D.~M. \& Sandford, E. 2016, \mnras, 463, 1323

\bibitem[{Kley \& Dirksen(2006)}]{Kley2006}
Kley, W. \& Dirksen, G. 2006, \aap, 447, 369

\bibitem[{Kokubo \& Ida(1998)}]{Kokubo1998}
Kokubo, E. \& Ida, S. 1998, Icarus, 131, 171

\bibitem[{Kokubo \& Ida(2002)}]{Kokubo2002}
Kokubo, E. \& Ida, S. 2002, \apj, 581, 666

\bibitem[{Kreidberg {et~al.}(2014)Kreidberg, Bean, D{\'{e}}sert, Benneke,
  Deming, Stevenson, Seager, Berta-Thompson, Seifahrt, \&
  Homeier}]{Kreidberg2014}
Kreidberg, L., Bean, J.~L., D{\'{e}}sert, J.~M., {et~al.} 2014, Nature, 505, 69

\bibitem[{{Leleu} {et~al.}(2019){Leleu}, {Coleman}, \&
  {Ataiee}}]{Leleu2019stability}
{Leleu}, A., {Coleman}, G. A.~L., \& {Ataiee}, S. 2019, \aap, 631, A6

\bibitem[{{Lissauer} {et~al.}(2011){Lissauer}, {Ragozzine}, {Fabrycky},
  {Steffen}, {Ford}, {Jenkins}, {Shporer}, {Holman}, {Rowe}, {Quintana},
  {Batalha}, {Borucki}, {Bryson}, {Caldwell}, {Carter}, {Ciardi}, {Dunham},
  {Fortney}, {Gautier}, {Howell}, {Koch}, {Latham}, {Marcy}, {Morehead}, \&
  {Sasselov}}]{Lissauer2011}
{Lissauer}, J.~J., {Ragozzine}, D., {Fabrycky}, D.~C., {et~al.} 2011, \apjs,
  197, 8

\bibitem[{Lodders(2003)}]{Lodders2003}
Lodders, K. 2003, \apj, 591, 1220

\bibitem[{{L{\"u}st}(1952)}]{Luest1952}
{L{\"u}st}, R. 1952, Zeitschrift Naturforschung Teil A, 7, 87

\bibitem[{Lynden-Bell \& Pringle(1974)}]{Lynden-Bell1974}
Lynden-Bell, D. \& Pringle, J.~E. 1974, \mnras, 168, 603

\bibitem[{MacDonald {et~al.}(2020)MacDonald, Dawson, Morrison, Lee, \&
  Khandelwal}]{MacDonald2020}
MacDonald, M.~G., Dawson, R.~I., Morrison, S.~J., Lee, E.~J., \& Khandelwal, A.
  2020, \apj, 891, 20

\bibitem[{Manara {et~al.}(2019)Manara, Mordasini, Testi, Williams, Miotello,
  Lodato, \& Emsenhuber}]{Manara2019}
Manara, C.~F., Mordasini, C., Testi, L., {et~al.} 2019, \aap, 631, L2

\bibitem[{Marboeuf {et~al.}(2014)Marboeuf, Thiabaud, Alibert, Cabral, \&
  Benz}]{Marboeuf2014}
Marboeuf, U., Thiabaud, A., Alibert, Y., Cabral, N., \& Benz, W. 2014, \aap,
  570 [\eprint[arXiv]{1407.7271}]

\bibitem[{Matsuyama {et~al.}(2003)Matsuyama, Johnstone, \&
  Murray}]{Matsuyama2003}
Matsuyama, I., Johnstone, D., \& Murray, N. 2003, \apj, 585, L143

\bibitem[{{Mayor} {et~al.}(2011){Mayor}, {Marmier}, {Lovis}, {Udry},
  {S{\'e}gransan}, {Pepe}, {Benz}, {Bertaux}, {Bouchy}, {Dumusque}, {Lo Curto},
  {Mordasini}, {Queloz}, \& {Santos}}]{Mayor2011}
{Mayor}, M., {Marmier}, M., {Lovis}, C., {et~al.} 2011, arXiv e-prints,
  arXiv:1109.2497

\bibitem[{Mayor \& Queloz(1995)}]{Mayor1995}
Mayor, M. \& Queloz, D. 1995, Nature, 378, 355

\bibitem[{Millholland {et~al.}(2017)Millholland, Wang, \&
  Laughlin}]{Millholland2017}
Millholland, S., Wang, S., \& Laughlin, G. 2017, \apj, 849, L33

\bibitem[{{Mishra} {et~al.}(2019){Mishra}, {Alibert}, \& {Udry}}]{Mishra2019}
{Mishra}, L., {Alibert}, Y., \& {Udry}, S. 2019, in EPSC-DPS Joint Meeting
  2019, Vol. 2019, EPSC--DPS2019--1616

\bibitem[{{Mordasini}(2018)}]{Mordasini2018}
{Mordasini}, C. 2018, {Planetary Population Synthesis}, ed. H.~J. {Deeg} \&
  J.~A. {Belmonte}, 143

\bibitem[{Mordasini {et~al.}(2008)Mordasini, Alibert, Benz, Klahr, \&
  Henning}]{Mordasini2008}
Mordasini, C., Alibert, Y., Benz, W., Klahr, H., \& Henning, T. 2008, PhD
  thesis

\bibitem[{Mordasini {et~al.}(2012{\natexlab{a}})Mordasini, Alibert, Benz,
  Klahr, \& Henning}]{Mordasini2012(correlations)}
Mordasini, C., Alibert, Y., Benz, W., Klahr, H., \& Henning, T.
  2012{\natexlab{a}}, \aap, 541, A97

\bibitem[{Mordasini {et~al.}(2009)Mordasini, Alibert, Benz, \&
  Naef}]{Mordasini2009}
Mordasini, C., Alibert, Y., Benz, W., \& Naef, D. 2009, \aap, 501, 1161

\bibitem[{Mordasini {et~al.}(2012{\natexlab{b}})Mordasini, Alibert, Georgy,
  Dittkrist, Klahr, \& Henning}]{Mordasini2012(MR)}
Mordasini, C., Alibert, Y., Georgy, C., {et~al.} 2012{\natexlab{b}}, \aap, 547,
  A112

\bibitem[{Mordasini {et~al.}(2012{\natexlab{c}})Mordasini, Alibert, Klahr, \&
  Henning}]{Mordasini2012(models)}
Mordasini, C., Alibert, Y., Klahr, H., \& Henning, T. 2012{\natexlab{c}}, \aap,
  547, A111

\bibitem[{Mordasini {et~al.}(2017)Mordasini, Marleau, \&
  Molli{\`{e}}re}]{Mordasini2017}
Mordasini, C., Marleau, G.~D., \& Molli{\`{e}}re, P. 2017, \aap, 608
  [\eprint[arXiv]{1708.00868}]

\bibitem[{Mordasini {et~al.}(2015)Mordasini, Molli{\`{e}}re, Dittkrist, Jin, \&
  Alibert}]{Mordasini2015}
Mordasini, C., Molli{\`{e}}re, P., Dittkrist, K.-M., Jin, S., \& Alibert, Y.
  2015, International Journal of Astrobiology, 14, 201

\bibitem[{Mulders {et~al.}(2019)Mulders, Mordasini, Pascucci, Ciesla,
  Emsenhuber, \& Apai}]{Mulders2019}
Mulders, G.~D., Mordasini, C., Pascucci, I., {et~al.} 2019, \apj, 887, 157

\bibitem[{Mulders {et~al.}(2020)Mulders, O'brien, Ciesla, Apai, \&
  Pascucci}]{Mulders2020}
Mulders, G.~D., O'brien, D.~P., Ciesla, F.~J., Apai, D., \& Pascucci, I. 2020

\bibitem[{Mulders {et~al.}(2018)Mulders, Pascucci, Apai, \&
  Ciesla}]{Mulders2018}
Mulders, G.~D., Pascucci, I., Apai, D., \& Ciesla, F.~J. 2018, \aj, 156, 24

\bibitem[{{Murchikova} \& {Tremaine}(2020)}]{Murchikova2020}
{Murchikova}, L. \& {Tremaine}, S. 2020, \aj, 160, 160

\bibitem[{{Murray} \& {Correia}(2010)}]{Murray2010}
{Murray}, C.~D. \& {Correia}, A.~C.~M. 2010, {Keplerian Orbits and Dynamics of
  Exoplanets}, ed. S.~{Seager}, 15--23

\bibitem[{Nakamoto \& Nakagawa(1994)}]{Nakamoto1994}
Nakamoto, T. \& Nakagawa, Y. 1994, \apj, 421, 640

\bibitem[{Oliphant(2006)}]{numpy}
Oliphant, T.~E. 2006, A guide to NumPy, Vol.~1 (Trelgol Publishing USA)

\bibitem[{Paardekooper {et~al.}(2011)Paardekooper, Baruteau, \&
  Kley}]{Paardekooper2011}
Paardekooper, S.~J., Baruteau, C., \& Kley, W. 2011, \mnras, 410, 293

\bibitem[{pandas~development team(2020)}]{pandas}
pandas~development team, T. 2020, pandas-dev/pandas: Pandas

\bibitem[{{Pecaut} \& {Mamajek}(2013)}]{Pecaut2013}
{Pecaut}, M.~J. \& {Mamajek}, E.~E. 2013, \apjs, 208, 9

\bibitem[{Petigura {et~al.}(2017)Petigura, Howard, Marcy, Johnson, Isaacson,
  Cargile, Hebb, Fulton, Weiss, Morton, Winn, Rogers, Sinukoff, Hirsch, \&
  Crossfield}]{Petigura2017}
Petigura, E.~A., Howard, A.~W., Marcy, G.~W., {et~al.} 2017, \aj, 154, 107

\bibitem[{Pollack {et~al.}(1996)Pollack, Hubickyj, Bodenheimer, Lissauer,
  Podolak, \& Greenzweig}]{Pollack1996}
Pollack, J.~B., Hubickyj, O., Bodenheimer, P., {et~al.} 1996, Icarus, 124, 62

\bibitem[{Rauer {et~al.}(2014)Rauer, Catala, Aerts, Appourchaux, Benz,
  Brandeker, Christensen-Dalsgaard, Deleuil, Gizon, Goupil, G{\"{u}}del,
  Janot-Pacheco, Mas-Hesse, Pagano, Piotto, Pollacco, Santos, Smith,
  Su{\'{a}}rez, Szab{\'{o}}, Udry, Adibekyan, Alibert, Almenara, Amaro-Seoane,
  Eiff, Asplund, Antonello, Barnes, Baudin, Belkacem, Bergemann, Bihain, Birch,
  Bonfils, Boisse, Bonomo, Borsa, Brand{\~{a}}o, Brocato, Brun, Burleigh,
  Burston, Cabrera, Cassisi, Chaplin, Charpinet, Chiappini, Church, Csizmadia,
  Cunha, Damasso, Davies, Deeg, D{\'{i}}az, Dreizler, Dreyer, Eggenberger,
  Ehrenreich, Eigm{\"{u}}ller, Erikson, Farmer, Feltzing, {de Oliveira Fialho},
  Figueira, Forveille, Fridlund, Garc{\'{i}}a, Giommi, Giuffrida, Godolt,
  da~Silva, Granzer, Grenfell, Grotsch-Noels, G{\"{u}}nther, Haswell, Hatzes,
  H{\'{e}}brard, Hekker, Helled, Heng, Jenkins, Johansen, Khodachenko,
  Kislyakova, Kley, Kolb, Krivova, Kupka, Lammer, Lanza, Lebreton, Magrin,
  Marcos-Arenal, Marrese, Marques, Martins, Mathis, Mathur, Messina, Miglio,
  Montalban, Montalto, {P. F. G. Monteiro}, Moradi, Moravveji, Mordasini,
  Morel, Mortier, Nascimbeni, Nelson, Nielsen, Noack, Norton, Ofir, Oshagh,
  Ouazzani, P{\'{a}}pics, Parro, Petit, Plez, Poretti, Quirrenbach, Ragazzoni,
  Raimondo, Rainer, Reese, Redmer, Reffert, Rojas-Ayala, Roxburgh, Salmon,
  Santerne, Schneider, Schou, Schuh, Schunker, Silva-Valio, Silvotti, Skillen,
  Snellen, Sohl, Sousa, Sozzetti, Stello, Strassmeier, {\v{S}}vanda,
  Szab{\'{o}}, Tkachenko, Valencia, {Van Grootel}, Vauclair, Ventura, Wagner,
  Walton, Weingrill, Werner, Wheatley, \& Zwintz}]{Rauer2014}
Rauer, H., Catala, C., Aerts, C., {et~al.} 2014, Experimental Astronomy, 38,
  249

\bibitem[{Ricker {et~al.}(2014)Ricker, Winn, Vanderspek, Latham, Bakos, Bean,
  Berta-Thompson, Brown, Buchhave, Butler, Butler, Chaplin, Charbonneau,
  Christensen-Dalsgaard, Clampin, Deming, Doty, {De Lee}, Dressing, Dunham,
  Endl, Fressin, Ge, Henning, Holman, Howard, Ida, Jenkins, Jernigan, Johnson,
  Kaltenegger, Kawai, Kjeldsen, Laughlin, Levine, Lin, Lissauer, MacQueen,
  Marcy, McCullough, Morton, Narita, Paegert, Palle, Pepe, Pepper, Quirrenbach,
  Rinehart, Sasselov, Sato, Seager, Sozzetti, Stassun, Sullivan, Szentgyorgyi,
  Torres, Udry, \& Villasenor}]{Ricker2014}
Ricker, G.~R., Winn, J.~N., Vanderspek, R., {et~al.} 2014, Journal of
  Astronomical Telescopes, Instruments, and Systems, 1, 014003

\bibitem[{{Sandford} {et~al.}(2019){Sandford}, {Kipping}, \&
  {Collins}}]{Sandford2019}
{Sandford}, E., {Kipping}, D., \& {Collins}, M. 2019, \mnras, 489, 3162

\bibitem[{Santerne {et~al.}(2019)Santerne, Malavolta, Kosiarek, Dai, Dressing,
  Dumusque, Hara, Lopez, Mortier, Vanderburg, Adibekyan, Armstrong, Barrado,
  Barros, Bayliss, Berardo, Boisse, Bonomo, Bouchy, Brown, Buchhave, Butler,
  Cameron, Cosentino, Crane, Crossfield, Damasso, Deleuil, Mena, Demangeon,
  D{\'{i}}az, Donati, Figueira, Fulton, Ghedina, Harutyunyan, H{\'{e}}brard,
  Hirsch, Hojjatpanah, Howard, Isaacson, Latham, Lillo-Box,
  L{\'{o}}pez-Morales, Lovis, Fiorenzano, Molinari, Mousis, Moutou, Nava,
  Nielsen, Osborn, Petigura, Phillips, Pollacco, Poretti, Rice, Santos,
  S{\'{e}}gransan, Shectman, Sinukoff, Sousa, Sozzetti, Teske, Udry, Vigan,
  Wang, Watson, Weiss, Wheatley, \& Winn}]{Santerne2019}
Santerne, A., Malavolta, L., Kosiarek, M.~R., {et~al.} 2019, 1

\bibitem[{Santos {et~al.}(2005)Santos, Israelian, Mayor, Bento, Almeida, Sousa,
  \& Ecuvillon}]{Santos2005}
Santos, N.~C., Israelian, G., Mayor, M., {et~al.} 2005, \aap, 437, 1127

\bibitem[{{Sarkis} {et~al.}(2021){Sarkis}, {Mordasini}, {Henning}, {Marleau},
  \& {Molli{\`e}re}}]{Sarkis2020}
{Sarkis}, P., {Mordasini}, C., {Henning}, T., {Marleau}, G.~D., \&
  {Molli{\`e}re}, P. 2021, \aap, 645, A79

\bibitem[{Schneider {et~al.}(2011)Schneider, Dedieu, {Le Sidaner}, Savalle, \&
  Zolotukhin}]{Schneider2011}
Schneider, J., Dedieu, C., {Le Sidaner}, P., Savalle, R., \& Zolotukhin, I.
  2011, A{\&}A, 532

\bibitem[{Shakura \& Sunyaev(1973)}]{Shakura1973}
Shakura, N.~I. \& Sunyaev, R.~A. 1973, \aap, 24, 337

\bibitem[{Sing {et~al.}(2016)Sing, Fortney, Nikolov, Wakeford, Kataria, Evans,
  Aigrain, Ballester, Burrows, Deming, D{\'{e}}sert, Gibson, Henry, Huitson,
  Knutson, Etangs, Pont, Showman, Vidal-Madjar, Williamson, \&
  Wilson}]{Sing2016}
Sing, D.~K., Fortney, J.~J., Nikolov, N., {et~al.} 2016, Nature, 529, 59

\bibitem[{Thiabaud {et~al.}(2014)Thiabaud, Marboeuf, Alibert, Cabral, Leya, \&
  Mezger}]{Thiabaud2014}
Thiabaud, A., Marboeuf, U., Alibert, Y., {et~al.} 2014, \aap, 562
  [\eprint{1312.3085}]

\bibitem[{Thompson {et~al.}(2018)Thompson, Coughlin, Hoffman, Mullally,
  Christiansen, Burke, Bryson, Batalha, Haas, Catanzarite, Rowe, Barentsen,
  Caldwell, Clarke, Jenkins, Li, Latham, Lissauer, Mathur, Morris, Seader,
  Smith, Klaus, Twicken, {Van Cleve}, Wohler, Akeson, Ciardi, Cochran, Henze,
  Howell, Huber, Pr{\v{s}}a, Ram{\'{i}}rez, Morton, Barclay, Campbell, Chaplin,
  Charbonneau, Christensen-Dalsgaard, Dotson, Doyle, Dunham, Dupree, Ford,
  Geary, Girouard, Isaacson, Kjeldsen, Quintana, Ragozzine, Shabram, Shporer,
  Aguirre, Steffen, Still, Tenenbaum, Welsh, Wolfgang, Zamudio, Koch, \&
  Borucki}]{Thompson2018}
Thompson, S.~E., Coughlin, J.~L., Hoffman, K., {et~al.} 2018, \apjs, 235, 38

\bibitem[{Twicken {et~al.}(2018)Twicken, Catanzarite, Clarke, Girouard,
  Jenkins, Klaus, Li, McCauliff, Seader, Tenenbaum, Wohler, Bryson, Burke,
  Caldwell, Haas, Henze, \& Sanderfer}]{Twicken2018}
Twicken, J.~D., Catanzarite, J.~H., Clarke, B.~D., {et~al.} 2018, \pasp, 130,
  064502

\bibitem[{Twicken {et~al.}(2016)Twicken, Jenkins, Seader, Tenenbaum, Smith,
  Brownston, Burke, Catanzarite, Clarke, Cote, Girouard, Klaus, Li, McCauliff,
  Morris, Wohler, Campbell, Uddin, Zamudio, Sabale, Bryson, Caldwell,
  Christiansen, Coughlin, Haas, Henze, Sanderfer, \& Thompson}]{Twicken2016}
Twicken, J.~D., Jenkins, J.~M., Seader, S.~E., {et~al.} 2016, \aj, 152, 158

\bibitem[{Tychoniec {et~al.}(2018)Tychoniec, Tobin, Karska, Chandler, Dunham,
  Harris, Kratter, Li, Looney, Melis, P{\'{e}}rez, Sadavoy, Segura-Cox, \& van
  Dishoeck}]{Tychoniec2018}
Tychoniec, {\L}., Tobin, J.~J., Karska, A., {et~al.} 2018, \apjs, 238, 19

\bibitem[{{Udry} \& {Santos}(2007)}]{Udry2007}
{Udry}, S. \& {Santos}, N.~C. 2007, \araa, 45, 397

\bibitem[{Van~Rossum \& Drake(2009)}]{python3}
Van~Rossum, G. \& Drake, F.~L. 2009, Python 3 Reference Manual (Scotts Valley,
  CA: CreateSpace)

\bibitem[{{Venturini} {et~al.}(2020){Venturini}, {Guilera}, {Haldemann},
  {Ronco}, \& {Mordasini}}]{Julia2020}
{Venturini}, J., {Guilera}, O.~M., {Haldemann}, J., {Ronco}, M.~P., \&
  {Mordasini}, C. 2020, \aap, 643, L1

\bibitem[{Venuti {et~al.}(2017)Venuti, Bouvier, Cody, Stauffer, Micela, Rebull,
  Alencar, Sousa, Hillenbrand, \& Flaccomio}]{Venuti2017}
Venuti, L., Bouvier, J., Cody, A.~M., {et~al.} 2017, \aap, 599, 1

\bibitem[{Veras \& Armitage(2004)}]{Veras2004}
Veras, D. \& Armitage, P.~J. 2004, \mnras, 347, 613

\bibitem[{{Virtanen} {et~al.}(2020){Virtanen}, {Gommers}, {Oliphant},
  {Haberland}, {Reddy}, {Cournapeau}, {Burovski}, {Peterson}, {Weckesser},
  {Bright}, {van der Walt}, {Brett}, {Wilson}, {Jarrod Millman}, {Mayorov},
  {Nelson}, {Jones}, {Kern}, {Larson}, {Carey}, {Polat}, {Feng}, {Moore}, {Vand
  erPlas}, {Laxalde}, {Perktold}, {Cimrman}, {Henriksen}, {Quintero}, {Harris},
  {Archibald}, {Ribeiro}, {Pedregosa}, {van Mulbregt}, \&
  {Contributors}}]{scipy}
{Virtanen}, P., {Gommers}, R., {Oliphant}, T.~E., {et~al.} 2020, Nature Methods

\bibitem[{Wang(2017)}]{Wang2017}
Wang, S. 2017, Research Notes of the American Astronomical Society, 1, 26

\bibitem[{Waskom \& the seaborn~development team(2020)}]{seaborn}
Waskom, M. \& the seaborn~development team. 2020, mwaskom/seaborn

\bibitem[{Weiss {et~al.}(2018)Weiss, Marcy, Petigura, Fulton, Howard, Winn,
  Isaacson, Morton, Hirsch, Sinukoff, Cumming, Hebb, \& Cargile}]{Weiss2018}
Weiss, L.~M., Marcy, G.~W., Petigura, E.~A., {et~al.} 2018, \aj, 155, 48

\bibitem[{{Weiss} \& {Petigura}(2020)}]{Weiss2019}
{Weiss}, L.~M. \& {Petigura}, E.~A. 2020, \apjl, 893, L1

\bibitem[{{Williams} \& {Cieza}(2011)}]{Williams2011}
{Williams}, J.~P. \& {Cieza}, L.~A. 2011, \araa, 49, 67

\bibitem[{{Winn}(2010)}]{Winn2010}
{Winn}, J.~N. 2010, arXiv e-prints, arXiv:1001.2010

\bibitem[{{Winn} \& {Fabrycky}(2015)}]{Winn2015}
{Winn}, J.~N. \& {Fabrycky}, D.~C. 2015, \araa, 53, 409

\bibitem[{{Winn} {et~al.}(2018){Winn}, {Sanchis-Ojeda}, \&
  {Rappaport}}]{Winn2018}
{Winn}, J.~N., {Sanchis-Ojeda}, R., \& {Rappaport}, S. 2018, \nar, 83, 37

\bibitem[{Xu {et~al.}(2018)Xu, Lai, \& Morbidelli}]{Xu2018}
Xu, W., Lai, D., \& Morbidelli, A. 2018, \mnras, 481, 1538

\bibitem[{Youdin(2008)}]{Youdin2008}
Youdin, A. 2008, in EAS Publications Series, ed. T.~Montmerle, D.~Ehrenreich,
  \& A.~M. Lagrange (EAS Publications Series)

\bibitem[{Zar(2014)}]{Zar2014}
Zar, J.~H. 2014, {Biostatistical Analysis}, 5th edn. (Essex: Pearson), 761

\bibitem[{{Zhu}(2020)}]{Zhu2019}
{Zhu}, W. 2020, \aj, 159, 188

\end{thebibliography}
        
        \begin{appendix}
                
                \section{Detection biases explain the negative correlation between mutual separation and average sizes and masses}
                \label{sec:mutual_hill}
                
                \begin{figure*}
                        \centering
                        \includegraphics[width=6cm]{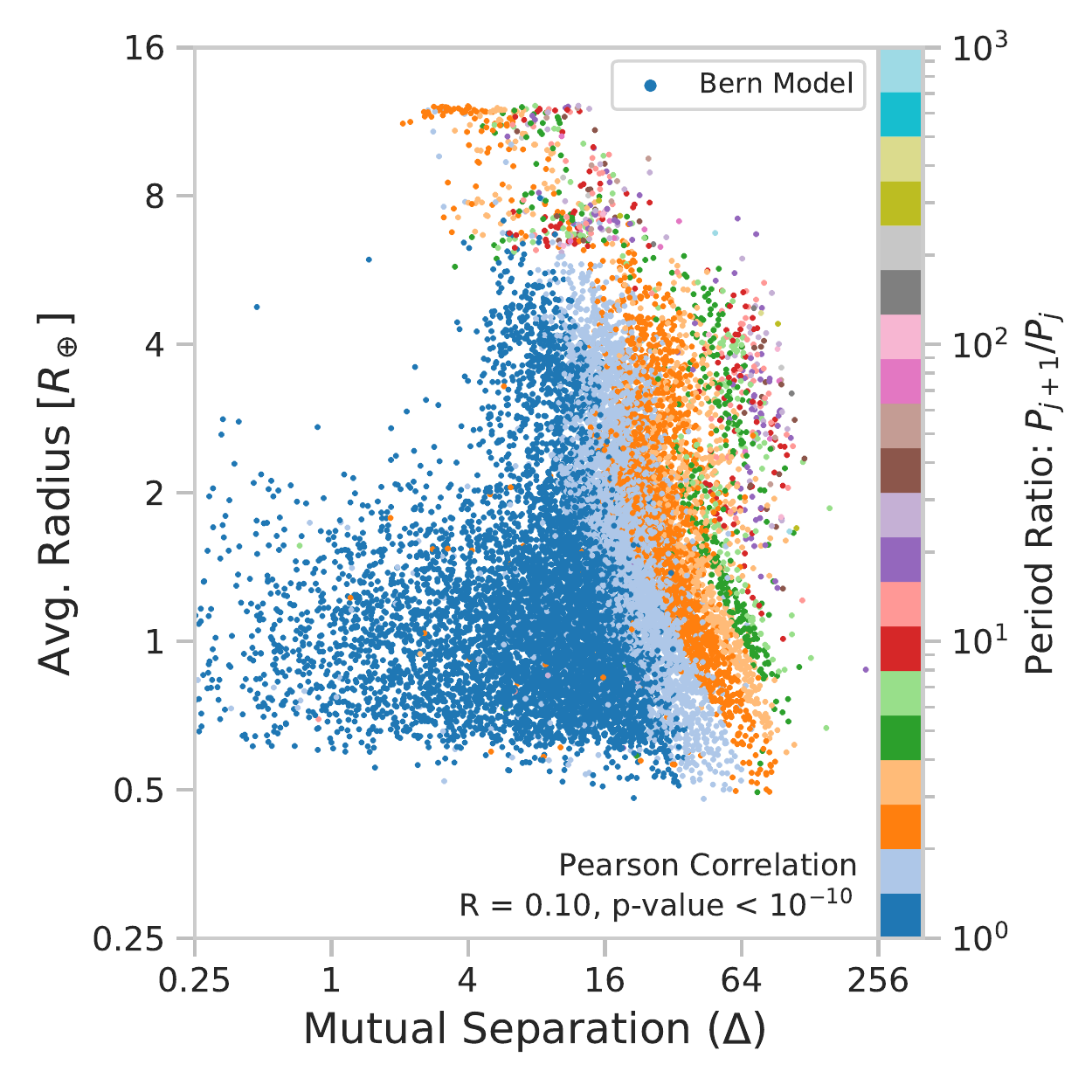}
                        \includegraphics[width=6cm]{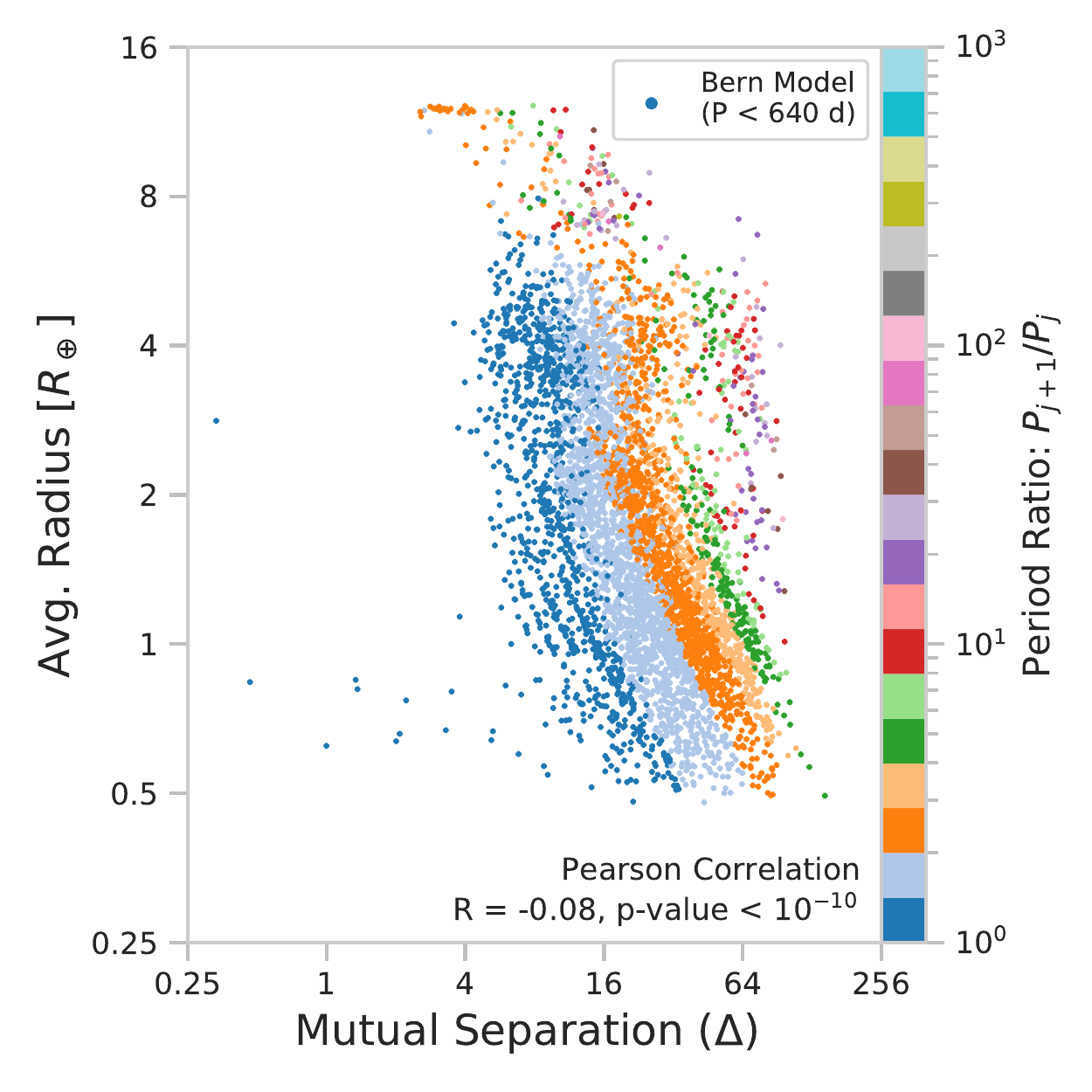}
                        \includegraphics[width=6cm]{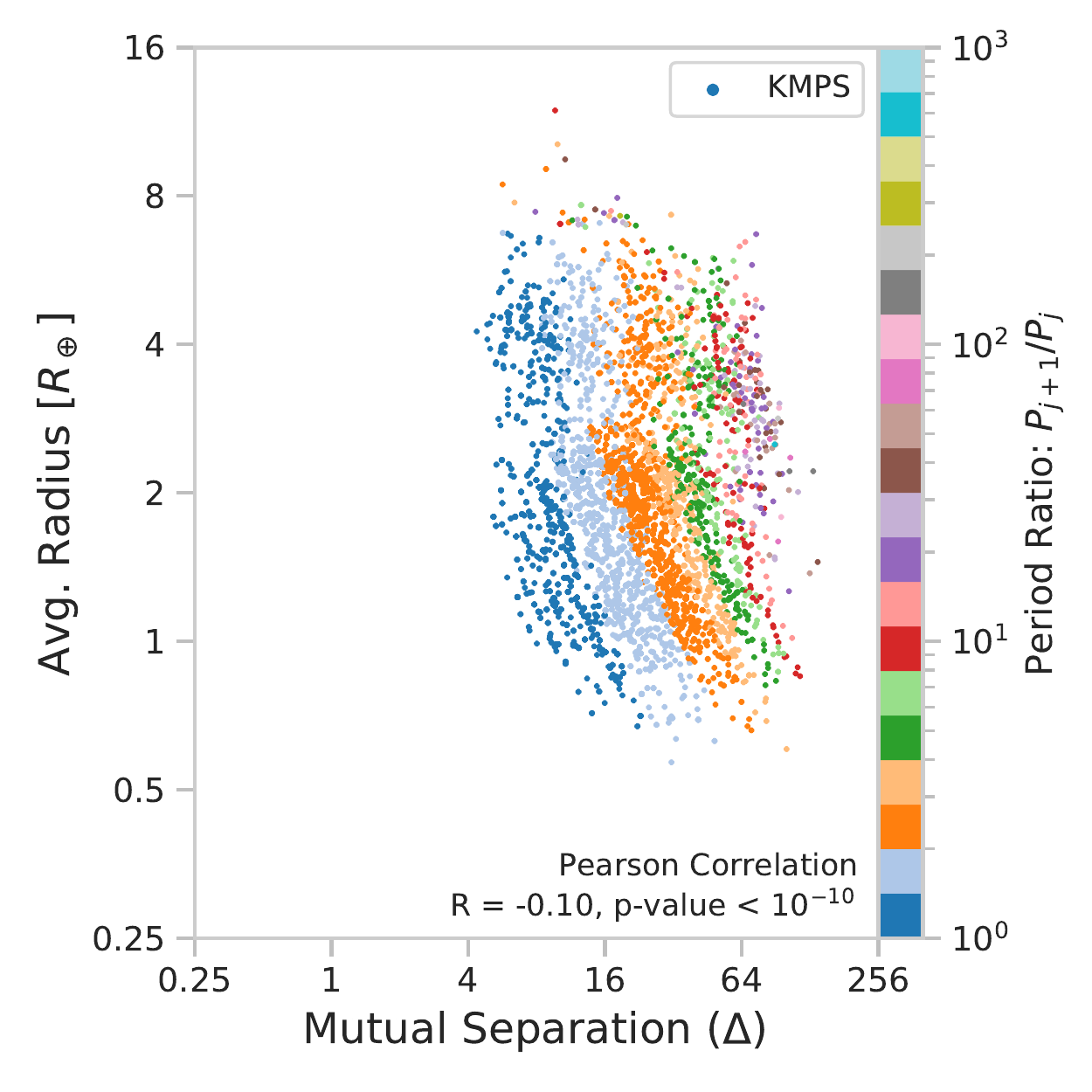}
                        \\
                        \includegraphics[width=6cm]{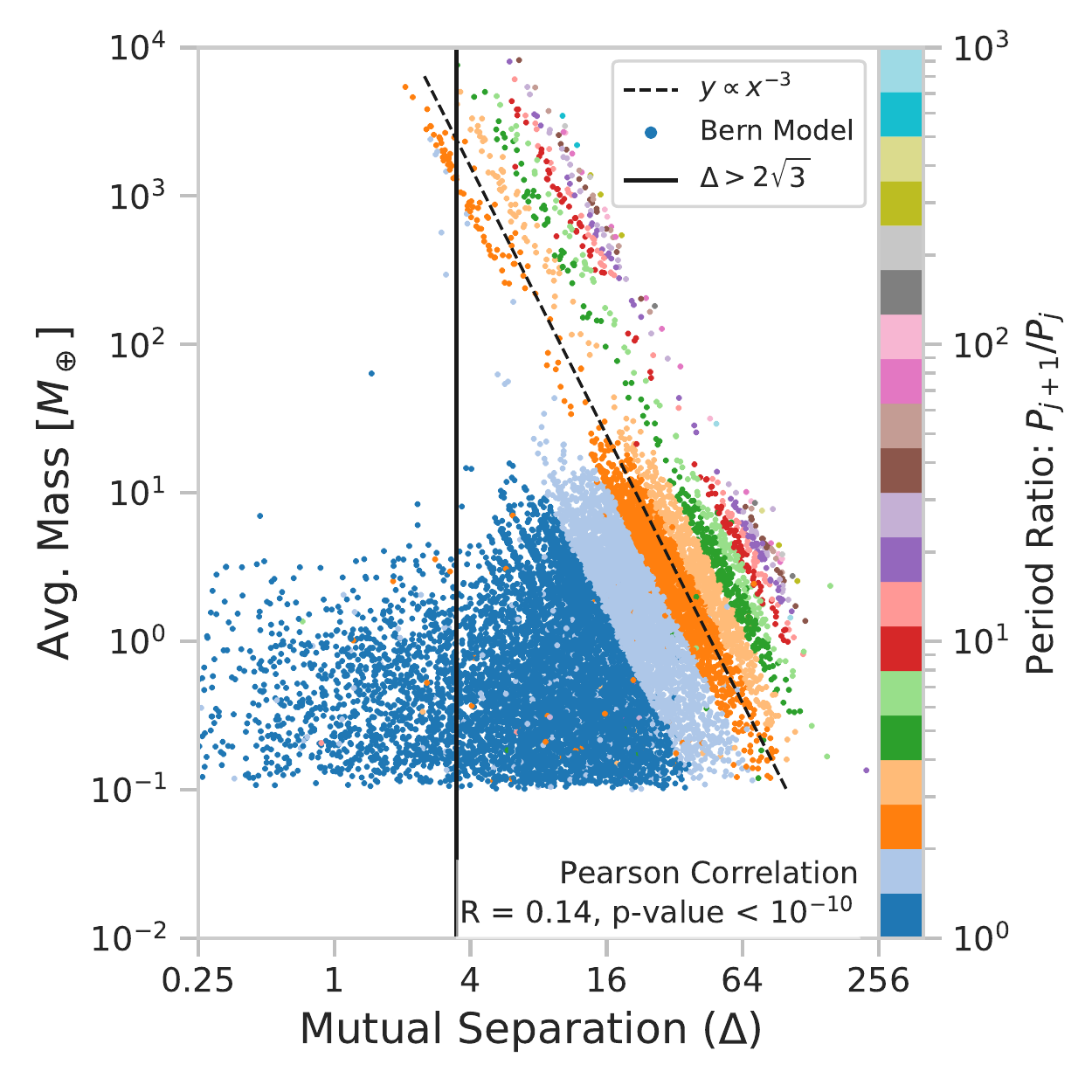}
                        \includegraphics[width=6cm]{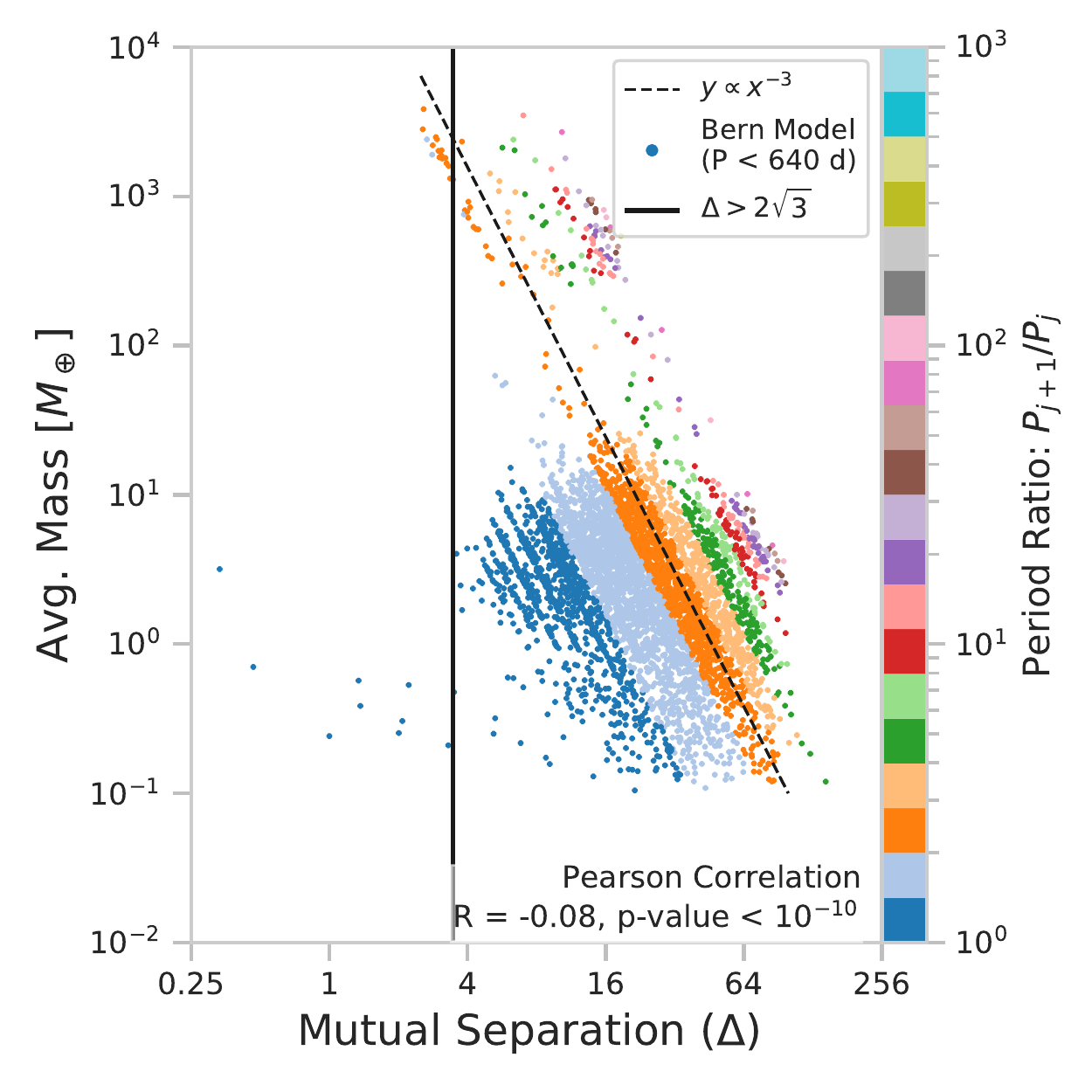}
                        \includegraphics[width=6cm]{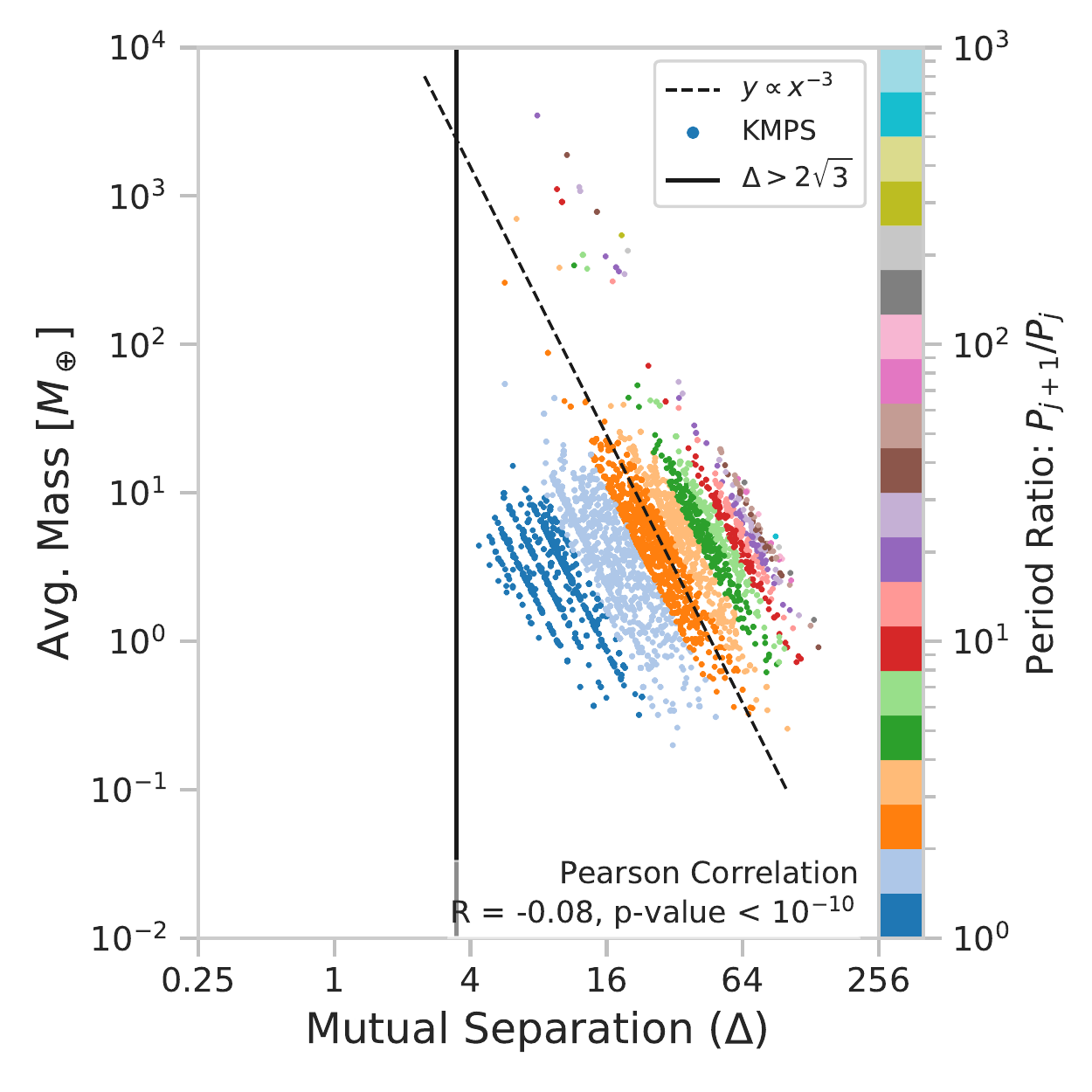}                   
                        \caption{Dependence of mutual separation of adjacent planets on their average sizes (top) and average masses (bottom). Observed exoplanetary populations (CKSM by \W18) show a negative correlation between average sizes and mutual separation of adjacent planets, which is well reproduced here (top right). However, this negative correlation arises from the dependence of mutual Hill radii on the $1/3$ power of planetary masses. This is shown by the dashed line in the plots with average masses. The vertical line marks the Hill stability criterion from \cite{Chambers1996}. Points on the right side of this line are Hill stable ($\Delta > 2\sqrt{3}$). } 
                        \label{fig:pip_hill}
                \end{figure*}
                
                In their effort to study the connection between planetary sizes and their orbital spacing, \W18 explored the mutual separation (in units of mutual Hill radii)  between planets. The mutual separation between two adjacent planets is defined using their mutual Hill radius $\rhill^\mathrm{mutual}$. The mutual Hill radius of two adjacent planets (with masses $m_{j}$ and $m_{j+1}$) can be thought of as the Hill sphere radius of a single planet located between the two adjacent planets and whose mass is the sum of the two planets. This gives
                \begin{equation}
                \label{eq:mutualhill}
                \rhill^\mathrm{mutual} = \Bigg( \frac{m_{j} + m_{j+1}}{3~\mstar} \Bigg)^{1/3} \Bigg(\frac{a_j + a_{j+1}}{2}\Bigg)
                .\end{equation}
                Here, $\mstar$ is the stellar mass, while $a_j$ and $a_{j+1}$ are the star--planet distances.  The mutual separation $\Delta$ between two planets is the ratio of their orbital separation to their mutual Hill radius. It measures the dynamical spacing between two planets in units of their mutual Hill radius:  
                \begin{equation}
                \Delta = \frac{a_{j+1} - a_{j}}{\rhill^\mathrm{mutual}}
                \end{equation}
                
                \cite{Weiss2018} find that there is a significant negative correlation between mutual separation and the average size of adjacent planets in the CKSM catalogue ($R = -0.2$). They state that the mutual Hill radius incorporates the mass of the planet, and thus the computation of mutual Hill radius separation should ideally remove the contribution of planet size. They add that this correlation is driven by an absence of points in the lower left corner of the plot, which means that the absence of very small planets at close dynamical spacings is not due to detection bias.
                
                \editbold{Here, we show that the dependence of mutual Hill radius on average size is affected by the relationship between mutual Hill radii and mass (through eq. \ref{eq:mutualhill}).  In addition, since transit surveys detect planets only in the inner region of a planetary system (close to the host star), adjacent planetary pairs with small dynamical spacing will be missing from observations.}
                
                In Fig. \ref{fig:pip_hill} (top right) the average radius of two adjacent planets (in \kmps/) is plotted as a function of their mutual Hill separation. The period ratio of the adjacent planets is shown in colour. In good agreement with \W18, there is a significant negative correlation, which is confirmed by a Pearson correlation test ($R = -0.1$).  \editbold{The bottom plots show the average mass of adjacent planets as a function of their mutual Hill radii. 
                We note that the scatter of points in the top panel closely resembles the scatter of points in the bottom panel. In these plots, adjacent planets of constant period ratios lie on straight lines (i.e. points of the same colour). In the plot with average sizes, adjacent planets of constant period ratios tend to be on straight lines, but show considerable scatter for average sizes $>3~\rearth$. This scatter can be traced to the intrinsic scatter in the radius of planets with size $>3~\rearth$ in the mass-radius diagram in Fig. \ref{fig:kmps_pip_m_r}. The slope of the average mass versus the mutual Hill plot comes from the $\tfrac{1}{3}$ power on the mass term in eq. \ref{eq:mutualhill}. It gives average masses a slope of $-3$ on the log. average mass versus log. mutual Hill plane. To visualize this, the $y \propto x^{-3}$ dependence is plotted as a dashed line in Fig. \ref{fig:pip_hill} (right). The slope of this line closely matches  the underlying points. This indicates that the incorporation of planet mass in the calculation of mutual Hill radius may influence the relationship between planet sizes and mutual Hill radii.}
                
                \editbold{To test whether this correlation arises from limitations of the transit method, the same correlation test was also done for the theoretical underlying population. Figure \ref{fig:pip_hill}  shows the average radius versus mutual Hill separation. The lower left corner of the plot, which was empty for the \kmps/ population, is filled for the underlying population, but remains empty for the underlying population of detectable planets. This shows that the inner regions of planetary systems do not have planetary pairs with small mutual separation. Since transit surveys can only detect planets in the inner region, this explains why observations will find a negative correlation (which is due to the absence of points in the lower left corner). This suggests that the negative correlation seen in CKSM is probably also due to detection biases and limitations of the transit method.} This behaviour is similar to the correlations between average masses and mutual separation.
                
                The question arises regarding why   the theoretically observed population (\kmps/) or the underlying population of detectable planets shows no adjacent pairs with small size and mass and low mutual separation? This is understandable due to limitations and detection biases of the transit method.  
                
                The Hill sphere radius of a planet defines the region around a planet in which the gravitation field of the planet dominates. As the star--planet distance increases, the influence of the star diminishes and $\rhill$ of a planet increases. This can be seen through eq. \ref{eq:rhill}. This is also true for the mutual Hill radii, $\rhill^\mathrm{mutual}$. As two adjacent planets move further out, their mutual Hill radii increases. The inverse dependence of $\Delta$ on $\rhill^\mathrm{mutual}$ implies that as the mutual Hill separation between adjacent planets increases, their mutual separation decreases. \editbold{Thus, adjacent planets in the outer regions of a planetary system will have a large mutual Hill radius and consequently a small dynamical separation. This explains the absence of points in the lower left corner for the plot showing underlying population of detectable planets. Since detection biases of the transit method disfavour the discovery of small planets further out from their host star, it also explains why planets with small average sizes and masses are missing from the lower left corner in the \kmps/ plots. This provides a plausible explanation for the negative correlation, reported in observations by \W18, between mutual separation and average sizes.}

                \section{\textit{The Bern Model}: Additional details}
                \label{bernmodelappendix_description}                
                Additional details of the Bern Model are presented here. First, we describe the physical processes occurring before planets are born. These processes involve the host star and the protoplanetary disk. Then we describe the processes which model planet formation, i.e. the accretion of solids and gases. 

                \subsection{Before planet formation begins}
                \label{bernmodelappendix_beforeplanetformation}

        \subsubsection{Stellar Evolution}
The model includes the evolution of a fixed mass star ($\mstar = \SI{1}{\msun}$) by incorporating the stellar evolution tracks from \cite{Baraffe2015}. The evolving stellar properties influence the behaviour of the disk and the growing planets in multiple ways. For example, stellar irradiation and temperature ($\lstar, \tstar$) affects the thermodynamical aspects of  the disks and the planets. Stellar radius $\rstar$ strongly affects the transit signal generated by a transiting planet and also allows the tracking of collisions between any object and the star.

\subsubsection{Protoplanetary disk: Gaseous phase}
The gas disk plays a crucial role in the growth of planets and shaping the planetary system architectures. Accretion of this nebular gas \editminor{may} lead to gaseous envelopes around many planets. Additionally, the gas disk interacts with planetesimals, embryos, and protoplanets through effects such as gas drag, migration, and eccentricity damping. 

The model follows the evolution of an axisymmetric geometrically thin gas disk in a time-independent gravitational potential. Gas is accreted by the star and growing planets, and is lost via photoevaporation. Meanwhile, the outer regions of the disk are pushed away to conserve angular momentum until the disk is completely dissipated. The disk evolution is computed in the region from $r_\mathrm{in}$ \editminor{up to} $r_\text{max} = \SI{1000}{\au}$. Here, $r$ is the radial distance from the star and $r_\mathrm{in}$ is a Monte Carlo variable for the model (discussed in Sect. \ref{sec:ngpps}). The \editminor{vertically integrated and azimuthally averaged} surface density of gas, $\sigmag$, evolves as 
\begin{equation}
\sigmadotg(r) = \frac{1}{r} \frac{\partial}{\partial r} F(r) - \sigmadotgphoto(r) - \sigmadotgplan(r),
\label{eq:gasdiskevolution}
\end{equation}
where $F$ is radial flux of gases from viscous angular momentum transport \citep{Luest1952,Lynden-Bell1974} $F(r) = 3 r^{1/2} \frac{\partial}{\partial r} (r^{1/2} \sigmag \nu)$.
Effective turbulent viscosity, $\nu$, is parametrized by a dimensionless parameter $\alpha$ as $ \nu = \alpha c_s h$ \citep[following][]{Shakura1973}\footnote{$\alpha$ measures efficiency of transport due to turbulence. Since random isotropic motions do not have length scales larger than the local disk scale height, $\alpha$ is usually $< 1$ \citep{King2009}.}. Here, $c_s$ is the isothermal speed of sound (which depends on temperature and mean molecular mass of gas) and $h$ is the local disk pressure scale height ($\sim$ half of disk thickness). Following observations \citep{Manara2019, Flaherty2020}, in this work $\alpha = \num{2e-3}$.

The extreme UV and far-UV radiation from the host star and neighbouring stars,  respectively, heat the disk such that thermal motion can overcome gravitational potential resulting in the dispersal of the disk \citep{Clarke2001, Matsuyama2003}. Following \cite{Mordasini2012(MR)}, internal and external photoevaporation losses are included in $\sigmadotgphoto$. These mechanisms control the disk lifetime via the mass loss rate, $\mwind$, which is a Monte Carlo variable for the model and is discussed in Sect. \ref{sec:ngpps}. The disk also loses \editminor{some} amount of gas to planetary accretion, which is represented by $\sigmadotgplan$\footnote{In the first $\sim \SI{e5}{years}$ the disk gains mass from the molecular cloud collapse. This can be modelled by adding a source term to eq. \ref{eq:gasdiskevolution}, as is done in \cite{Hueso2005}.}. 

The model begins with an initial surface density profile for the gas disk given by \citep{Veras2004}
\begin{equation}
\begin{split}
\sigmag(r) &= \sigmanorm \left( \frac{r}{\SI{5.2}{\au}}\right)^{-\betag}    \exp{\left(\left(\frac{r}{\rcutg}\right)^{\betag-2}\right)}   \left(1-\sqrt{\frac{r_\mathrm{in}}{r}}\right),
\\
\betag  &= 0.9 \ \ \text{(power law index).}
\end{split}
\end{equation}
Here, $\sigmanorm$ (normalization constant) and $\rcutg$ (\editminor{characteristic radius}) are governed by a Monte Carlo variable, $\mgasdisk$, the initial gas disk mass through the relations (see \papertwo/): 
\begin{equation}
\label{eq:mgasdisk_influences}
\begin{split}
\mgasdisk &= \sigmanorm \left[ \left(\frac{2 \pi}{2 - \betag}\right) \left(\frac{1}{\SI{5.2}{\au}}\right)^{-\betag} \left(\frac{1}{\rcutg}\right)^{2-\betag}  \right],
\\
\mgasdisk &= \SI{2e-3}{\msun} \left( \frac{\rcutg}{\SI{10}{\au}} \right)^{1.6}.  
\end{split}
\end{equation}

The flaring disk gains thermal energy from the host star, viscous dissipation and a background thermal radiation (at \SI{10}{\kelvin}). For thermal equilibrium the disk cools down by radiating away this energy from its surface \citep{Alibert2005, Fortier2013}. 
Assuming the disk is in hydrostatic equilibrium, the local energy produced due to viscosity is removed through radiative flux, which has to diffuse towards the disk surface\footnote{This is because (a) the disk is geometrically thin and (b) the disk is assumed to be optically thick along the radial direction.}. 
Considering radiative transfer through optically thick and thin regions allows the evaluation of the mid-plane disk temperature \citep[following][]{Nakamoto1994, Hueso2005}. The disk temperature impacts planetary interiors and their growth rates.

\subsubsection{Protoplanetary disk: Planetesimals}
Planetary embryos\editmajor{ initially} grow by accreting planetesimals from the solid disk to become planetary cores. The planetesimal disk is modelled as a fluid with surface density $\sigmasol$ \citep{Fortier2013}. This disk evolves via planetary accretion, aerodynamic interaction with nebular gas, and viscous stirring from planets or planetesimals. The interactions excite planetesimal eccentricity and inclination, possibly resulting in ejection of some solids, which influences the rate at which they are accreted by embryos. For simplicity, two kinds of planetesimals are assumed: rocky (refractory materials) with $\rhoplan = \SI{3.2}{\gram\per\cubic\centi\meter}$ and icy (volatile rich) with $\rhoplan = \SI{1.0}{\gram\per\cubic\centi\meter}$ of equal and fixed size of $300~\text{m}$. The initial surface density profile is (\editminor{see \papertwo/ for details})
\begin{equation}
\begin{split}
\sigmasol(r) &= \sigmas0 \left[ f_\text{s}(r)  \left(\frac{r}{\SI{5.2}{\au}}\right)^{-\betas} \exp{\left(-\left(\frac{r}{\rcuts}\right)^{\betas-2}\right)}  \right],
\\
\betas  &= 1.5 \ \ \text{(power law index)}
\\
\rcuts &= \frac{\rcutg}{2}.
\end{split}
\end{equation}
Here, $\sigmas0$ allows the mass of solid disk to be fixed as $\msoliddisk = \mgasdisk \ \fpg $. The dust-to-gas ratio, $\fpg$, is a Monte Carlo variable (see Sect. \ref{sec:ngpps}). The rock-to-ice ratio, $f_\text{s}(r)$, is the ratio of condensed solid to total solids \citep[following][]{Thiabaud2014}.


                \subsection{Planet formation}
                \label{bernmodelappendix_planetformation}

        \subsubsection{Accretion of solids}
Protoplanetary embryos accrete planetesimals from their feeding zone, which is defined as an annulus \editminor{on each side of the embryos' orbit}. The width of \editminor{the feeding zone} is given in terms of the Hill radius $\rhill$:
\begin{equation}
\label{eq:rhill}
\begin{split}
r_\text{feed} & = \mu \ \rhill, \hspace{5em} \text{($\mu$ = 5, \editminor{\cite{Fortier2013}})}
\\
\text{where,} \hspace{2em} \rhill   & =  a_M \left( \frac{\mplanet}{3\mstar} \right)^{1/3}.
\end{split}
\end{equation}
Competition for solids occurs when the feeding zone of multiple planets overlap \citep{Alibert2013}.
The overlapping feeding zone, with surface density $\sigmamean$, is separated into individual regions for each planet (see \paperone/). The accretion rate ($\mdotcore$) of planetesimals of spherical radius $\rplan$ by an embryo of core mass $\mcore$ depends on the probability of a protoplanet--planetesimal collision $\pcoll$, angular velocity $\Omega$ (which is $\sqrt{G \mstar/ r^3}$), $\sigmamean$, and $\rhill$:
\begin{equation}
\label{eq:coreaccretion}
\mdotcore = \Omega \ \sigmamean \ \rhill^2  \ \pcoll(\rplan, \rhill,  r_\mathrm{capture}, \vrel)
.\end{equation}
The collision rates, in turn, depend on the dynamical evolution of the solid disk. Planetesimals experience aerodynamic drag forces from the gas, and interact gravitationally with the protoplanets and amongst themselves. These interactions influence the relative velocity, $\vrel$, between the two colliding bodies. Additionally, when planets become massive ($\gtrsim \SI{1}{\mearth}$) their gas envelope affects the dynamics of penetrating solids. This results in an enhancement of the planetesimal capture radius, $r_\mathrm{capture}$. 

\subsubsection{Accretion of gases}
The accretion of gas by a planet depends on the local thermodynamical state of the protoplanetary disk and, interestingly, on the planetary interior as well. The internal structure of the planet is obtained by demanding conservation of mass, hydrodynamic equilibrium, and that energy diffusion be either radiative or convective. The demand for energy conservation is 
implemented through an iterative scheme that searches for a solution that is consistent with the boundary conditions (see \paperone/; \cite{Mordasini2012(models)}).

Initially, the gaseous envelope around all planets transitions smoothly into the nebular gas, the so-called {attached} phase (see panel (d) in Fig. \ref{fig:bernmodel}). In the attached phase gas from the nebular disk flows into the planet to compensate for planetary contraction. Planets contract as they cool down by radiating away the energy gained through accretion. The surface pressure and temperature of the planet are balanced with those of disk mid-plane When a planet reaches a critical threshold mass, large radiative losses cannot be balanced by accretional energy, and further \editminor{contraction} of the envelope ensues. This results in even more gas accretion, further increasing the accretional energy, and  runaway accretion of gas is inevitable. Consequentially, planets gain a massive envelope and very rapidly become giant planets.

For planetary cores massive enough to undergo runaway gas accretion, the rate of gas accretion may exceed the ability of the disk to supply gas (the maximum gas accretion rate). Then the envelope detaches from the gas disk and the planet continues to accrete in this {detached} phase.
In the detached phase gas accretion does not depend on the planetary internal structure but on the protoplanetary gas disk. The planet's radius contracts very rapidly to $\sim \rjupiter$ as it adjusts to the new boundary conditions. 

The last phase of planetary evolution, which is common to all planets, is the {isolated} phase, which occurs after the gas disk has dissipated. Gas accretion comes to a halt and planets will now contract as they cool down. 

                \section{\kobe/:  Additional details}
                \label{kobeappendix}
        \subsection{\kobeshadows/}
\label{subsec:kobeshadows}

\begin{figure*}
        \centering
        \includegraphics[width=8cm]{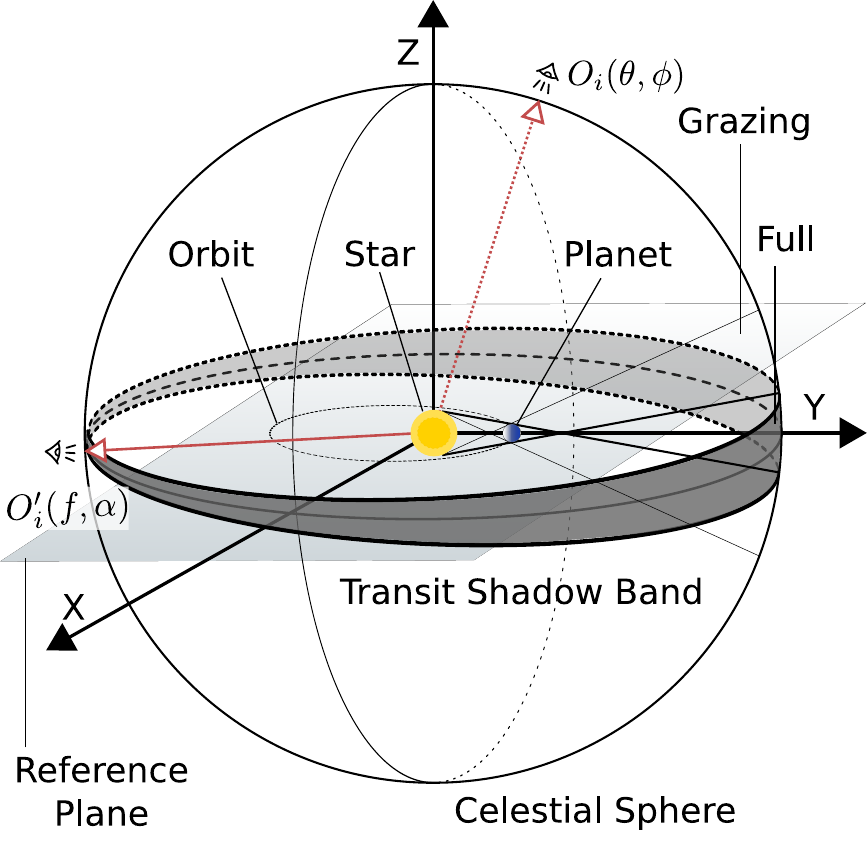}
        \includegraphics[width=8cm]{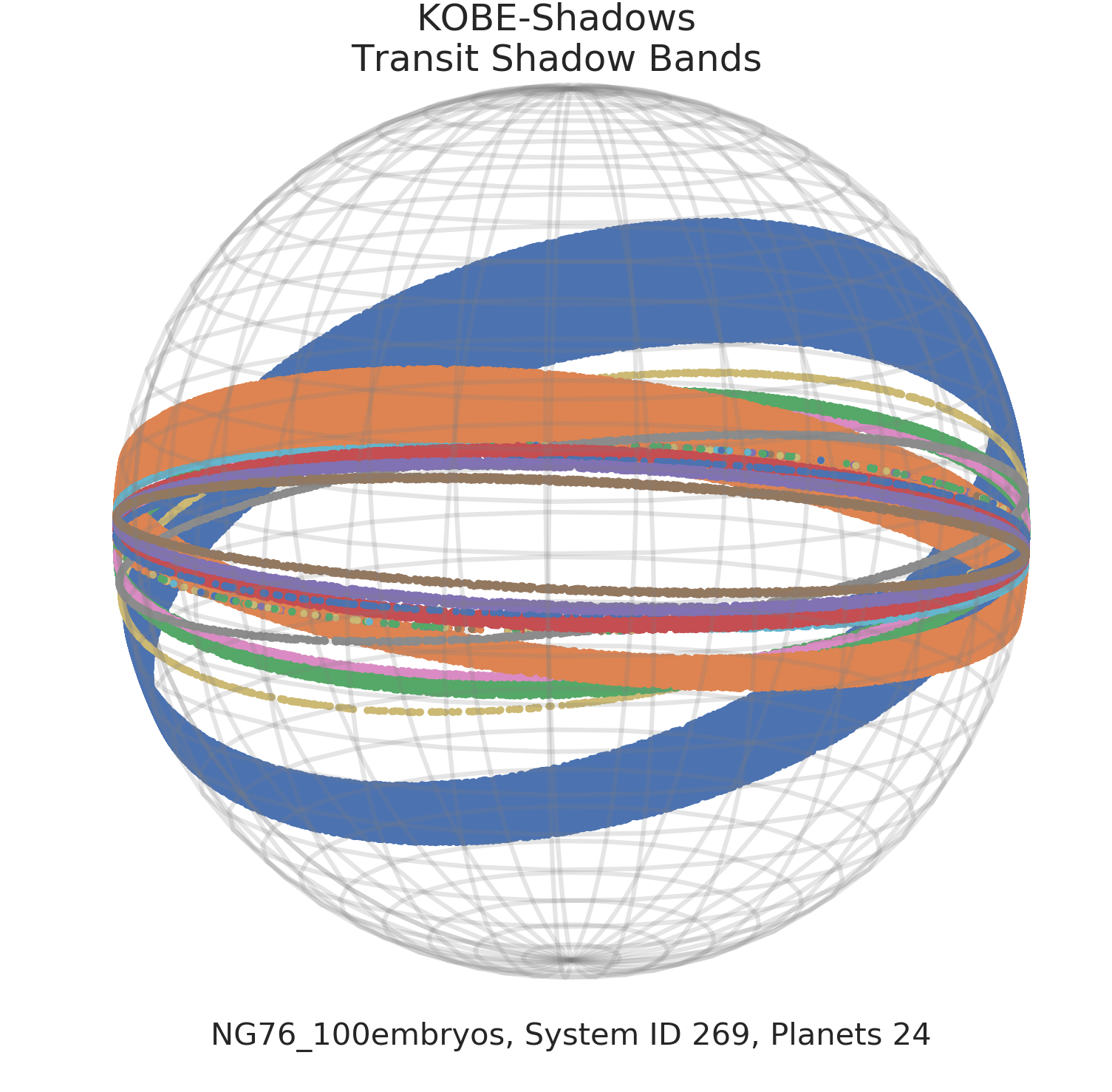}
        \caption{The geometry of transiting planets as implemented in \kobe/. Left: Transit geometry for a single star--planet system. The dashed red arrow represents the initial location of an observer $O_i (\theta, \phi)$, and the solid red arrow gives the rotated location of same observer $O_i (f, \alpha)$ (see text for details). 
                Right: Example of transit shadow bands for a 24-planet system (star and planets are not shown). \kobeshadows/ calculates whether an observer is present inside the full transit shadow band of any planet in this system. 
                The innermost planet has $P_\mathrm{tra} = 12 \%$ and is detectable by observers marked in blue. The second ($P_\mathrm{tra} = 6\%$) and the third innermost planet ($P_\mathrm{tra} = 2\%$) are detectable by observers (in orange and green, respectively).
                The probability to potentially detect any four planets simultaneously via transit for this system is $\sim 8 \%$, which is halved for simultaneously detecting any nine planets.
        }
        \label{fig:kobeshadows}
\end{figure*}

\begin{figure*}
	\centering
	\includegraphics[width=8cm]{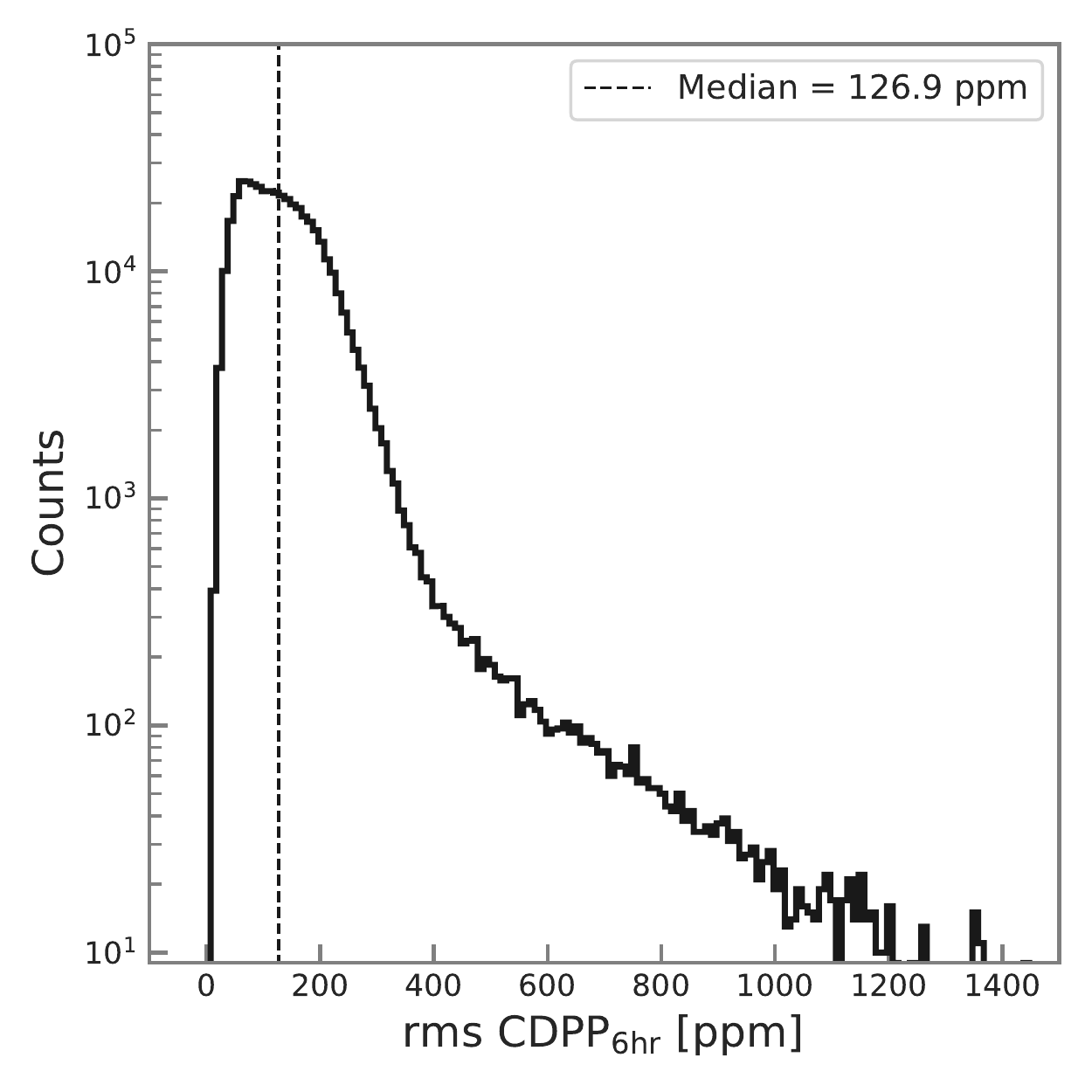}
	\includegraphics[width=8cm]{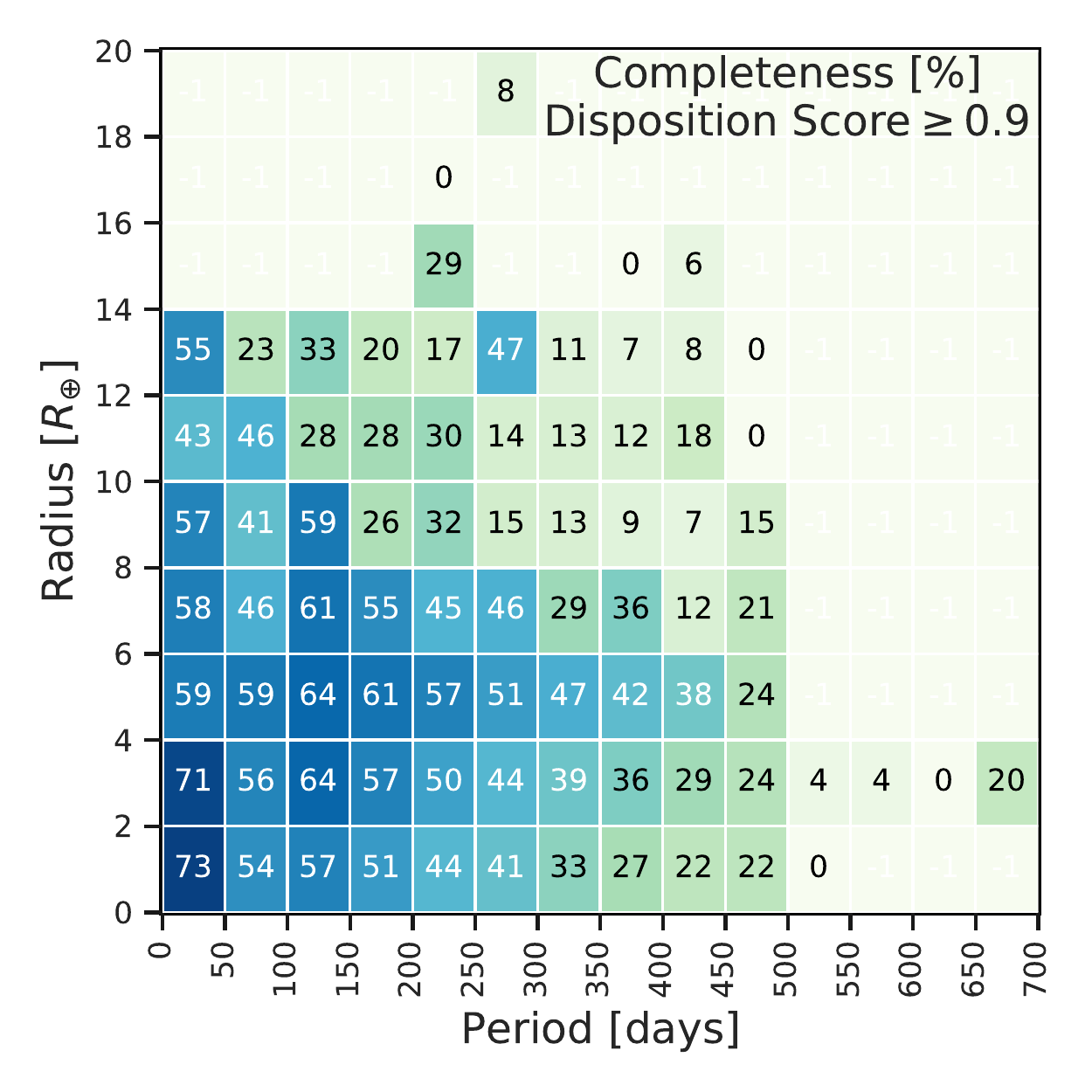}
	\caption{Inputs required by \kobe/ to simulate the Kepler transit survey. Left: Distribution of stellar noise, rms $\mathrm{CDPP}_{\SI{6}{\hour}}$, as seen by Kepler for FGK solar-type stars. Each individual \kobe/ system is assigned a CDPP value from this distribution.
		Right: Two-dimensional histogram depicting completeness of Kepler's Robovetter, as calculated by \kobevetter/. To emulate high reliability, a disposition score cut-off is placed. Labels on each bin indicate the completeness (in $\%$) for that bin. This calculation is based on the injected transit signals \citep{Coughlin2017}. Completeness is not calculated for bins with less than ten injected signals (not labelled). These bins are assigned a completeness of $0\%$.}
	\label{fig:cdppvetter}
\end{figure*}


Whether a planet can transit for a given observer is determined by \kobeshadows/. For this the {transit shadow band} (TSB) is determined by examining transit geometry, as shown in Fig. \ref{fig:kobeshadows} (left).
A planet orbiting a star will intercept some of the starlight, casting its shadow on a celestial sphere centred on this star. As the planet moves in its orbit the planet's shadow traces a shadow band on the celestial sphere. The area on the celestial sphere which falls inside a planet's shadow constitutes the TSB. 

All observers inside a planet's TSB can potentially detect this planet via its transit. This planet will not appear to be transiting to any observer who is outside the planet's TSB. \kobeshadows/ utilizes this distinction between transiting and non-transiting planets to detect the former. 

To compare with Kepler, simulating $\sim200\,000$ stars in NGPPS is computationally expensive. \kobeshadows/ offers a convenient solution.
Using transits, a single observer may never find all the planets in a planetary system. This could occur either when the TSBs from all the planets never overlap or when the location of an observer is outside the TSB of some planets. These are the basic geometrical limitations of the transit method. When the same system is observed by another observer located elsewhere, \editminor{they} may find different transiting planets than those found by the first observer. This means that two observers can view two different subsets of the same planetary system. By analysing the transit geometry of a planetary system for several observers, \kobeshadows/ generates multiple subsets of transiting systems (one from each observer). Subsequent modules in \kobe/ treat these subsystems, \kobe/ systems, as independent planetary systems. This allows \kobe/ to emulate observations of $200\,000$ stars from $1\,000$ NGPPS systems. Additional steps are taken in other modules to ensure that these system are treated independently\footnote{\editmajor{
                \kobe/ systems, being subsets of the NGPPS systems, are not truly independent. 
                In \kobe/ a single observer observing $1\,000$ synthetic planetary systems may detect $\sim100$ transiting planets around $\sim50$ stars. The strategy of using multiple observers in \kobe/ grew out of a need to (a) find more transiting planets, (b) imprint the effect of transit probability (via TSB calculation), and (c) systematize \kobe/'s approach. 
}}. \kobe/ systems with at least one transiting planet form the \kobeshadows/ catalogue, which is analysed in subsequent modules.

To compute the TSB for multiple planets in a system, \kobeshadows/ requires three things: 
First, the stellar and the planetary radii, $\left(\rstar, \rplanet\right)$. These are provided by the Bern Model and they evolve with time. Second, the orbital elements $\left(a, e, i, \Omega, \omega, \overline{f}\right)$. The six orbital elements are semi-major axis $a$, eccentricity $e$, inclination $i$, longitude of ascending node $\Omega$, argument of periapse $\omega$, and true anomaly $\overline{f}$. The first two elements describe the 2D shape of an elliptical orbit. The next three orbital elements define the relative orientation of an orbit in 3D space. The last element gives the position of a planet on its orbit with respect to the periastron\footnote{A planet's position on the orbit, $\overline{f}$, is not required to calculate its TSB because the calculation is done for \editminor{virtually} all positions on the orbit.}. The Bern Model provides all of these elements for all planets. However, an orbit's inclination as calculated in the model is with reference to the original protoplanetary disk, as opposed to any observer's reference sky-plane. These values are calculated only during the \editminor{\textit{N}-body} stage. A planet's period $P$ is calculated as $P^2 = \tfrac{4 \pi^2}{G \mstar} a^3 $. Finally, third, the location of an observer, $O_i (\theta, \phi)$. The location of an observer is specified in a standard coordinate system $(X, Y, Z)$. The origin is at the star (see Fig. \ref{fig:kobeshadows}, left). The  $X$-axis is along a reference line and $(X,Y)$ define a reference plane ($Z$ is perpendicular to this plane). An observer's location is specified in polar co-ordinates by an azimuth angle $\theta \in [0,2\pi]$ and a polar angle $\phi \in [0,\pi]$. The subscript $i$ is used to distinguish between multiple observers, and two observers are not allowed to be at the same location. The Bern Model does not provide the location of any special observer. It is assumed that the location of observers does not evolve with time. 

The  calculation procedure implemented in \kobeshadows/ is as follows: 

\begin{enumerate}
        \item The number of observers, $n_\mathrm{obs}$,  is fixed by the number of stars observed by a survey, $n_{\star, \mathrm{survey}}$, and the number of synthetic systems (\editminor{at time $t$}) in a population, $n_\mathrm{system} (t)$. The relation is
        \begin{equation}
        n_\mathrm{obs} = \frac{n_{\star, \mathrm{survey}}}{n_\mathrm{system} (t)}.
        \end{equation}
        Fixing $n_{\star, \mathrm{survey}}$ to $200\,000$ for Kepler, a population with $n_\mathrm{system} (t) = 1\,000,$ would require $200$ observers. Although all three populations
        simulate $1\,000$ systems, $n_\mathrm{system} (t)$ may diminish with time. For the nominal populations, at $\SI{4}{\giga\year}$, NG74 has 998, NG75 has 999, and NG76 has $1\,000$ systems. 
        \\
        \item The locations of the observers  
          are distributed uniformly and homogeneously around the celestial sphere in the $(X,Y,Z)$ coordinate system. This is done by generating two different random numbers $u_1, u_2 \in [0,1)$ and using inverse transform sampling to obtain polar coordinates:
        \begin{equation}
        \theta = 2 \pi u_1, \hspace{1em} \text{and} \hspace{1em} \phi = \mathrm{cos}^{-1}(1 - 2 u_2).
        \end{equation}
        This initial location of observers remains fixed for one NGPPS system. Spherical coordinates are converted to Cartesian giving $\vec{v}_{(X,Y,Z)}$, the \editminor{unit-normed} vectorial location of an observer. 
        \\
        \item To align the  transit geometry, the 
        Bern Model gives the position of a planet and its orbit in a different coordinate system, $(x,y,z)$ with origin at the star. Here, $x$ is along the periastron, $(x,y)$ are in the orbital plane, and $z$ points along the angular momentum vector of the planet. To proceed, the two coordinate systems must be aligned. This is done by rotating the initial location of all observers (given in $X,Y,Z$ coordinates) with respect to each planet's orbit (given in $x,y,z$ coordinates) via three rotations (for details see \cite{Murray2010})\footnote{Equivalently, the planet's orbit could be rotated, keeping the location of observers fixed.}:
        \begin{equation}
        \vec{v'}_{(x,y,z)} = R^{-1}_Z (\omega)\ R^{-1}_X (i)\ R^{-1}_Z (\Omega)\ \ \vec{v}_{(X,Y,Z)}.
        \end{equation}
        Here, $R^{-1}_X, R^{-1}_Y$, and $R^{-1}_Z$ are the inverses of the standard SO(3) rotation matrices. $R^{-1}_Z (\Omega)$ aligns the line of nodes with the $X$-axis. $R^{-1}_X (i)$ aligns the reference plane with the orbital plane. $R^{-1}_Z (\omega)$ rotates the reference plane such that the $X$-axis points along the periastron point of the orbit. The new location of the observer is given by the vector $\vec{v'}_{(x,y,z)}$ and in polar coordinates $O'_i (f, \alpha)$. In Fig. \ref{fig:kobeshadows} (left) an example of such rotation is shown. 
        
        Now the location of observers and the planetary orbits is known in the same coordinate system. This step ensures that the relative orientation between different planets is maintained, and information coming from all orbital elements is used.
        \\
        \item \editle{Next, we estimate the angular width of the TSB.} The semi-cone angle $\psi$, subtended by the planet's shadow as it blocks starlight at the azimuthal location of observer $O'_i (f, \alpha),$ is given by \citep{Winn2010}
        \begin{equation}
        \begin{split}
        \mathrm{sin} (\psi ) &= \left(\frac{\rstar \pm \rplanet}{\rplanetstardistance} \right), 
        \hspace{1em}
        \begin{cases}
        + \hspace{1em}\text{include grazing transits}
        \\
        - \hspace{1em}\text{exclude grazing transits}
        \end{cases}
        \\
        \rplanetstardistance &= \frac{a~(1 - e^2)}{1 + e\  \mathrm{cos}(f)}.
        \end{split}
        \end{equation}
        Here, $\rplanetstardistance$ is the star--planet distance when the planet's true anomaly is $\overline{f} = f$\footnote{For this paper, grazing transits are excluded.}. Finally, it is checked whether the observer $O'_i (f, \alpha)$ is inside the TSB through the following condition:
        \begin{equation}
        \begin{split}
        \alpha \in \left[\frac{\pi}{2} - \psi, \frac{\pi}{2} + \psi \right] &\rightarrow \text{Inside TSB},
        \\
        \text{otherwise} &\rightarrow \text{Outside TSB}.
        \end{split}
        \end{equation}
        This information is now stored for a planet at the original location of the observer $O_i (\theta, \phi)$.
        \\ 
        \item Repeat from step 4 for all $n_\mathrm{obs}$ observers. 
        \item Repeat from step 3 for all planets in a system. 
        \item Repeat from step 2 for all systems in a population.
        
\end{enumerate}

Figure \ref{fig:kobeshadows} (right) displays the result from \kobeshadows/ for a system with 24 planets in NG76. This calculation, done using $10^6$ observers and considering only full transits, takes about 1s. 

These calculations undergo two major consistency checks. Firstly, the area under a planet's TSB over the area of the celestial sphere gives the transit probability for this planet. Here, the number of observers found inside a TSB over $n_\mathrm{obs}$ gives a numerical proxy for the same. The analytical expression for transit probability, $P_\mathrm{tra}$, of a planet can be easily derived \citep{Barnes2007}:
\begin{equation}
\label{eq:transitprobability}
\begin{split}
P_\mathrm{tra} &= \frac{\rstar \pm \rplanet}{a \  (1 - e^2)}
\hspace{1em}
\begin{cases}
+ \hspace{1em}\text{include grazing transits}
\\
- \hspace{1em}\text{exclude grazing transits}
\end{cases}
\end{split}.
\end{equation}
The numerical and analytical values of $P_\mathrm{tra}$, with varying $n_\mathrm{obs}$ from \numrange{e5}{e8}, are found to be in good agreement. In addition, the transit probability for multiple planets is available for free via the numerical recipe described above. Finding the same through analytical approaches is a difficult problem. 

Secondly, the impact parameter of a transiting planet for all observers inside the TSB should be less than $1+ (\rplanet/\rstar)$.  The impact parameter, $b$, is the sky-projected star--planet distance expressed in units of stellar radii, and is given by
\begin{equation}
b = \frac{\rplanetstardistance}{\rstar} \ \mathrm{cos}(i).
\end{equation}
The impact parameter for an observer $O'_i (f, \alpha)$, is calculated by identifying $i=\alpha$. This condition is satisfied by all observers that are inside the TSB, for all planets, for all systems, in all three populations. 

The above procedure produces the \kobeshadows/ catalogue, which consists of \kobe/ systems containing at least one transiting planet. Although all of the planets in this catalogue will transit,  not all of them will be detected. \kobetransits/ takes care of this problem. 

\subsection{\kobetransits/}
\label{subsec:kobetransits}

\kobetransits/ examines the transit signal for all transiting planets found by \kobeshadows/. To be detected a transiting planet has to produce a signal that is strong enough to be detected by observers.  To check this, \kobetransits/ calculates the transit S/N. An estimate of the noise as seen by an observer is required for this.

In \kobetransits/ noise for a \kobe/ System is sampled from the distribution of rms combined differential photometric precision (CDPP). Produced by the Kepler pipeline (DR25) for individual target stars, CDPP is an empirical measure of the stellar photometric noise \citep{Christiansen2012}\footnote{This data is available in tabular format from the \hyperref{https://exoplanetarchive.ipac.caltech.edu/docs/Kepler_completeness_reliability.html}{category}{name}{NASA Exoplanet Archive}.}. Three cuts are placed on this distribution to ensure that FGK solar-type stars are sampled. These are on mass $\mstar \in [0.7, 1.3] \msun $, on radius $\rstar \le \SI{5}{\rsun}$, and on stellar temperature $\tstar \in [3880, 7200] \si{\kelvin}$ \citep{Pecaut2013}. 

The amount of stellar flux blocked by a planet is proportional to its area. The transit signal generated by a single transit is $\delta = (\rplanet^2/\rstar^2)$. The single transit S/N is
\begin{equation}
\begin{split}
S/N  &= \frac{\delta}{\cdppeff}, \hspace{1em} \text{single transit}
\\
\cdppeff &= \mathrm{CDPP}_{\ttrial} \Bigg( \frac{\ttrial}{\tdur} \Bigg)^{\frac{1}{2}}.
\end{split}
\end{equation}
Here, $\cdppeff$ is the effective stellar noise seen by an observer during a transit of duration $\tdur$. $\mathrm{CDPP}_{\ttrial}$ is the rms CDPP calculated by the  Kepler pipeline for different trials of transit durations (varying from \SIrange{1.5}{15}{\hour}). Following \W18, here $\ttrial = \SI{6}{\hour}$. Figure \ref{fig:cdppvetter} (left) shows the distribution of rms $\mathrm{CDPP}_{\SI{6}{\hour}}$ after placing the above cuts. To enhance the independent treatment of \kobe/ systems, a noise value is drawn randomly from the rms $\mathrm{CDPP}_{\SI{6}{\hour}}$ distribution for every star in the \kobeshadows/ catalogue.

The transit duration for a planet is estimated by the time taken by a planet to cross the stellar disk. \editminor{Following \W18, circular orbits and $b=0$ are assumed. This gives}
\begin{equation}
\tdur = \frac{2\rstar}{\big(\frac{2 \pi a}{P}\big)} = \frac{\rstar P}{\pi a}.
\end{equation}
When a planet's transit is observed $\ntransit$ times, the multi-transit S/N improves by a factor of $\sqrt{\ntransit}$. For a planet with period $P$ and transit survey of duration $t_\mathrm{survey}$, the average number of transits can be estimated as $\ntransit = t_\mathrm{survey} / P$. Then the multi-transit S/N generated by a transiting planet for $t_\mathrm{survey} = \tkepler = \SI{3.5}{\year}$ is 
\begin{equation}
\label{eq:transitsnr}
S/N =  \underbrace{\bigg(\frac{\rplanet}{\rstar}\bigg)^2}_{= \delta} \
\underbrace{\bigg(\frac{\rstar}{a}\bigg)^{\frac{1}{2}}}_{= P_\mathrm{tra}} \ 
\bigg[ \bigg(\frac{\tkepler}{\pi \ \ttrial}\bigg)^{\frac{1}{2}} \ \frac{1}{\mathrm{CDPP}_{\ttrial}}\bigg].
\end{equation}
This equation shows that the transit $S/N$ is given by the transit signal generated by a planet, \editbold{multiplied by a factor $\sqrt{\frac{\rstar}{a}}$}, scaled with instrument- and survey-related constants. Thus, a large planet closely orbiting a small quiet star will produce a high $S/N$. These are some of the detection biases of the transit method. 

In \kobetransits/ transiting planets that have $S/N \ge 7.1$ and $\ntransit \ge 2$ constitute the \kobe/-periodic threshold crossing event (\periodictce/) catalogue. For the Kepler pipeline, the threshold for detection of a TCE was multiple event statistic (MES, analogous to multi-transit S/N) $\ge 7.1 \sigma$, and $\ntransit \ge 3$ \citep{Twicken2016,Twicken2018, Christiansen2012, Thompson2018}. Following \W18, the minimum $\ntransit$ is fixed to 2.


\subsection{\kobevetter/}
\label{subsec:kobevetter}


In \cite{Thompson2018}, the DR25 TCE are further reviewed by an automatic program, the Robovetter. The Robovetter examines several metrics, and identifies consistent TCEs as Kepler objects of interest (KOIs). Through further analysis, the Robovetter eventually vets KOIs as either planet candidates or false positives. However, not all candidates are true exoplanets and some false positives may have been genuine signals. 

The catalogue produced by the Robovetter was characterized for completeness and reliability. Completeness measures the fraction of true planets that are absent in the catalogue, and reliability measures the fraction of planetary candidates that are truly exoplanets \citep{Coughlin2017}. These measurements are done by injecting simulated transit signals in 
the Robovetter. The Robovetter completeness is given by the fraction of injected transit signals that are characterized as planetary candidates. 
The confidence of vetting a TCE as a planetary candidate is expressed with a disposition score (ranging from \numrange{0}{1}, a higher value implying higher confidence that a TCE is a candidate). Selecting only high scoring candidates produces a catalogue that has less completeness, but is highly reliable.\footnote{\cite{Thompson2018} suggested using a cut on disposition score for occurrence rate studies. This approach has been used by \cite{Mulders2018} and \cite{Hsu2018}, and extensively studied by \cite{Bryson2019}.} 

\kobevetter/ calculates the Robovetter completeness (for disposition score $\ge 0.9$) using the results of these injection tests \citep{Coughlin2017}\footnote{The Robovetter Disposition Table is available from the \hyperref{https://exoplanetarchive.ipac.caltech.edu/docs/KeplerSimulated.html}{category}{name}{NASA Exoplanet Archive}.}. Completeness is calculated as a function of planetary radius (bin size $\SI{2}{\rearth}$) and period (bin size $\SI{50}{\day}$). Figure \ref{fig:cdppvetter} (right) shows a 2D histogram of completeness (including reliability). These values are applied to the \periodictce/ catalogue in a straightforward manner. If a 2D bin has completeness of $C\%$, then for all planets in the \periodictce/ catalogue that fall in this bin $C\%$ are randomly vetted as candidates and make the planetary candidates catalogue. The rest are rejected as false positives. For example, if there are 100 planets with $\rplanet \in [0,2) \rearth$ and $P \in [300,350)~\si{\day}$ in the \periodictce/ catalogue, then 33 planets will be randomly marked as candidates and the remaining 66 planets are vetted as false positives.

        \end{appendix}
        
\end{document}